\pdfoutput=1
\documentclass[11pt]{article}
\usepackage{cite}
 \usepackage {soul}
\usepackage{graphicx}
\usepackage{latexsym}   
\usepackage{mathrsfs}
\usepackage[scr=rsfs,cal=boondox]{mathalfa}
\usepackage[overload]{textcase}

\font\grande=cmr9.5 scaled \magstep4
\font\medio=cmr9.5 scaled \magstep2
\outer\def\beginsection#1\par{\medbreak\bigskip
      \message{#1}\leftline{\bf#1}\nobreak\medskip
\vskip-\parskip
      \noindent}

\begin{document}
\bibliographystyle{unsrt}

\titlepage

\vspace{1cm}
\begin{center}
{\grande The multiplicity distributions}\\
\vspace{0.5 cm}
{\grande of the high frequency gravitons}\\
\vspace{1.5 cm}
Massimo Giovannini\footnote{e-mail address: massimo.giovannini@cern.ch}\\
\vspace{0.5cm}
{{\sl Department of Physics, CERN, 1211 Geneva 23, Switzerland }}\\
\vspace{0.5cm}
{{\sl INFN, Section of Milan-Bicocca, 20126 Milan, Italy}}
\vspace*{1cm}
\end{center}
\vskip 0.3cm
\centerline{\medio  Abstract}
\vskip 0.5cm
When the gravitons are created thanks to the variation of the space-time curvature 
the multiplicity distributions quantify the probability of producing a given number of species around the average occupation numbers of the underlying multiparticle states. The multiplicities of the relic gravitons are infinitely divisible and their analytical expressions encompass, in different regions of the physical parameters, the Poisson, Bose-Einstein, Gamma and Pascal probability distributions. Depending upon the post-inflationary timelines, the averaged multiplicities are controlled by the range of the comoving frequencies and they can be much larger in the GHz region than in the aHz domain where the temperature and polarization anisotropies of the microwave background set strict limits on the spectral energy density of the relic gravitons. Thanks to the absolute upper bound on the maximal frequency of the gravitons we can acknowledge that the averaged multiplicities are exponentially suppressed above the THz, where only few pairs of gravitons are independently created from the vacuum. The statistical properties of the produced particles can be finally interpreted in the light of the second-order interference effects and this perspective is quantitatively scrutinized by analyzing the associated quantum sensitivities.  For putative instruments reaching spectral amplitudes ${\mathcal O}(10^{-35})/\sqrt{\mathrm{Hz}}$ in the MHz or GHz bands, the Bose-Einstein correlations could be used to probe the properties of cosmic gravitons and their super-Poissonian statistics. Both from the theoretical and observational viewpoint it is unjustified to require the same sensitivities (e.g. ${\mathcal O}(10^{-21})/\sqrt{\mathrm{Hz}}$) in the kHz and GHz domains.
\noindent
\vspace{5mm}
\vfill
\newpage

\renewcommand{\theequation}{1.\arabic{equation}}
\setcounter{equation}{0}
\section{Motivations and goals}
\label{sec1}
In the current cosmological lore (dubbed, by some, concordance paradigm) the determinations of the pivotal observables from different experiments have been so far mutually compatible. After the first measurements of the temperature and polarization anisotropies of the Cosmic Microwave Background (CMB in what follows) \cite{LL0}, in the last twenty years the inferred parameters converged to their present values \cite{LL1,LL2,LL3,LL4} (see also the recent perspective of Ref. \cite{LL5}). In its simplest formulation the concordance paradigm assumes the presence of an early inflationary stage of expansion so that the temperature and the polarization anisotropies of the CMB are ultimately the result of adiabatic and Gaussian initial conditions both for the curvature inhomogeneities and for the tensor modes of the geometry. The tensor contribution (still to be directly observed) does not only affect the polarization anisotropies (as sometimes inaccurately stated) but it also modifies the temperature inhomogeneities and the cross-correlations between the temperature and the polarization power spectra. For this reason the concurrent analyses of the different observables set important limits on the ratio between the power spectra of the scalar and tensor modes of the geometry ($r_{T}$ in what follows) at the (conventional) pivot scale $k_{p}$ which we shall take to be $0.002\, \mathrm{Mpc}^{-1}$. This scale actually corresponds to typical comoving frequencies of the order of $\nu_{p} = k_{p}/(2\pi) = 3.09 \, \mathrm{aHz}$ (where $1\, \mathrm{aHz} = 10^{-18}$ Hz) and the relative weight of the tensor and scalar power spectra (denoted hereunder by $r_{T}$) constrains the conventional inflationary scenarios. For the present ends, according to the current observations we should always require that $r_{T} \leq {\mathcal O}(0.03)$ for $\nu = {\mathcal O}(\nu_{p})$ \cite{LL1,LL2,LL3,LL4,LL5}. 

On a general ground the random backgrounds of gravitational radiation may have either a classical or a quantum origin but while classical fluctuations are given once forever (on a given space-like hypersurface) quantum fluctuations keep on reappearing during the inflationary stage that takes place during the early stages of the concordance scenario (see, for instance, \cite{AS1,AS2} and references therein). As the 
accelerated expansion develops, the classical inhomogeneities get progressively suppressed \cite{AS3,AS4} and the initial conditions for the evolution of the scalar and tensor modes of the geometry are eventually set by quantum mechanics. The result of this process is not a generic random (or stochastic) background of gravitational radiation but rather the creation of gravitons that are excited by the evolution of the space-time curvature, as argued long ago even before the formulation of the inflationary scenarios \cite{AS5,AS6,AS7,AS8}. In spite of the inaccurate terminologies what ultimately counts are the underlying physical features: the random backgrounds of cosmological origin are non-stationary \cite{AS9} and the relic gravitons 
are the result of a quantum process leading to  
spectral energy densities that vanish in the limit $\hbar\to 0$ \cite{AS10}. 
When many particles are created the key observables are provided by the multiplicity distribution of the produced species and this is true in different contexts ranging from quantum optics \cite{MD1,MD2} to high-energy physics \cite{MD3,MD4}. The multiplicity distribution ultimately estimates the probability of producing $n$ particles in a given process and it is typically expressed as a function of the averaged multiplicity of the species under consideration. For a given multiplicity distribution all the associated factorial moments 
are uniquely defined \cite{KAS}. Furthermore, because of its specific construction the multiplicity distribution is directly related to the properties of the production mechanism as firstly suggested by Parker \cite{MD5,MD6} both in the case of gravitational and in strong interactions \cite{MD7}. In this paper we are going to analyze the multiplicity distributions of the relic gravitons and demonstrate that their properties eventually permit in principle (if not yet in practice) to distinguish between random backgrounds of quantum and classical origins.

The multiplicity distributions of the relic gravitons are characterized by the average occupation number and by the dispersion of the underlying quantum state. However the averaged multiplicity, the dispersion 
and all the factorial moments of the distribution are determined by the properties of the initial state, by the specific interval of the comoving frequency and, ultimately, by the timeline of the expansion rate. There is actually a direct connection between the averaged multiplicity and the spectral energy density in critical units evaluated at the present time [$\Omega_{gw}(\nu,\tau_{0})$ in what follows]. In the concordance paradigm $\Omega_{gw}(\nu,\tau_{0})$ is quasi-flat \cite{FL1,FL2,FL3} except for a low-frequency  region where $\Omega_{gw}(\nu,\tau_{0})\propto (\nu_{\mathrm{eq}}/\nu)^{2}$ and $\nu_{\mathrm{eq}} = {\mathcal O}(50)\,\mathrm{aHz}$ \cite{FL4}. The approximate scale invariance of $\Omega_{gw}(\nu,\tau_{0})$ realized between few aHz and ${\mathcal O}(200)$ MHz provided conventional inflationary stage 
is supplemented by a post-inflationary timeline {\em always dominated by radiation} until matter-radiation equality: if the post-inflationary expansion rate is slower than radiation $\Omega_{gw}(\nu, \tau_{0})$ can be sharply increasing \cite{FL5}.
The currently operating wide-band interferometers set limits on $\Omega_{gw}(\nu, \tau_{0})$ 
for typical frequencies in the audio range (i.e. between few Hz and $10$ kHz) 
\cite{AU1,AU2,AU3a,AU3} and these bounds roughly demand $\Omega_{gw}(\nu, \tau_{0})< {\mathcal O}(10^{-9})$ \cite{AU4} for a quasi-flat spectrum whose 
amplitude can be theoretically estimated as ${\mathcal O}(10^{-17})$ \cite{AU5}.
Depending on the post-inflationary evolution it can however happen that 
the spectral energy density is larger than in the 
conventional paradigm \cite{AU5} for frequencies higher than the kHz 
but always smaller than the THz which is the upper bound 
on the frequency of relic gravitons \cite{AU6}. This is why it is particularly 
interesting to consider the properties of the multiplicity distributions 
in the high and ultra-high frequency regions where, incidentally, 
waveguides \cite{WG1,WG2,WG3}, microwave cavities with superconducting walls
\cite{CAV1,CAV2,CAV3,CAV4,CAV5,CAV6,CAV7} laser interferometers \cite{WL1,WL2} can be used for the detection of relic gravitons between the MHz and the THz, as suggested long ago \cite{QQ1} (see also \cite{QQ3,QQ4}). The high frequency instruments might be relevant for detecting single gravitons \cite{FDD1} in the context of what it is often dubbed quantum sensing \cite{QS1,QS2,QS3} even if, as we shall see, the sensitivities required for the chirp amplitudes and for the spectral amplitudes are $10$ or even $15$ orders of magnitude smaller than the ones currently assumed at high frequencies for building prototype detectors.

In short the layout of this investigation is the following. In section \ref{sec2} the descriptions of the tensor modes of the geometry are reviewed in the quantum mechanical perspective motivated by the concordance scenario. Section \ref{sec3} is devoted to the general derivation of the multiplicity distributions of the relic gravitons. The statistical properties deduced  in section \ref{sec3} are scrutinized  
in section \ref{sec4} with the purpose of showing that the multiplicity distributions are infinitely divisible. The connection between the multiplicity 
distributions and the second-order interference effects is also outlined. The concrete examples are collected in section  
\ref{sec5} where the discussion is focused on the interplay between the shapes of the multiplicity distributions and the different domains of comoving frequencies. Particular attention is paid to the ranges that are close to the maximal frequency (i.e. $\nu_{\mathrm{max}} = {\mathcal O}(\mathrm{THz})$). It is argued that the statistical features discussed in section \ref{sec4} also emerge by considering  the frequency dependence of  the averaged multiplicities and of the dispersions in a more phenomenological perspective. We also analyze the correlated and uncorrelated limits as a function of the parameters of the post-inflationary evolution.  We finally compute the minimal chirp and spectral amplitudes required by the potential detection of high frequency gravitons obeying the multiplicity distributions deduced here. The  concluding remarks are collected in section \ref{sec6}. For the sake of conciseness (and to avoid lengthy 
digressions) two technical results (both relevant for the analyses) have been relegated to the appendices \ref{APPA} and \ref{APPB} that could be however useful for the potentially interested readers.

\renewcommand{\theequation}{2.\arabic{equation}}
\setcounter{equation}{0}
\section{Relic gravitons, random backgrounds  and their description}
\label{sec2}
In the concordance paradigm the dominant source of large-scale inhomogeneities is provided by adiabatic and Gaussian fluctuations of the spatial curvature and this observation is compatible with a quantum mechanical origin of the curvature perturbations during an inflationary epoch \cite{LL0,LL1,LL2,LL3,LL4,LL5}. Indeed, a stage of accelerated expansion irons the classical inhomogeneities \cite{AS3,AS4} so that only quantum mechanics may provide the relevant initial conditions of the gravitational fluctuations. The spatial gradients and the curvature possibly present at the onset of the inflationary 
stage are suppressed during an accelerated stage although they may increase when the background decelerates \cite{BK1,BK2}. This approximate timeline effectively realizes a suggestion originally due to Sakharov \cite{BK3} stipulating that the initial conditions of large-scale inhomogeneities should be related with quantum fluctuations.  The same kind of comment holds also in the case of the tensor modes of the geometry \cite{FL1,FL2,FL3} that are efficiently excited in curved backgrounds \cite{AS5,AS6,AS7}.  However, depending upon the specific value of $r_{T}$, the tensor power spectra are ${\mathcal O}(100)$ times smaller than the one of the scalar modes for typical pivot scale $k_{p} = 0.002 \, \mathrm{Mpc}^{-1}$ corresponding to comoving frequencies $\nu_{p} = k_{p}/(2\pi) = {\mathcal O}(3)\, \mathrm{aHz}$. The wavelengths associated with $k_{p}$, i.e. $\lambda_{p} = 2\pi/k_{p}$ 
left the Hubble radius approximately $N_{k} = {\mathcal O}(60)$ $e$-folds before the end of inflation: the number of $e$-folds controls at once the suppression of the spatial gradients and the suppression of $r_{T}$. The value of $N_{k}$ also depends on the post-inflationary evolution: when the post-inflationary timeline is faster than radiation 
$N_{k} < 60$ \cite{KK1,KK2} while $N_{k} > 60$ is the expansion rate is slower than radiation \cite{KK3}. Overall there is an indetermination of ${\mathcal O}(15)$ $e$-folds on the value of $N_{k}$ and this observation also affects the high frequency slope of the spectrum of relic gravitons and the explicit expression of the maximal frequency (see, in this respect, the discussion of section \ref{sec5}). In the present section, for the sake of completeness, we are going to summarize the essentials 
of the quantum description of the relic gravitons with particular attention to the subjects that are directly relevant for the derivation of the multiplicity distributions. Natural units $\hbar = c = k_{B} =1$ 
(where $k_{B}$ is the Boltzmann constant) are employed throughout; in these units $M_{P} = 1.22\times 10^{19} \mathrm{GeV}$. We shall also use the notation $\overline{M}_{P}= M_{P}/\sqrt{8\pi} = 1/\ell_{P}$ where $\ell_{P}$ indicates the Planck length. Finally, the scale factor at the present time $\tau_{0}$ is normalized as $a_{0} =1$ implying that, at $\tau_{0}$, comoving and physical frequencies coincide.

\subsection{The tensor actions}
In Einstein-Hilbert gravity the second-order 
action for the tensor modes of the geometry can be expressed as \cite{AS7} (see also Refs. \cite{AU5} and \cite{BK4,BK5})
\begin{equation}
S_{g} = \frac{1}{8 \ell_{P}^2} \int d^{4} x \, \sqrt{- g} \, \overline{g}^{\mu\nu} \, \partial_{\mu} h_{i\,j} 
\partial_{\nu} h^{i\, j},
\label{AC1}
\end{equation}
where $h_{i\, j}$ denotes the (solenoidal and traceless) tensor modes of the geometry (i.e. $\partial_{i}\, h^{i}_{\,\,j} = h_{i}^{\,\,i} =0$ in the spatially flat case).  Indeed, the current observational data suggest a spatially flat  geometry \cite{LL0,LL1,LL2,LL3,LL4,LL5} and Eq. (\ref{AC1}) applies to case of a conformally flat background metric $\overline{g}_{\mu\nu} = a^2(\tau) \eta_{\mu\nu}$ where $\eta_{\mu\nu}$ is the Minkowski metric, $a(\tau)$ indicates the scale factor and $\tau$ denotes the conformal time coordinate. In this situation the first-order (tensor) fluctuations of $\overline{g}_{\mu\nu}$ are $\delta^{(1)}_{t} g_{i\, j} = - a^2(\tau) \,\,h_{i\,j}$. Taking into account these specifications, a more explicit form of Eq. (\ref{AC1}) can be written as:
\begin{equation}
S_{g} = \frac{1}{8 \ell_{P}^2} \int d^{4} x \, a^2(\tau) \eta^{\mu\nu}\, \partial_{\mu} h_{i\,j} 
\partial_{\nu} h^{i\, j} \to \frac{1}{8 \ell_{P}^2} \int d^{4} x \, a^2(\tau) \biggl[\partial_{\tau} h_{i\,j} \partial_{\tau} h^{i\, j} - \partial_{k} h_{i\,j} \partial_{k} h^{i\, j}\biggr].
\label{AC1b}
\end{equation}
The actions of Eqs. (\ref{AC1})--(\ref{AC1b}) make perfect sense within the concordance scenario,
the results discussed hereunder hold also in more general situations. We shall examine, in particular, 
the case of conformally related frames and the potential presence of a refractive 
index. 

\subsection{Potential generalizations}
We are now going to argue that the general form of Eqs. (\ref{AC1})--(\ref{AC1b})
is preserved under a conformal rescaling of the original metric. This means, in particular, 
that the analysis of the tensor modes of the geometry is not frame dependent (see also \cite{AU5}). For a conformal rescaling of the four-dimensional geometry we have 
\begin{equation}
g_{\mu\nu}(x) \to G_{\mu\nu}(x) = q(x) g_{\mu\nu}, \qquad \sqrt{-g} \to \sqrt{- G} = q^2(x)\, \sqrt{-g},
\label{CC1a}
\end{equation}
where $x$ now denotes a generic space-time point $x= (\tau, \, \vec{x})$.
Equation (\ref{CC1a}) can be separately written for the background metric and for its tensor fluctuations since $G_{\mu\nu}= \overline{G}_{\mu\nu} + \delta^{(1)}_{t} G_{\mu\nu}$ and 
$g_{\mu\nu}= \overline{g}_{\mu\nu} + \delta^{(1)}_{t} g_{\mu\nu}$. In terms of original metric the explicit expressions of $\overline{G}_{\mu\nu} $ and  $\delta^{(1)}_{t} G_{\mu\nu}$ are, respectively, 
$\overline{G}_{\mu\nu}= q(\tau) \overline{g}_{\mu\nu}$ and $\delta^{(1)}_{t} G_{i\,j}= q(\tau) \delta^{(1)}_{t} g_{i\,j}$. But since in the conformally related frame we have that $\overline{G}_{\mu\nu} = \widetilde{a}^2(\tau) \eta_{\mu\nu}$ [and $\delta^{(1)}_{t} G_{i\, j} = - \widetilde{a}^2(\tau)\,\, \widetilde{h}_{i\,j}$],  the action (\ref{AC1}) transforms as
\begin{eqnarray}
S_{g} \to \widetilde{S}_{g}  = \frac{1}{8\,\ell_{P}^2}\,\int d^{4} x \, \frac{\sqrt{- \overline{G}}}{q(\tau)} \,\,\overline{G}^{\mu\nu} \,\,\partial_{\mu} \widetilde{h}_{i\,j} \partial_{\nu} \widetilde{h}^{i\,j} =  \frac{1}{8 \ell_{P}^2} \int d^{4} x \,\widetilde{a}^2(\tau)\eta^{\mu\nu}\, \partial_{\mu} \widetilde{h}_{i\,j} 
\partial_{\nu} \widetilde{h}^{i\, j}.
\label{AC1bb}
\end{eqnarray}
The result of Eq. (\ref{AC1bb}) follows since, at the background level, $\widetilde{a}(\tau)=\sqrt{q(\tau)} \, a(\tau)$
while for the first-order fluctuations  $\widetilde{h}_{i\,j} =  h_{i\,j}$. Therefore Eqs. (\ref{AC1b}) and (\ref{AC1bb}) not only share the same form but they are 
also physically equivalent\footnote{It also follows from Eq. 
(\ref{CC1a}) that  $\delta^{(1)}_{t} G_{i\, j} = q(\tau) \delta^{(1)}_{t} g_{i\, j}$ and if we recall the expression of the first-order fluctuations in the original frame (i.e. $\delta^{(1)}_{t} g_{i\, j} = - a^2(\tau) \,\,h_{i\,j}$) we get, as stressed after Eq. (\ref{CC1a}),  $\delta^{(1)}_{t} G_{i\, j} = - \widetilde{a}^2(\tau)\,\, \widetilde{h}_{i\,j}$. The tensor modes defined in the two frames coincide (i.e. $\widetilde{h}_{i\,j} =  h_{i\,j}$) and their description is frame-independent.}.

An action formally similar to Eq. (\ref{AC1}) may also arise from a completely different
physical situations not necessarily corresponding to the Einstein-Hilbert theory. For instance when the gravitons inherit an effective refractive index coming from 
the interactions with the background curvature the action (\ref{AC1}) gets modified as \cite{SZ1,SZ2,SZ3}
\begin{equation}
S_{g} = \frac{1}{8 \ell_{P}^2} \int d^{3} x\, \int d\tau \biggl[ a_{1}^2(\tau)\, \partial_{\tau} h_{i\,j} \partial_{\tau} h^{i\,j} - a_{2}^2(\tau) \,\partial_{k} h_{i\,j} \partial^{k} h^{i\,j} \biggr].
\label{AC2}
\end{equation}
In writing Eq. (\ref{AC2}) we assumed that the interactions with the curvature do not 
introduce higher-order time derivatives.  The presence of quadratic corrections to the Einstein-Hilbert action in the Gauss-Bonnet form may lead to $a_{1}(\tau) \neq a_{2}(\tau)$ \cite{SZ3}. In the case $a_{1}(\tau) = a_{2}(\tau) = a(\tau)$ we recover the standard form of Eq. (\ref{AC1}). We want now to show that, although $a_{1}(\tau) \neq a_{2}(\tau)$,  the action (\ref{AC2}) can be recast in the same form of Eq. (\ref{AC1b}) provided the conformal time coordinate $\tau$ is replaced by a more general 
time parametrization. To clarify this point we first note that
Eq. (\ref{AC2}) can be expressed as:
\begin{equation}
S_{g} = \frac{1}{8 \ell_{P}^2} \int d^{3} x\, \int d\tau \,  a_{1}^2(\tau)\,\biggl[ \partial_{\tau} h_{i\,j} \partial_{\tau} h^{i\,j} - \frac{1}{n^2(\tau)} \,\partial_{k} h_{i\,j} \partial^{k} h^{i\,j} \biggr],
\label{AC3}
\end{equation}
where now $n(\tau) = a_{1}(\tau)/a_{2}(\tau)$ plays the r\^ole of an effective refractive 
index accounting for the interactions with the space-time curvature. Equation (\ref{AC3})
can be rewritten in the same form of the original Ford-Parker action (\ref{AC1b}) but with a modified 
time coordinate (conventionally referred to hereunder as the $\eta$-time) \cite{SZ3}:
\begin{equation}
S_{g} = \frac{1}{8 \ell_{P}^2} \int d^{3} x\, \int d\eta  \,  b^2(\eta) \biggl[ \partial_{\eta} h_{i\,j} \partial_{\eta} h^{i\,j} - \partial_{k} h_{i\,j} \partial^{k} h^{i\,j} \biggr].
\label{AC4}
\end{equation}
The $\eta$-time is introduced from the relation $n(\eta) \,d\eta = d\tau$ and 
in Eq. (\ref{AC4})  $b(\eta) = a_{1}(\eta)/\sqrt{n(\eta)} = \sqrt{a_{1}(\eta)\, a_{2}(\eta)}$. 
Equation (\ref{AC4})  encompasses various physical situations
and the results obtained in this framework can be reduced to the case of Eq. (\ref{AC1}) by 
replacing $\eta \to \tau$, $n(\eta) \to 1$ and $b(\eta) \to a(\tau)$.
The canonical form of Eq. (\ref{AC4}) is obtained after a redefinition of the tensor modes as $b(\eta)\, h_{i\, j} = \mu_{i\, j}$ 
\begin{equation}
S_{g} = \int d^{3} x \, \int d \eta \, {\mathcal L}_{g}(\vec{x}, \eta),
\label{AC5}
\end{equation}
where the Lagrangian density ${\mathcal L}_{g}(\vec{x}, \eta)$ is given by:
\begin{equation}
{\mathcal L}_{g}(\vec{x},\eta) = \frac{1}{8 \ell_{P}^2}\biggl[ \partial_{\eta} \mu_{i j}\,\, \partial_{\eta} \mu^{i j} -  {\mathcal F} \partial_{\eta} \mu_{i j} \, \mu^{i j}  - {\mathcal F} \partial_{\eta} \mu^{i\, j} \, \mu_{i\, j} + {\mathcal F}^2 \mu_{i j} \, \mu^{i j} - \partial_{k} \mu_{i j} \,\partial^{k} \mu^{i j} \biggr].
\label{AC6}
\end{equation}
In Eq. (\ref{AC6}) we have that ${\mathcal F} = (\partial_{\eta}b)/b$ and $\partial_{\eta}$ denotes throughout a derivation with respect to the $\eta$-time coordinate. The canonical momenta $\pi_{i j}  = [\partial_{\eta} \mu_{i j} - {\mathcal F} \mu_{i j}]$ 
and $\pi^{i j}  = [\partial_{\eta} \mu^{i j} - {\mathcal F} \mu^{i j}]$ follow from Eq. (\ref{AC6}) so that  the canonical Hamiltonian becomes
\begin{eqnarray}
H_{g}(\eta) &=& \int d^3 x \biggl[ \pi_{i j} \,\, \mu^{i j} + \mu_{i j}\,\, \pi^{i\,j} - {\mathcal L}_{g}(\vec{x},\eta)\biggr],
\label{AC7}\\
&=& \int d^{3} x\, \biggl[ 8 \ell_{P}^2 \, \pi_{i j} \,\pi^{i j} + {\mathcal F}\biggl( \mu_{i j} \,\pi^{i j}
+  \mu^{i j} \,\,\pi_{i j} \biggr)+ \frac{1}{8\ell_{P}^2} \partial_{k} \mu_{i j} \,\,\partial^{k} \mu^{i j} \biggr].
\label{AC8}
\end{eqnarray}
In Fourier space the canonical fields and the associated momenta are:
\begin{equation}
\mu_{i j}(\vec{x}, \eta) =\frac{\sqrt{2}\, \ell_{P}}{(2\pi)^{3/2}} \int d^{3} k \, \mu_{i\, j}(\vec{k}, \eta) \, e^{- i \vec{k}\cdot\vec{x}}, \qquad \pi_{i j}(\vec{x}, \eta) = \frac{1}{4 \, \sqrt{2} \, \ell_{P} (2\pi)^{3/2}}  \int d^{3} k \, \pi_{i\, j}(\vec{k}, \eta) \, e^{- i \vec{k}\cdot\vec{x}}.
\label{AC9aa}
\end{equation}
If we now introduce a triplet of orthogonal vectors $\hat{m}$, $\hat{n}$ and $\hat{k}$ (with 
$\hat{m} \times \hat{n} = \hat{k}$) the Fourier amplitudes can be expanded 
in the basis of the two tensor polarizations as: 
\begin{eqnarray}
\mu_{i j}(\vec{k}, \eta) = \sum_{\beta}\,\,e^{(\beta)}_{i\,j}(\hat{k}) \, \mu_{\vec{k}, \,\beta}(\eta),\qquad
\pi_{i j}(\vec{k}, \eta) = \sum_{\beta}\,\, e^{(\beta)}_{i\,j}(\hat{k}) \, \pi_{\vec{k}, \,\beta}(\eta),
\label{AC10}
\end{eqnarray}
where $\beta = \oplus,\, \otimes$ denote the (two) tensor polarizations; in terms of $\hat{m}$, $\hat{n}$  we have that $e^{(\oplus)}_{i\,j}$ and 
$e^{(\otimes)}_{i\,j}$ are: 
\begin{equation}
e^{(\oplus)}_{i\,j}(\hat{k}) = \hat{m}_{i} \, \hat{m}_{j} - \hat{n}_{i} \, \hat{n}_{j}, \qquad 
e^{(\otimes)}_{i\,j}(\hat{k}) = \hat{m}_{i} \, \hat{n}_{j} + \hat{n}_{i} \, \hat{m}_{j}.
\label{AC11}
\end{equation}
Note that, by definition,  $\hat{k}^{i}\,e^{(\alpha)}_{i\,j} =  \hat{k}^{j}\,e^{(\alpha)}_{i\,j} =0$ and that $e^{(\alpha)}_{i\,j}\, \, e^{i\,j}_{(\alpha^{\prime})} = 2 \,\,\delta^{\alpha}_{\,\,\alpha^{\prime}}$.

\subsection{Classical and quantum fields}
Once the classical fields introduced in Eqs. (\ref{AC9aa})--(\ref{AC10}) are promoted 
to the status of Hermitian operators [i.e.  $\mu_{i j}(\vec{x}, \eta) \to \widehat{\mu}_{i\, j}(\vec{x},\eta)$ and $\pi_{i j}(\vec{x}, \eta) \to \widehat{\pi}_{i\, j}(\vec{x},\eta)$)] the Fourier 
amplitudes $\widehat{\mu}_{\vec{k},\alpha}$ and $\widehat{\pi}_{\vec{k},\alpha}$ can be expressed in terms of the corresponding creation and annihilation operators as: 
\begin{equation}
\widehat{\mu}_{\vec{k},\, \alpha} = \frac{1}{\sqrt{2 k}} \biggl( \widehat{a}_{\vec{k},\, \alpha} + \widehat{a}_{-\vec{k},\, \alpha}^{\dagger}\biggr), \qquad \widehat{\pi}_{\vec{k},\, \alpha} =  i \,\sqrt{\frac{k}{2}} \biggl( \widehat{a}_{-\vec{k},\, \alpha}^{\dagger} - \widehat{a}_{\vec{k},\, \alpha} \biggr).
\label{AC12}
\end{equation}
Since, by definition, $[ \widehat{a}_{\vec{k}, \, \alpha}, \,  \widehat{a}_{\vec{p}, \, \beta}^{\dagger}] = \delta^{(3)}(\vec{k} - \vec{p})$, from Eq. (\ref{AC12}) it also follows that the commutation relations at equal time are
\begin{equation}
\bigl[\widehat{\mu}_{\vec{k},\,\alpha}, \, \widehat{\pi}_{\vec{p},\,\beta}\bigr]= i \, \delta_{\alpha\beta} \, \delta^{(3)}(\vec{k} + \vec{p}),
\label{AC13}
\end{equation}
where $\delta_{\alpha\,\beta}$ denotes the Kr\"oneker delta over the polarizations whereas  $\delta^{(3)}(\vec{k} + \vec{p})$ indicates the Dirac delta distribution over the momenta. 
From Eq. (\ref{AC13}) we also have that the operators corresponding to the Fourier components of Eq. (\ref{AC10}) obey
\begin{equation}
\bigl[ \widehat{\mu}_{i j}(\vec{k}, \eta), \, \widehat{\pi}_{m n}(\vec{p}, \eta)\bigr] = i \,  {\mathcal S}_{i j m n}(\hat{k}) \delta^{(3)}(\vec{k}+ \vec{p}).
\label{AC14}
\end{equation}
In Eq. (\ref{AC14}) the term ${\mathcal S}_{i j m n}(\hat{k})$ comes from the sum over the polarizations and it is given by:
\begin{equation}
{\mathcal S}_{i j m n}(\hat{k}) = \bigl[ p_{i m}(\hat{k})\, p_{j n}(\hat{k}) +  p_{i n}(\hat{k})\, p_{j m}(\hat{k}) - p_{i j}(\hat{k})\, p_{m n}(\hat{k}) \bigr]/4,
\label{AC15}
\end{equation}
where $p_{i j}(\hat{k}) = (\delta_{i j} - \hat{k}_{i} \, \hat{k}_{j})$ denote the standard transverse projectors.

\subsection{The Hamiltonian operator}
After inserting Eq. (\ref{AC9aa}) into Eq. (\ref{AC8}), the 
Hamiltonian operator becomes:
\begin{eqnarray}
\widehat{H}_{g}(\eta) &=& \frac{1}{2} \int d^{3} p \sum_{\alpha=\oplus,\otimes} \biggl\{ \widehat{\pi}_{-\vec{p},\,\alpha}\, \widehat{\pi}_{\vec{p},\,\alpha} + p^2 \widehat{\mu}_{-\vec{p},\,\alpha}\, \widehat{\mu}_{\vec{p},\,\alpha} 
+ {\mathcal F} \biggl[ \widehat{\pi}_{-\vec{p},\,\alpha}\, \widehat{\mu}_{\vec{p},\,\alpha} + 
\widehat{\mu}_{-\vec{p},\,\alpha}\, \widehat{\pi}_{\vec{p},\,\alpha} \biggr] \biggr\},
\label{QA1}
\end{eqnarray}
and if  we now insert Eq. (\ref{AC12}) into Eq. (\ref{QA1}) the quantum Hamiltonian is expressed in terms of the creation and annihilation operators:
\begin{eqnarray}
\widehat{H}_{g}(\eta) &=& \frac{1}{2} \,\sum_{\alpha=\oplus,\, \otimes}\, \int d^{3} p  \biggl\{ p \,\,\biggl[ \widehat{a}^{\dagger}_{\vec{p},\,\alpha} \widehat{a}_{\vec{p},\,\alpha} 
+ \widehat{a}_{- \vec{p},\,\alpha} \widehat{a}^{\dagger}_{-\vec{p},\,\alpha} \biggr] 
+ \lambda \,\,\widehat{a}^{\dagger}_{-\vec{p},\,\alpha} \widehat{a}_{\vec{p},\,\alpha}^{\dagger} 
+ \lambda^{\ast} \,\,\widehat{a}_{\vec{p},\,\alpha} \widehat{a}_{-\vec{p},\,\alpha} \biggr\}.
\label{QA2}
\end{eqnarray}
The equations obeyed by $\widehat{a}_{\vec{p}}$ and $\widehat{a}_{-\vec{p},\,\alpha}^{\dagger}$ in the Heisenberg description follow from Eq. (\ref{QA2}) and they are: 
\begin{equation}
\partial_{\eta} \widehat{a}_{\vec{p},\,\alpha}=  - i\, p \, \widehat{a}_{\vec{p},\,\alpha} -  i \, \lambda \widehat{a}_{- \vec{p},\,\alpha}^{\dagger},\qquad\qquad
\partial_{\eta}\widehat{a}_{-\vec{p},\,\alpha}^{\dagger} =  i\, p \, \widehat{a}_{-\vec{p},\,\alpha}^{\dagger} +  i \, \lambda^{\ast} \widehat{a}_{\vec{p},\,\alpha}.
\label{DNT5}
\end{equation}
The solution of Eq. (\ref{DNT5}) can be expressed through two (complex) functions $u_{p,\,\alpha}(\eta,\eta_{i})$ and $v_{p,\,\alpha}(\eta,\eta_{i})$. According to this logic, the operators at a generic conformal time $\eta$ are related by unitarity to the ones that annihilate the initial vacuum state:
\begin{eqnarray}
\widehat{a}_{\vec{p},\,\alpha}(\eta) &=& u_{p,\alpha}(\eta,\, \eta_{i}) \, \widehat{b}_{\vec{p},\,\alpha}(\eta_{i}) - v_{p,\,\alpha}(\eta,\eta_{i}) \, \widehat{b}_{- \vec{p},\, \alpha}^{\,\,\dagger}(\eta_{i}) ,
\label{DR2}\\
\widehat{a}_{- \vec{p},\, \alpha}^{\,\,\dagger}(\eta) &=& u^{\ast}_{p,\alpha}(\eta,\, \eta_{i}) \, \widehat{b}_{-\vec{p},\, \alpha}^{\,\,\dagger}(\eta_{i})  - v_{p,\,\alpha}^{\ast}(\eta,\eta_{i})  \, \widehat{b}_{\vec{p},\,\alpha}(\eta_{i}) .
\label{DR3}
\end{eqnarray}
The functions $u_{p,\,\alpha}(\eta,\eta_{i})$ and $v_{p,\,\alpha}(\eta,\eta_{i})$ appearing 
in Eqs. (\ref{DR2})--(\ref{DR3}) are subjected to the condition $|u_{p,\,\alpha}(\eta,\eta_{i})|^2 - |v_{p\,\alpha}(\eta,\eta_{i})|^2 =1$ that is a consequence of the unitary evolution.
Inserting Eqs. (\ref{DR2})--(\ref{DR3}) into Eq. (\ref{DNT5}) the evolution equations for $u_{p,\,\alpha}(\eta, \eta_{i})$ and $v_{p,\,\alpha}(\eta, \eta_{i})$ become:
\begin{equation}
\partial_{\eta} \,u_{p,\,\alpha} = - i p\, u_{p,\,\alpha}  + i \lambda\,\,  v_{p,\,\alpha}^{\ast}, 
\qquad \partial_{\eta}\, v_{p,\,\alpha}= - i p\, v_{p,\,\alpha} + i \lambda \,\,u_{p,\,\alpha}^{\ast}.
\label{DNT8}
\end{equation}

\subsection{Discrete mode representation}
In the commutators introduced before [see e.g. Eqs. (\ref{AC13})--(\ref{AC14})] the Dirac delta distribution is a consequence of Eqs. (\ref{AC9a})--(\ref{AC10a}) where the field operators are represented as a continuous set of $\vec{k}$-modes.
While the continuous mode representation has been employed so far, 
it is sometimes more practical to deal with a discrete set of modes especially when discussing the explicit forms of the multiparticle states. For a discrete set of modes the operators in Fourier space are given by:
\begin{equation}
\widehat{\mu}_{i j}(\vec{x}, \eta) = \frac{\sqrt{2} \ell_{P}}{L^{3/2}} \sum_{\vec{k}}\,\, \widehat{\mu}_{i\, j}(\vec{k}, \eta) \, e^{- i \vec{k}\cdot\vec{x}}, \qquad \widehat{\pi}_{i j}(\vec{x}, \eta) = \frac{1}{4 \,\sqrt{2} \, \ell_{P} \, L^{3/2}} \sum_{\vec{k}} \widehat{\pi}_{i\, j}(\vec{k}, \eta) \, e^{- i \vec{k}\cdot\vec{x}},
\label{AC9a}
\end{equation}
where the three-vector $\vec{k}$ has components $(k_{x}, \, k_{y},\, k_{z}) = 2\pi (n_{1}, \, n_{2},\, n_{3})/L$ 
with $n_{i} = 0, \, \pm 1, \pm 2,\,.\,.\,.$ and $i = 1, \, 2,\, 3$.  The set of the modes is infinite 
but because we are considering a finite volume (and a discrete set of modes) it is 
countably infinite. The Hamiltonian operator is given by the sum of the Hamiltonians valid for each $\vec{k}$-mode: 
\begin{equation}
\widehat{H}_{g} = \sum_{\vec{k}} \sum_{\alpha=\oplus,\otimes}\, \widehat{H}_{\vec{k},\,\alpha}, \qquad \widehat{H}_{\vec{k}}(\tau) = \omega_{k} \biggl( \widehat{a}_{\vec{k},\,\alpha}^{\,\,\dagger} \,\widehat{a}_{\vec{k},\,\alpha} + 
\widehat{a}_{-\vec{k},\,\alpha}\, \widehat{a}_{-\vec{k},\,\alpha}^{\,\,\dagger} \biggr) + g\,\widehat{a}_{\vec{k},\,\alpha}^{\,\,\dagger}\,\widehat{a}_{-\vec{k},\,\alpha}^{\,\,\dagger} + g^{\ast} \,\widehat{a}_{\vec{k},\,\alpha}\,\widehat{a}_{-\vec{k},\,\alpha}.
\label{DR1}
\end{equation}
where we  renamed, for convenience, $\omega_{k}= \sqrt{k/2}$ and $g = \lambda/2$. 
If we let $L\to \infty$ in Eq. (\ref{AC9a})  any 
discrete sum over the $\vec{k}$-vectors becomes an integral according to the rule 
\begin{equation}
\sum_{\vec{k}} \to \frac{V}{(2\pi)^{3}} \int d^{3}k, \qquad \widehat{\mu}_{i\, j}(\vec{k}, \eta) \to \frac{(2\pi)^{3/2}}{\sqrt{V}} \widehat{\mu}_{i\, j}(\vec{k}, \eta), \qquad \widehat{\pi}_{i\, j}(\vec{k}, \eta) \to \frac{(2\pi)^{3/2}}{\sqrt{V}} \widehat{\pi}_{i\, j}(\vec{k}, \eta),
\label{AC10a}
\end{equation}
where $V = L^3$. The same rescalings of Eq. (\ref{AC10a}) also hold for the creation and annihilation operators and we will have, in particular 
\begin{equation}
\widehat{a}_{\vec{k},\,\alpha} \to \frac{\sqrt{V}}{(2\pi)^{3/2}} \,\,\widehat{a}_{\vec{k},\,\alpha}, \qquad \delta^{(3)}(\vec{k} - \vec{p}) \delta_{\alpha\beta} \to \frac{V}{(2\pi)^{3}} \, \delta_{\vec{k},\,\vec{p}} \, \delta_{\alpha\beta}.
\label{AC10b}
\end{equation}
In Eq. (\ref{AC10b}) $\delta_{\vec{k},\,\vec{p}}$ indicates the Kr\"oneker symbol replacing the Dirac delta distribution typical of the continuous mode representation. For a discrete set of modes  the commutation relations can then be written as $[\widehat{a}_{\vec{k},\,\alpha}, \, \widehat{a}_{\vec{k},\,\beta}^{\,\dagger}] = \delta_{\vec{k},\, \vec{p}} \delta_{\alpha\beta}$. Thanks to Eqs. (\ref{AC10a})--(\ref{AC10b}) we can switch, when needed, from the continuous mode representation to the discrete one or vice versa. 

\renewcommand{\theequation}{3.\arabic{equation}}
\setcounter{equation}{0}
\section{The multiplicity distributions of the relic gravitons}
\label{sec3}
\subsection{The multiparticle final states}
The probability amplitudes of observing $n$ pairs of gravitons with opposite (comoving) three-momenta follows 
from the multiparticle final state $|\{ s\}\rangle$ whose general 
expression  can be written as:
\begin{equation}
|\{ s\}\rangle = \prod_{\vec{k},\, \alpha} | s_{\vec{k},\, \alpha}(\eta_{i}, \, \eta_{f}) \rangle, \qquad \widehat{a}_{\vec{k},\,\alpha}(\eta_{f}) | s_{\vec{k},\,\alpha}(\eta_{i}, \, \eta_{f}) \,\rangle = \widehat{a}_{-\vec{k},\,\alpha}(\eta_{f}) | s_{\vec{k},\, \alpha}(\eta_{i}, \, \eta_{f}) \,\rangle=0.
\label{state1}
\end{equation}
For each $\vec{k}$-mode of the field the final state of Eq. (\ref{state1}) is annihilated by  
the operators $\widehat{a}_{\vec{k},\,\alpha}$ and $\widehat{a}_{-\vec{k},\,\alpha}$ of Eqs. (\ref{DR2})-(\ref{DR3}). In view of a more explicit form of Eq. (\ref{state1}) it is useful to 
acknowledge that  Eqs. (\ref{DR2})-(\ref{DR3}) can be rewritten in terms of two unitary operators denoted, respectively,  by 
$\widehat{{\mathcal R}}(\delta_{k,\, \alpha})$ and $\widehat{\Sigma}(z_{k,\, \alpha})$:
\begin{eqnarray}
\widehat{a}_{\vec{k},\,\alpha} &=&  \widehat{S}^{\,\dagger}(z_{k,\,\alpha}) \, \widehat{{\mathcal R}}^{\,\dagger}(\delta_{k,\,\alpha}) \, \widehat{b}_{\vec{k},\, \alpha}  \, \widehat{{\mathcal R}}(\delta_{k,\,\alpha}) \, \widehat{S}(z_{k,\,\alpha}),
\nonumber\\
\widehat{a}_{-\vec{k},\,\alpha} &=&  \widehat{S}^{\,\dagger}(z_{k,\,\alpha}) \, \widehat{{\mathcal R}}^{\,\dagger}(\delta_{k,\,\alpha}) \, \widehat{b}_{-\vec{k},\, \alpha}  \, \widehat{{\mathcal R}}(\delta_{k,\,\alpha}) \, \widehat{S}(z_{k,\,\alpha}).
\label{SD1}
\end{eqnarray}
For the sake of conciseness we used a shorthand notation\footnote{We set, for instance,  $\widehat{a}_{\vec{k},\,\alpha}(\eta)\equiv \widehat{a}_{\vec{k},\,\alpha}$ and $\widehat{b}_{\vec{k},\,\alpha}(\eta_{i})= \widehat{b}_{\vec{k},\,\alpha}$ an similarly for the Hermitian conjugates. Unless strictly necessary, the same shorthand notation is also adopted hereunder for the multiparticle final states of Eq. (\ref{state1}) that are preferentially indicated as $| s_{\vec{k},\, \alpha}(\eta_{i}, \, \eta_{f}) \rangle \equiv | s_{\vec{k},\, \alpha}\rangle$ by dropping the corresponding time dependence.} by dropping the unessential 
arguments from the expressions of the operators. The explicit form of the operators is Eq. (\ref{SD1}) is:
\begin{eqnarray}
\widehat{{\mathcal R}}(\delta_{k,\,\alpha}) &=& e^{\,\, - i\, \delta_{k,\,\alpha} \, \widehat{b}_{\vec{k},\,\alpha}^{\, \dagger} \widehat{b}_{\vec{k},\,\alpha} - i\, \delta_{k,\, \alpha}\, \widehat{b}_{-\vec{k},\alpha}^{\, \dagger} \widehat{b}_{-\vec{k},\,\alpha}},
\label{SD2}\\
 \widehat{S}(z_{k,\,\alpha}) &=& e^{\,\,z_{k,\,\alpha}^{\ast} \, \widehat{b}_{\vec{k},\alpha} \widehat{b}_{- \vec{k},\,\alpha} - z_{k,\alpha} \widehat{b}_{\vec{k},\,\alpha}^{\dagger}  \widehat{b}_{- \vec{k},\,\alpha}^{\dagger}},
 \label{SD3}
 \end{eqnarray}
 where $z_{k,\,\alpha} = r_{k,\,\alpha}\, e^{i \theta_{k,\,\alpha}}$; both $r_{k,\,\alpha}$ and 
 $\theta_{k,\,\alpha}$ have a simple relation with $v_{k,\,\alpha}$, namely 
 $r_{k,\,\alpha} = \mathrm{arcsinh}{|v_{k,\,\alpha}|}$ and $e^{i \, \theta_{k,\,\alpha}} = v_{k,\,\alpha}/|v_{k,\,\alpha}|$. The multiparticle final state given in Eq. (\ref{state1}) can be formally expressed as:
 \begin{equation}
 |\{ s\}\rangle = \prod_{\vec{k},\, \alpha} \,\widehat{{\mathcal R}}(\delta_{k,\,\alpha}) \, \widehat{S}(z_{k,\,\alpha})  \,| m_{\vec{k},\, \alpha}\,;\,m_{-\vec{k},\, \alpha} \rangle.
\label{state2}
\end{equation}
The Fock states of Eq. (\ref{state2}) are defined in terms of the initial vacuum state 
annihilated by $\widehat{b}_{\vec{k},\alpha}$ and  $\widehat{b}_{- \vec{k},\,\alpha}$ and their explicit expression is:
\begin{equation}
| m_{\vec{k},\, \alpha}\,;\,m_{-\vec{k},\, \alpha} \rangle = \frac{\bigl(\,\widehat{b}_{\vec{k},\alpha}^{\,\,\dagger}\,\bigr)^{m_{\vec{k}, \alpha}}}{ \sqrt{ m_{\vec{k}, \, \alpha}\,!}}\,\,  \frac{\bigl(\,\widehat{b}_{-\vec{k},\alpha}^{\,\,\dagger}\,\bigr)^{m_{-\vec{k}, \alpha}}}{ \sqrt{ m_{-\vec{k}, \, \alpha}\,!}}\,\, |0_{\vec{k},\, \alpha} \,\,  0_{-\vec{k},\, \alpha}\rangle.
\label{state3}
\end{equation}
With all these ingredients the explicit expression of $| s_{\vec{k},\, \alpha}\rangle$ appearing at the right hand side of Eq. (\ref{state1}) follows from Eq. (\ref{state2}) after expressing $\widehat{S}(z_{k,\,\alpha})$ in a factorized form through the Baker-Campbell-Hausdorff decomposition of the $SU(1,1)$ group \cite{BKH1,BKH2,BKH3}:
\begin{equation}
\widehat{S}(z_{k,\,\alpha}) = \widehat{S}_{+}(z_{k,\,\alpha})\,\, \widehat{S}_{0}(z_{k,\,\alpha})\,\, \widehat{S}_{-}(z_{k,\,\alpha}),
\label{state4}
\end{equation}
where $\widehat{S}_{\pm}(z_{k,\,\alpha})$ and $\widehat{S}_{0}(z_{k,\,\alpha})$ 
are the exponentials of the group generators in the $SU(1,1)$ case with appropriate coefficients depending upon $u_{k,\,\alpha}$ and $v_{k,\,\alpha}$:
\begin{eqnarray}
\widehat{S}_{-}(z_{k,\,\alpha}) &=& e^{\,  (v_{k,\,\alpha}^{\ast}/u_{k,\, \alpha})\, \widehat{b}_{\vec{k},\,\alpha}\, \widehat{b}_{- \vec{k},\,\alpha}},
\nonumber\\
 \widehat{S}_{0}(z_{k,\,\alpha}) &=& e^{ - \ln{|u_{k,\, \alpha}|}\, \bigl(\,\widehat{b}_{\vec{k},\, \alpha}^{\,\dagger}\,\widehat{b}_{\vec{k},\, \alpha} + \widehat{b}_{-\vec{k},\, \alpha}\,\widehat{b}_{-\vec{k},\, \alpha}^{\,\dagger}  \,\bigr)},
 \nonumber\\
 \widehat{S}_{+}(z_{k,\,\alpha}) &=& e^{ \,- (v_{k,\,\alpha}/u_{k,\, \alpha})\, \widehat{b}_{\vec{k},\,\alpha}^{\,\dagger}\,\,\widehat{b}_{- \vec{k},\,\alpha}^{\,\dagger}}.
 \label{state5}
 \end{eqnarray}
 The explicit form of $|\,s_{\vec{k},\, \alpha}  \rangle$ is then obtained by applying, in the order given by Eq. (\ref{state4}), the three 
 operators of Eq. (\ref{state5}) to the Fock state (\ref{state3}). After some lengthy but straightforward algebra the result of this twofold step for the multiparticle final state of Eq. (\ref{state1}) becomes:
 \begin{eqnarray}
  |\,s_{\vec{k},\, \alpha}  \rangle &=& \sum_{\ell_{\vec{k}, \alpha} =0}^{\infty}\,\,  \sum_{j_{\vec{k}, \alpha} =0}^{j_{\vec{k},\,\alpha}^{\mathrm{max}}}
  {\mathcal P}(u_{k,\, \alpha}, v_{k,\, \alpha},m_{\pm\vec{k}, \alpha}, \ell_{\vec{k}, \, \alpha}, j_{\vec{k}, \, \alpha})
\nonumber\\
&\times& {\mathcal Q}(m_{\pm\vec{k}, \alpha}, \ell_{\vec{k}, \, \alpha}, j_{\vec{k}, \, \alpha})
 | \ell_{\vec{k}, \, \alpha} + m_{\vec{k}, \, \alpha} - j_{\vec{k}, \, \alpha}\,;\, \ell_{\vec{k}, \, \alpha} + m_{-\vec{k}, \, \alpha} - j_{\vec{k}, \, \alpha} \rangle,
 \label{state6}
 \end{eqnarray}
 where $j_{\vec{k},\,\alpha}^{\mathrm{max}} = \mathrm{min}[m_{\vec{k}, \, \alpha},\,m_{-\vec{k}, \, \alpha}]$ is the smallest integer between $m_{\vec{k}, \, \alpha}$ and $m_{-\vec{k}, \, \alpha}$.  In Eq. (\ref{state6}) the 
 two functions ${\mathcal P}(u_{k,\, \alpha}, v_{k,\, \alpha},\, m_{\pm\,\vec{k}, \alpha}, \ell_{\vec{k}, \, \alpha}, j_{\vec{k}, \, \alpha})$
 and ${\mathcal Q}(m_{\pm\,\vec{k}, \alpha}, \ell_{\vec{k}, \, \alpha}, j_{\vec{k}, \, \alpha})$
 have been introduced for practical reasons; between the two expressions the former depends explicitly on $u_{k,\, \alpha}$ and $v_{k,\, \alpha}$ while the latter is controlled 
 by  $m_{\pm\,\vec{k}, \alpha}$,  $\ell_{\vec{k}, \, \alpha}$ and  $j_{\vec{k}, \, \alpha}$:
 \begin{eqnarray}
&&{\mathcal P}(u_{k,\, \alpha}, v_{k,\, \alpha},m_{\pm\,\vec{k}, \alpha}, \ell_{\vec{k}, \, \alpha}, j_{\vec{k}, \, \alpha}) =  \frac{ e^{- i\, \gamma_{\vec{k},\,\alpha}}}{|u_{k,\,\alpha}|^{(m_{\vec{k}, \alpha} + m_{-\vec{k}, \alpha}- 2 j_{\vec{k}, \alpha}+1)}}
  \biggl( - \frac{v_{k,\,\alpha}}{u_{k,\, \alpha}}\biggr)^{\ell_{\vec{k}, \, \alpha}} \,  \biggl(\frac{v_{k,\,\alpha}^{\ast}}{u_{k,\, \alpha}}\biggr)^{j_{\vec{k}, \, \alpha}},
 \nonumber\\
&& {\mathcal Q}(m_{\pm\vec{k}, \alpha}, \ell_{\vec{k}, \, \alpha}, j_{\vec{k}, \, \alpha}) = \frac{\sqrt{ m_{\vec{k}, \, \alpha}!\,  m_{-\vec{k}, \, \alpha}! \, ( \ell_{\vec{k}, \, \alpha} +  m_{\vec{k}, \, \alpha} -  j_{\vec{k}, \, \alpha})! \,( \ell_{\vec{k}, \, \alpha} +  m_{-\vec{k}, \, \alpha} -  j_{\vec{k}, \, \alpha})! }}{ \ell_{\vec{k}, \, \alpha}!\,  j_{\vec{k}, \, \alpha}!\, (m_{\vec{k}, \, \alpha} -  j_{\vec{k}, \, \alpha})!\, (m_{-\vec{k}, \, \alpha} -  j_{\vec{k}, \, \alpha})!}.
 \label{state7}
 \end{eqnarray}
In Eq. (\ref{state7}) the phases appearing throughout the derivation have been grouped in the term $\gamma_{\vec{k},\,\alpha} = [\delta_{\vec{k}, \, \alpha} (m_{\vec{k}, \, \alpha}+ m_{-{k}, \, \alpha}) + 2 \delta_{\vec{k},\alpha}(\ell_{\vec{k},\, \alpha} - j_{\vec{k}, \, \alpha})]$. The final multiparticle state (\ref{state7}) and the corresponding multiplicity distribution will now be analyzed in different physical limits.  Before proceeding further 
we just wish to stress that the sum over $j_{\vec{k}, \alpha}$ in Eq. (\ref{state6}) can also be defined between $0$ and $\infty$, as the one over $\ell_{\vec{k}, \, \alpha}$. However, all the terms of the series beyond $j_{\vec{k},\,\alpha}^{\mathrm{max}} = \mathrm{min}[m_{\vec{k}, \, \alpha},\,m_{-\vec{k}, \, \alpha}]$, are actually zero. This happens since 
 the sum over $j_{\vec{k}, \alpha}$ is the result of the action of $\widehat{S}_{-}(z_{k,\,\alpha})$ over the initial multiparticle state; since this operator
 implies an infinite series of destruction operators applied to a finite number of species, it should be clear that the sum must actually stop at a certain  $j_{\vec{k},\,\alpha}^{\mathrm{max}}$.

\subsection{Bose-Einstein distribution and spontaneous emission}
In case the inflationary stage of expansion is sufficiently long (i.e. $N_{k} \gg 60$), it is commonly assumed that all modes of the field are in the vacuum as if it would be possible 
to define a single vacuum state in curved backgrounds \cite{AS2}. The form of the time-dependent Hamiltonians of Eqs. (\ref{QA1})--(\ref{QA2}) may however change for a canonical transformation and while at the classical level this possibility  does not play any r\^ole, it becomes relevant form the quantum mechanical viewpoint since the vacuum minimizes a specific Hamiltonian operator\footnote{The relation between canonical transformation of classical system and unitary transformations of quantum systems \cite{AS2} implies that the differences induced by different vacua affect certain corrections of the tensor (and scalar) power spectra that have been claimed to be \cite{AS2} unobservable and have not been observed so far in spite of their repeated scrutiny \cite{TP1,TP2,TP3}.}.  In what follows we are going to consider the treatment based on  Eqs. (\ref{QA1})--(\ref{QA2}) and acknowledge that, for an exceedingly long inflationary stage, the initial particle concentrations are washed out so that, effectively, in Eq. (\ref{state6})  we could set $m_{\vec{k}, \, \alpha} \to 0$ and $m_{-\vec{k}, \, \alpha} \to 0$. But since $j_{\vec{k},\,\alpha}^{\mathrm{max}} = \mathrm{min}[m_{\vec{k}, \, \alpha},\,m_{-\vec{k}, \, \alpha}]$ we shall also have $j_{\vec{k},\,\alpha}^{\mathrm{max}} \to 0$ so that the multiparticle final state of Eq. (\ref{state1}) becomes 
\begin{equation}
|\{ s\}\rangle = \prod_{\vec{k},\, \alpha} \frac{e^{- i\, \gamma_{\vec{k},\,\alpha}}}{|u_{k,\,\alpha}|} \, \sum_{\ell_{\vec{k}, \alpha} =0}^{\infty}
\biggl(- \frac{v_{k,\, \alpha}}{u_{k,\,\alpha}}\biggr)^{\ell_{\vec{k},\, \alpha}} \, | \ell_{\vec{k}, \alpha} ;\,
\ell_{\vec{k}, \alpha} \rangle.
\label{BE1}
\end{equation}
From Eq. (\ref{BE1}) the probability amplitude of finding $n$ pairs of gravitons with opposite (comoving) three-momenta in the final multiplarticle state is given by:
\begin{equation}
\langle n_{\vec{k},\, \beta};\, n_{- \vec{k},\, \beta} | s_{\vec{k},\, \alpha}\rangle = \frac{e^{- i\, \gamma_{\vec{k},\,\alpha}}}{|u_{k,\,\alpha}|} \biggl(- \frac{v_{k,\, \alpha}}{u_{k,\,\alpha}}\biggr)^{n_{\vec{k},\, \alpha}} \, \delta_{\alpha\beta}.
\label{BE2}
\end{equation}
Consequently the probability distribution $p(n_{\vec{k},\,\alpha})$ deduced from Eq. (\ref{BE2}) is: 
\begin{equation}
p_{n_{\vec{k},\,\alpha}}= \frac{|v_{k,\,\alpha}|^{2\, n_{\vec{k}, \alpha}}}{|u_{k,\alpha}|^{ 2 ( n_{\vec{k},\, \alpha} +1)}}
, \qquad\qquad p(\{n\}) = \prod_{\vec{k}, \alpha}\, p_{\,n_{\vec{k},\,\alpha}},
\label{BE3}
\end{equation}
where $p(\{n\})$  indicates the joint probability distribution. It is worth mentioning 
that the result of Eq. (\ref{BE3}) has been first deduced, with slightly different notations, by Parker \cite{MD5,MD6} (see also Refs. \cite{PTOMS,PARKTH}) in the case of the production of scalar phonons in curved backgrounds. From the results of Eq. (\ref{BE3})  all the factorial moments of the distribution can be determined. The first-order moment gives the averaged multiplicity, the second-order moment is related to the dispersion and so on and so forth. Instead of determining the moments one by one we may obtain them at once from the factorial moment generating function
\begin{equation}
G(z) = \langle (1 +z)^{\widehat{n}_{\vec{k},\,\alpha}} \rangle = \sum_{q=0}^{\infty}\,\, \frac{z^{q}}{q!} \,\, \langle \widehat{n}^{(q)}_{\vec{k},\,\alpha} \rangle,
\label{BE4}
\end{equation}
where $\widehat{n}_{\vec{k},\,\alpha}= \widehat{b}_{\vec{k},\,\alpha}^{\dagger} \, \widehat{b}_{\vec{k},\,\alpha}$. In Eq. (\ref{BE4}) the coefficient of $z^{q}/q!$  is actually the factorial moment of order $q$, i.e. $\langle \widehat{n}^{(q)}_{\vec{k},\,\alpha} \rangle =
\langle \widehat{n}_{\vec{k},\,\alpha} ( \widehat{n}_{\vec{k},\,\alpha} -1)  (\widehat{n}_{\vec{k},\,\alpha} -2)\, .\,.\,. ( \widehat{n}_{\vec{k},\,\alpha} -q +1) \rangle$. The explicit form of $G(z)$ can be computed once the probability distribution is 
known and, for instance, from Eq. (\ref{BE3}) we have obtain
\begin{equation}
G(z)= \sum_{n_{\vec{k},\,\alpha}=0}^{\infty} (1 + z)^{n_{\vec{k},\,\alpha}}\, p_{n_{\vec{k},\,\alpha}} = \frac{1}{1 -  z\, |v_{k,\,\alpha}|^2}.
\label{BE5}
\end{equation}
Thus, thanks to the result of Eq. (\ref{BE5}) the factorial moment of order $q$ is given by
$\langle \widehat{n}^{(q)}_{\vec{k},\,\alpha} \rangle = q!\,\, |v_{k,\,\alpha}|^{2 q}$. These results define the Bose-Einstein (geometric) distribution whose lowest factorial moments are 
\begin{equation}
\langle \widehat{n}^{(1)}_{\vec{k},\,\alpha} \rangle = |v_{k,\,\alpha}|^{2} = \overline{n}_{k,\,\alpha}, \qquad \langle \widehat{n}^{(2)}_{\vec{k},\,\alpha} \rangle = 2 \,\overline{n}_{k,\,\alpha}^{2}.
\label{BE6}
\end{equation}
From Eq. (\ref{BE6}) it is also possible to deduce the dispersion $D_{k,\,\alpha}^2$
\begin{equation}
D_{k,\,\alpha}^2 = \langle \widehat{n}^{2}_{\vec{k},\,\alpha} \rangle -  \langle \widehat{n}_{\vec{k},\,\alpha} \rangle^2 = 
\overline{n}_{k,\,\alpha} ( 1 + \overline{n}_{k,\,\alpha}).
\label{BE7}
\end{equation}
Equation (\ref{BE3}) can also be expressed as a function of the averaged 
multiplicity of the pairs $\overline{n}_{k,\,\alpha}$ since the unitary evolution implies that, at any time,
$|u_{k,\,\alpha}|^2 = 1 + |v_{k,\,\alpha}|^2$:
\begin{equation}
p_{n_{\vec{k},\,\alpha}}(\overline{n}_{k,\,\alpha}) = \frac{\overline{n}_{k,\,\alpha}^{n_{\vec{k},\,\alpha}}}{(\overline{n}_{k,\,\alpha} + 1)^{n_{\vec{k},\,\alpha}+1}}, \qquad\qquad \sum_{n_{\vec{k},\,\alpha}=0}^{\infty} p_{n_{\vec{k},\,\alpha}} =1.
\label{BE8}
\end{equation}
From Eq. (\ref{BE8}) it is possible to deduce 
the probability generating function $P(s_{k,\,\alpha})$ (see e.g. \cite{KAS}):
\begin{equation}
P(s_{k,\,\alpha}, \overline{n}_{k,\,\alpha}) = \sum_{n_{\vec{k},\,\alpha}=0}^{\infty} s_{k,\,\alpha}^{n_{\vec{k},\,\alpha}} \, p_{n_{\vec{k},\,\alpha}}(\overline{n}_{k,\,\alpha}) = \frac{1}{1 + (1 - s_{k,\,\alpha}) \, \overline{n}_{k,\,\alpha}}\,,
\label{BE9}
\end{equation}
implying that $P(s_{k,\,\alpha},\overline{n}_{k,\,\alpha})\to 1$ for $s_{k,\,\alpha}\to 1$. From the derivatives of $P(s_{k,\,\alpha},\,\overline{n}_{k,\,\alpha})$ with respect to $s_{k,\,\alpha}$ we can again compute the moments of the distribution\footnote{The probability generating function defined in Eq. (\ref{BE9}) and the generating function of the factorial moments introduced in Eq. (\ref{BE5}) are in fact similar.}.  We finally remark that  Eqs. (\ref{BE7})--(\ref{BE8}) hold independently for each polarization; thus 
 the index $\alpha$ can be suppressed by only keeping the $k$-dependence in the averaged multiplicity. This simplification will be often adopted hereunder for the sake of convenience.

\subsection{Pascal distributions and stimulated emission}
 Since the duration of inflation is not observationally accessible, it might well happen that the initial state is not exactly the vacuum. In a reverse perspective the statistical properties of the multiplicity distribution should reveal the features of the initial state. If the initial stage contains a finite number of gravitons we can suppose that only the species with momentum $\vec{k}$ are initially present and set $m_{- \vec{k},\, \alpha} \to 0$ while $m_{ \vec{k},\, \alpha} \neq 0$. The results shall remain the same if we would assume $m_{\vec{k},\, \alpha} \to 0$ and $m_{-\vec{k},\, \alpha} \neq 0$. From Eq. (\ref{state6}) in the limit $m_{- \vec{k},\, \alpha} \to 0$ we can conclude that
$j_{\vec{k},\,\alpha}^{\mathrm{max}} = \mathrm{min}[m_{\vec{k}, \, \alpha},\,m_{-\vec{k}, \, \alpha}]\to 0$ and the multiparticle final state becomes, in this limit,
\begin{equation}
|\{ s\}\rangle = \prod_{\vec{k},\, \alpha} \frac{e^{- i\, \gamma_{\vec{k},\,\alpha}}}{|u_{k,\,\alpha}|^{m_{\vec{k},\,\alpha}+1}} \, \sum_{\ell_{\vec{k}, \alpha} =0}^{\infty}
\biggl(- \frac{v_{k,\, \alpha}}{u_{k,\,\alpha}}\biggr)^{\ell_{\vec{k},\, \alpha}} \, \sqrt{\frac{( \ell_{\vec{k}, \, \alpha} +  m_{\vec{k}, \, \alpha})!}{\ell_{\vec{k}, \, \alpha} \,!\,\,m_{\vec{k}, \, \alpha}\,!}}\, \,| \ell_{\vec{k}, \alpha} + m_{\vec{k},\, \alpha}\,;\,\ell_{\vec{k}, \alpha} \rangle.
\label{NB1}
\end{equation}
The production probability of $n_{\vec{k}} = n_{\vec{k},\, \beta}= n_{-\vec{k}, \,\beta}$ pairs with opposite three-momentum (given the presence of $m_{\vec{k}}= m_{\vec{k},\,\alpha}$ gravitons in the initial state) follows after  squaring the corresponding amplitude\footnote{As already remarked in connection with Eqs. (\ref{BE7})--(\ref{BE8}),  since 
the production amplitude holds  independently for each polarization, 
 the index $\alpha$ can be suppressed by only keeping the $k$-dependence in the averaged multiplicity. }:
\begin{equation}
p_{n_{\vec{k}}}(m_{\vec{k}}) = \frac{(m_{\vec{k}}+ n_{\vec{k}})!}{m_{\vec{k}}!\, n_{\vec{k}}!}\, \frac{|v_{k}|^{2\, n_{\vec{k}}}}{|u_{k}|^{2 (m_{\vec{k}} +  n_{\vec{k}} + 1)}}.
\label{NB2}
\end{equation}
In the limit $m_{\vec{k}} \to 0$ the result of Eq. (\ref{NB2}) reproduces the Bose-Einstein 
distribution of Eq. (\ref{BE3}). Moreover from Eq. (\ref{NB2}) the generating function of Eq. (\ref{BE4}) and the associated factorial moments become, respectively
\begin{equation}
G(z) = \frac{1}{[1 - z\, |v_{k}|^2]^{m_{\vec{k}} +1}},\qquad \langle \widehat{n}^{(q)}_{\vec{k},\,\alpha} \rangle = \frac{( m_{\vec{k}} + q)!}{m_{\vec{k}} !} |v_{k}|^{2 q},
\label{NB3}
\end{equation}
that correctly reproduce the Bose-Einstein limit for $m_{\vec{k}} \to 0$. Furthermore, the 
averaged multiplicity computed from Eq. (\ref{NB3}) now becomes $\overline{n}_{k} 
= (1 + m_{\vec{k}}) |v_{k}|^{2}$ and, from this last expression, we can introduce 
directly the averaged multiplicity in Eq. (\ref{NB2}) by replacing $|v_{k}|^{2}$ 
with $\overline{n}_{k}/(m_{\vec{k}} +1)$:
\begin{equation}
p_{n_{\vec{k}}}(m_{\vec{k}}) =  \frac{(m_{\vec{k}} + n_{\vec{k}}) !}{m_{\vec{k}} !\, n_{\vec{k}} !} \, \frac{\overline{n}_{k}^{n_{\vec{k}}}}{(\overline{n}_{k} + m_{\vec{k}} +1)^{n_{\vec{k}}}} \, \biggl(\frac{m_{\vec{k}} + 1 }{\overline{n}_{k} + m_{\vec{k}} + 1}\biggr)^{m_{\vec{k}} +1}.
\label{DR14}
\end{equation}
If we now redefine $\kappa_{k} = m_{\vec{k}}+ 1$, Eq. (\ref{DR14}) becomes:
\begin{eqnarray}
p_{n_{\vec{k}}}(\overline{n}_{k},\,\kappa_{k}) &=& \left(\matrix{ \kappa_{k} + n_{\vec{k}} -1 \cr n_{\vec{k}} }\right) \, \biggl(\frac{\overline{n}}{\overline{n} + \kappa_{k}}\biggr)^{n_{\vec{k}}} \biggl(\frac{\kappa_{k}}{\overline{n}+ \kappa_{k}}\biggr)^{\kappa_{k}} 
\nonumber\\
&=& \frac{\kappa_{k} ( \kappa_{k} + 1)... ( \kappa_{k} + n_{\vec{k}} -1)}{n_{\vec{k}}!}  \biggl(\frac{\overline{n}_{k}}{\overline{n}_{k} + \kappa_{k}}\biggr)^{n_{\vec{k}}} \biggl(\frac{\kappa_{k}}{\overline{n}_{k}+ \kappa_{k}}\biggr)^{\kappa_{k}},
\label{DR15}
\end{eqnarray}
which is one the canonical forms of the Pascal distribution. 

\subsection{Mixed initial states and reduced density operators}
If the initial conditions contain 
a mixed thermal state the basic features of the multiplicity distributions do not drastically change.
To scrutinize this point in more detail we can go back to Eq. (\ref{NB1}) and formally consider the state $| s_{\vec{k},\,\alpha}\rangle$ as a function of the initial multiplicity:
\begin{eqnarray}
 | s_{\vec{k},\,\alpha}\,;\,m_{\vec{k},\,\alpha} \rangle &=& \frac{e^{- i\, \gamma_{\vec{k},\,\alpha}}}{|u_{k,\,\alpha}|^{m_{\vec{k},\,\alpha}+1}}\, \sum_{\ell_{\vec{k}, \alpha} =0}^{\infty}
\biggl(- \frac{v_{k,\, \alpha}}{u_{k,\,\alpha}}\biggr)^{\ell_{\vec{k},\, \alpha}} \, \sqrt{\frac{( \ell_{\vec{k}, \, \alpha} +  m_{\vec{k}, \, \alpha})!}{\ell_{\vec{k}, \, \alpha} \,!\,\,m_{\vec{k}, \, \alpha}\,!}}\, \,| \ell_{\vec{k}, \alpha} + m_{\vec{k},\, \alpha}\,;\,\ell_{\vec{k}, \alpha} \rangle.
 \label{M1}
 \end{eqnarray}
From Eq. (\ref{M1}) we can then construct the corresponding density matrix
\begin{equation}
\widehat{\rho}_{\vec{k},\,\alpha} = \sum_{m_{\vec{k}, \alpha} =0}^{\infty} \, p^{(\mathrm{stat})}_{m_{\vec{k},\, \alpha}} \, 
| s_{\vec{k},\,\alpha}\,;\,m_{\vec{k},\,\alpha} \rangle \langle \,m_{\vec{k},\,\alpha}\,;\,s_{\vec{k},\,\alpha}|,\qquad 
\widehat{\rho} = \prod_{\vec{k}, \, \alpha} \,\widehat{\rho}_{\vec{k},\,\alpha},
\label{M2}
\end{equation}
where $p^{(\mathrm{stat})}_{m_{\vec{k},\, \alpha}}$ now accounts for the statistical weights associated with the initial mixed state. An even  more explicit form 
of the density operator follows after inserting Eq. (\ref{M1}) into Eq. (\ref{M2}); we avoid 
this formula since it is conceptually straightforward but rather lengthy.
Thanks to the presence of $ p^{(\mathrm{stat})}_{m_{\vec{k},\, \alpha}}$ the traces of the total density matrix and of its square obey
\begin{equation}
\widehat{\rho}^2 \neq \widehat{\rho}, \qquad \mathrm{Tr} \widehat{\rho} = 1, \qquad \mathrm{Tr}\, 
\widehat{\rho}^2 \neq 1,
\label{M3}
\end{equation}
as expected in the case of a mixed state \cite{DM1}. A particularly relevant initial state is the one associated with a thermal 
mixture of gravitons and in this case\footnote{We stress that the multiplicity distribution of the relic gravitons is also geometric (i.e. Bose-Einstein) but strongly non-thermal.  The statistical weights of Eq. (\ref{M4}) are instead truly thermal since the corresponding average multiplicity coincides with the Bose-Einstein occupation number.} 
\begin{equation}
p^{(\mathrm{th})}_{m_{\vec{k},\, \alpha}} = \frac{\overline{n}_{\mathrm{th},\,k}^{m_{\vec{k},\,\alpha}}}{(\overline{n}_{\mathrm{th},\,k} + 1)^{m_{\vec{k},\,\alpha}+1}}, \qquad p^{(\mathrm{th})}(\{m\}) =  \prod_{\vec{k}, \, \alpha}\,p^{(\mathrm{th})}_{m_{\vec{k},\, \alpha}}.
\label{M4}
\end{equation}
For a detailed analysis of the multiplicity distributions  we can compute the reduced density operators \cite{DM1} by tracing, respectively, over the 
negative (positive) three-momenta:
\begin{equation}
\widehat{\rho}^{(+)} = \mathrm{Tr}_{-} \, \widehat{\rho} = \prod_{\vec{k},\, \alpha} \widehat{\rho}^{(+)}_{\vec{k},\, \alpha}, \qquad \widehat{\rho}^{(-)} = \mathrm{Tr}_{+} \, \widehat{\rho}= \prod_{\vec{k},\, \alpha} \widehat{\rho}^{(-)}_{\vec{k},\, \alpha},
\label{M5}
\end{equation}
where  $\mathrm{Tr}_{-}$ and $\mathrm{Tr}_{+}$ indicate, respectively, the traces over the negative and positive three-momenta. Let us start from $\widehat{\rho}^{(+)}_{\vec{k},\, \alpha}$ whose explicit form can be written as:
\begin{eqnarray}
\widehat{\rho}^{(+)}_{\vec{k},\, \alpha} &=& \sum_{n_{-\vec{k},\,\beta}=0}^{\infty}  \,\,\sum_{m_{\vec{k},\,\alpha}=0}^{\infty} \,\,\sum_{\ell_{\vec{k},\,\alpha}=0}^{\infty}\,\,
\sum_{\ell_{\vec{k},\,\alpha}^{\prime}=0}^{\infty}\, p^{(\mathrm{stat})}_{m_{\vec{k},\, \alpha}}\,\,\frac{e^{- i \delta_{\vec{k},\,\alpha}(\ell_{\vec{k}, \, \alpha} - \ell_{\vec{k},\, \alpha}^{\prime})}}{|u_{k,\,\alpha}|^{ 2 m_{\vec{k},\, \alpha} + 2}} 
\biggl( - \frac{v_{k,\,\alpha}}{u_{k,\,\alpha}}\biggr)^{\ell_{\vec{k},\,\alpha}}\,\,\biggl( - \frac{v_{k,\,\alpha}^{\ast}}{u_{k,\,\alpha}^{\ast}}\biggr)^{\ell_{\vec{k},\,\alpha}^{\prime}}
\nonumber\\
&\times& \frac{\sqrt{ (\ell_{\vec{k}, \, \alpha} +  m_{\vec{k}, \, \alpha})!\,  (\ell_{\vec{k}, \, \alpha}^{\prime} +  m_{\vec{k}, \, \alpha})!}}{ m_{\vec{k}, \, \alpha} ! \, \sqrt{\ell_{\vec{k}, \, \alpha}!\, \ell_{\vec{k}, \, \alpha}^{\prime}!} }
\langle n_{-\vec{k},\beta} | \ell_{\vec{k}, \, \alpha} + m_{\vec{k}, \, \alpha}\, ;\, \ell_{\vec{k},\, \alpha} \rangle \langle \ell_{\vec{k},\, \alpha}^{\prime}\, ;\,  m_{\vec{k}, \, \alpha}+ \ell_{\vec{k}, \, \alpha}^{\prime} | n_{-\vec{k},\, \beta} \rangle.
\label{M6}
\end{eqnarray}
The trace leads to two Kr\"oneker deltas that eliminate two sums and after renaming the 
summation indices we finally obtain that the phases disappear; the explicit 
form of Eq. (\ref{M6}) is then given by:
\begin{equation}
\widehat{\rho}^{(+)}_{\vec{k},\, \alpha} = \sum_{m_{\vec{k},\,\alpha}=0}^{\infty} \,\,\sum_{\ell_{\vec{k},\,\alpha}=0}^{\infty}
p^{(+)}_{\ell_{\vec{k},\, \alpha},\,m_{\vec{k},\, \alpha}}\,\,| \ell_{\vec{k}, \, \alpha} + m_{\vec{k},\, \alpha}\rangle \langle m_{\vec{k}, \, \alpha}+ \ell_{\vec{k}, \, \alpha} |,
\label{M7}
\end{equation}
where the statistical weights now take the following form:
\begin{equation}
p^{(+)}_{\ell_{\vec{k},\, \alpha},\,m_{\vec{k},\, \alpha}}= p^{(\mathrm{stat})}_{m_{\vec{k},\, \alpha}}\,\,\frac{(m_{\vec{k}, \, \alpha} + \ell_{\vec{k},\,\alpha})!}{m_{\vec{k}, \, \alpha} !\, \ell_{\vec{k},\,\alpha}!} \,\, \frac{|v_{k,\,\alpha}|^{2 \ell_{\vec{k},\,\alpha}}}{(1 + |v_{k,\,\alpha}|^{2})^{m_{\vec{k},\,\alpha} + \ell_{\vec{k},\, \alpha} +1}}.
\label{M8}
\end{equation}
With the same steps we can compute the explicit form of $\widehat{\rho}^{(-)}_{\vec{k},\, \alpha}$, namely
\begin{equation}
\widehat{\rho}^{(-)}_{\vec{k},\, \alpha} = \sum_{\ell_{\vec{k},\,\alpha}=0}^{\infty}\,\, p^{(-)}_{\ell_{\vec{k},\, \alpha}} \,\,| \ell_{\vec{k}, \, \alpha} \rangle \langle \ell_{\vec{k}, \, \alpha} |,
\label{M9}
\end{equation}
where, this time, $p^{(-)}_{\ell_{\vec{k},\, \alpha}}$ is:
\begin{equation}
p^{(-)}_{\ell_{\vec{k},\, \alpha}} = \sum_{m_{\vec{k},\,\alpha}=0}^{\infty} p^{(\mathrm{stat})}_{m_{\vec{k},\, \alpha}}\frac{(m_{\vec{k}, \, \alpha} + \ell_{\vec{k},\,\alpha})!}{m_{\vec{k}, \, \alpha} !\, \ell_{\vec{k},\,\alpha}!} \frac{|v_{k,\,\alpha}|^{2 \ell_{\vec{k},\,\alpha}}}{(1 + |v_{k,\,\alpha}|^{2})^{m_{\vec{k},\,\alpha} + \ell_{\vec{k},\, \alpha} +1}}.
\label{M10}
\end{equation}
When $p^{(\mathrm{stat})}_{m_{\vec{k},\, \alpha}}$ coincides with the $p^{(\mathrm{th})}_{m_{\vec{k},\, \alpha}}$ of 
Eq. (\ref{M4}) the explicit forms of $p^{(+)}_{\ell_{\vec{k},\, \alpha},\,m_{\vec{k},\, \alpha}}$ and $p^{(-)}_{\ell_{\vec{k},\, \alpha}}$ become:
\begin{eqnarray}
p^{(+)}_{\ell_{\vec{k},\, \alpha},\,m_{\vec{k},\, \alpha}} &=& \frac{(m_{\vec{k}, \, \alpha} + \ell_{\vec{k},\,\alpha})!}{m_{\vec{k}, \, \alpha} !\, \ell_{\vec{k},\,\alpha}!} \frac{\overline{n}_{\mathrm{th},\,k}^{m_{\vec{k},\,\alpha}}}{(\overline{n}_{\mathrm{th},\,k} + 1)^{m_{\vec{k},\,\alpha}+1}}\, \frac{|v_{k,\,\alpha}|^{2 \ell_{\vec{k},\,\alpha}}}{(1 + |v_{k,\,\alpha}|^{2})^{m_{\vec{k},\,\alpha} + \ell_{\vec{k},\, \alpha} +1}},
\label{M11}\\
p^{(-)}_{\ell_{\vec{k},\, \alpha}} &=& \frac{\bigl[|v_{k,\,\alpha}|^2 \,(\overline{n}_{\mathrm{th},\,k}+1)\bigr]^{\ell_{\vec{k},\,\alpha}}}{\bigl[ |v_{k,\,\alpha}|^2(\overline{n}_{\mathrm{th},\,k} + 1)\, +\,1 \bigr]^{\ell_{\vec{k},\,\alpha}+1}}.
\label{M12}
\end{eqnarray}
From Eq. (\ref{M12}) it follows that $p^{(-)}_{\ell_{\vec{k},\, \alpha}}$ is still a Bose-Einstein distribution 
where the average multiplicity is $\overline{N}_{k,\,\alpha} = |v_{k,\,\alpha}|^2 \,(\overline{n}_{\mathrm{th},\,k}+1)$. Also 
Eq. (\ref{M11}) belongs to the same class of distributions and to clarify this point we can write 
$p^{(+)}_{\ell_{\vec{k},\, \alpha},\,m_{\vec{k},\, \alpha}}$ as 
\begin{equation}
p_{\ell\, m} = \frac{(m + \ell)!}{m!\,\,\ell!} \frac{x^{m}}{(1+ x)^{m+1}} \, \frac{y^{\ell}}{(y + 1 )^{\ell + m +1}},
\label{M13}
\end{equation}
where, in practice, $x$ and $y$ correspond, respectively, to $\overline{n}_{\mathrm{th},\,k}$ and $ |v_{k,\,\alpha}|^2$. The probability generating function  deduced from Eq. (\ref{M13}) must contain two different arguments (i.e. $s$ and $w$ since two different random variable appear):
\begin{equation}
P(s,w) = \sum_{\ell=0}^{\infty} \sum_{m=0}^{\infty} s^{m} w^{\ell} p_{\ell\, m}  = \frac{1}{1 + (1-s) x + (1 -w) (1+ x) y}.
\label{M14}
\end{equation}
The two marginal probability distributions following from Eq. (\ref{M14}) are both of Bose-Einstein type with 
average multiplicities corresponding, respectively, to $x$ and to $y (1+ x)$. In other words we have that 
\begin{equation}
P(s,1) =  \frac{1}{1 + (1-s) x }, \qquad P(1,w) =  \frac{1}{1 + (1-w) (x+ 1) y },
\label{M14a}
\end{equation}
which are the probability generating functions for two Bose-Einstein distributions with averaged multiplicities 
given by $x\equiv \overline{n}_{\mathrm{th},\,k}$ and by $( x +1 ) y \equiv ( 1+ \overline{n}_{\mathrm{th},\,k})  |v_{k,\,\alpha}|^2$. 

\renewcommand{\theequation}{4.\arabic{equation}}
\setcounter{equation}{0}
\section{The statistical properties of the multiplicity distributions}
\label{sec4}
The multiplicity distributions deduced in the previous section are all infinitely divisible \cite{KAS}.  In this section we shall analyze some relevant physical limits of the obtained distributions and connect the obtained results with the production mechanisms. 

\subsection{Independent production}
As already mentioned, in connection Eq. (\ref{DR15}) it is useful to consider the properties of the probability generating function \cite{KAS,KAR} that we can write as\footnote{From the technical viewpoint we shall always refer to the averaged multiplicity of the pairs $\overline{n}_{k}$ so that the subscript associated with the polarization shall be omitted. Within this notation for instance the average multiplicity of the gravitons produced from the vacuum (and 
summed over the polarizations) will be given by $4 \, \overline{n}_{k}$ where the factor $4$ accounts for the two polarizations and for the opposite three-momenta of each pair.}:
\begin{equation}
P(s_{k}, \overline{n}_{k}, \kappa_{k}) = \sum_{n=0}^{\infty} s_{k}^{n}\, p_{n}(\kappa) =  \frac{{\kappa_{k}}^{\kappa_{k}}}{[\overline{n}_{k}(1 - s_{k}) + \kappa_{k}]^{\kappa_{k}}}.
\label{DR16}
\end{equation}
While for $\kappa_{k}\to 1$ we obtain the probability generating function of the Bose-Einstein distribution, the limit $s_{k}\to 1$ implies that all the discrete probability distributions must sum to $1$. From Eq. (\ref{DR16}) it follows that the dispersion $D_{k}^2$ is:
\begin{equation}
D_{k}^2 =  P^{\prime\prime}(1,\overline{n}_{k}, \kappa_{k}) + P^{\prime}(1,\overline{n}_{k}, \kappa_{k}) - [P^{\prime}(1,\overline{n}_{k}, \kappa_{k})]^2.
\label{DR17}
\end{equation}
In Eq. (\ref{DR17}) and in the remaining part of the present section the prime denotes a derivation\footnote{In section \ref{sec5} the prime denotes a derivation with respect to the conformal time coordinate. The two notations cannot be confused since they never appear in the same context.} with respect to $s_{k}$. After inserting Eq. (\ref{DR16}) into Eq. (\ref{DR17}) we obtain 
that the ratio between $D_{k}^2$ and $\overline{n}_{k}$ is 
\begin{equation}
\frac{D_{k}^2}{\overline{n}_{k}^2} =\frac{1}{\kappa_{k}} + \frac{1}{\overline{n}_{k}}.
\label{DR18}
\end{equation}
In the limit $\kappa_{k} \to 1$  Eq. (\ref{DR18}) implies $D_{k}^2 \to \overline{n}_{k} + \overline{n}_{k}^2$ as it follows from the Bose-Einstein (geometric) distribution \cite{KAR}.  A second meaningful limit is  $\kappa_{k} \to \infty$: from Eq. (\ref{DR18}) we have that the variance approximately equals the averaged multiplicity (i.e. $D_{k}^2\to \overline{n}_{k}$) and this is the characteristic signature of the Poisson distribution \cite{KAS,KAR}. 
We can actually show that if Eq. (\ref{DR16}) is evaluated for $\kappa_{k} \to \infty$ the probability generating function of Eq. (\ref{DR16}) tends to the one of a Poisson distribution. Let us first consider the inverse of $P(s_{k}, \overline{n}_{k}, \kappa_{k})$
\begin{equation}
\frac{1}{P(s_{k}, \overline{n}_{k}, \kappa_{k})}= \biggl[ 1 + \frac{ \overline{n}_{k} \,( 1 -s_{k})}{\kappa_{k}} \biggr]^{\kappa_{k}},
\label{DR19}
\end{equation}
and evaluate it in the limit $\kappa_{k}\to \infty$:
\begin{equation}
\lim_{\kappa_{k} \to \infty}  \biggl( 1 + \frac{ \overline{n}_{k} ( 1 -s)}{\kappa_{k}} \biggr)^{\kappa_{k}} \to e^{\overline{n}_{k}(1 -s_{k})} = \frac{1}{\overline{P}(s_{k},\overline{n}_{k})}.
\label{DR20}
\end{equation}
In Eq. (\ref{DR20}) $\overline{P}(s_{k}, \overline{n}_{k})$ is, by definition, the probability generating function associated 
with the Poisson distribution, i.e. 
\begin{equation}
\overline{P}(s_{k}, \overline{n}_{k}) = \sum_{n=0}^{\infty} p_{n}^{(P)} \, s_{k}^{n}, \qquad  p_{n}^{(P)} = \overline{n}_{k}^{n} e^{- \overline{n}_{k}}/n!.
\label{DR20a}
\end{equation}
The results of Eqs. (\ref{DR18}) and (\ref{DR20})--(\ref{DR20a}) show that for $\kappa_{k} \to \infty$ the pairs of gravitons are produced independently. This regime would also demand that the initial state dominates against the produced gravitons. From the physical viewpoint the  limit 
$\kappa_{k} \to \infty$ is then realized when $\overline{n}_{k} \ll \kappa_{k}$. If this 
happens the production of gravitons can be considered independent as in the Poisson case. As we are going to see in section \ref{sec5}, close to the maximal frequency of the spectrum which is ${\mathcal O}(\mathrm{THz})$ the production of the gravitons is approximately independent.

\subsection{The limit $\overline{n}_{k} \gg \kappa_{k}$} 
We expect that in the opposite case (i.e.  $\overline{n}_{k} \gg \kappa_{k}$)
the production of gravitons is not independent and the averaged multiplicity of the produced quanta greatly exceeds $\kappa_{k}$. This limit is realized, in practice, when the frequency range is small in comparison with the maximal frequency of the spectrum. In section \ref{sec5} we are going to see that 
$\overline{n}_{k}$ can actually be ${\mathcal O}(10^{20})$ 
so that the condition $\overline{n}_{k} \gg \kappa_{k}$ is realized together with $\overline{n}_{k} \gg n_{k}$, where $n_{k}$ is the multiplicity. From Eq. (\ref{DR15}) the limiting distribution can be derived by considering the product of the averaged multiplicity 
with the probability distribution $p_{n}(\overline{n}_{k}, \, \kappa_{k})$ in the range $\overline{n}_{k} \gg 1$, $n_{k} \gg1$ while the ratio 
$n_{k}/\overline{n}_{k}$ is kept fixed:
\begin{equation}
\overline{n}_{k} \,\,p_{n_{k}}(\overline{n}_{k}, \kappa_{k}) \simeq \psi_{n_{k}}(w_{k},\kappa_{k}), \qquad 
\psi_{n_{k}}(w_{k}, \kappa_{k})= \frac{\kappa_{k}^{\kappa_{k}}}{\Gamma(\kappa_{k})} w_{k}^{\kappa_{k}-1} e^{-\kappa_{k} w_{k}},\qquad w_{k}= n_{k}/\overline{n}_{k}.
\label{av14}
\end{equation}
In Eq. (\ref{av14}) $\psi_{n_{k}}(w_{k}, \kappa_{k})$ indicates the normalized Gamma distribution \cite{KAS,KAR} and some explicit numerical examples of the limit (\ref{av14}) will be provided in section \ref{sec5}. We stress that $\psi_{n_{k}}(w_{k}, \kappa_{k})$ can now be viewed as continuous 
function of $w_{k}$. Back in the 1970s it was suggested that the data of charged multiplicity distributions from hadronic reactions (and for different averaged multiplicities) should fall on the same curve when $\overline{n}_{k} \, p_{n}$ is plotted 
 against $n/\overline{n}_{k}$. The scaling limit is defined by the asymptotic behaviour of $p_{n}(\overline{n})$ as $n \to \infty$,
 $\overline{n}_{k} \to \infty$ while $n/\overline{n}_{k}$ is kept fixed. When expressed in terms of $n_{k}/\overline{n}_{k}$ it is also said, for short, that the distribution is given in the Koba-Nielsen-Olesen (KNO for short) variables \cite{MD3,MD4} (see also \cite{MD7}). Thus we can also say that, for fixed $\kappa_{k}$,  the distribution of Eq.  (\ref{DR15}) satisfies the KNO scaling with KNO function $\psi_{n_{k}}(w_{k},\, \kappa_{k})$. 

\subsection{Infinitely divisible distributions}
A probability distribution is infinitely divisible if, for any given non negative integer $k$, it is possible to find $k$ independent identically distributed random variables whose probability distributions sum up to the original distribution. Since the generating function of the sum of independent identically distributed random variables is given by the product of the generating functions  of each distribution of the sum, a distribution with probability generating function (pgf)  ${\mathcal P}(z)$ is infinitely divisible provided, for any integer $k$, there exist $k$ independent identically distributed random variables with generating function ${\mathcal Q}_{k}(z)$ such that ${\mathcal P}(z) = [{\mathcal Q}_{k}(z)]^{k}$. Since from Eq. (\ref{DR16}) we can write that  
\begin{equation}
P(s_{k}, \overline{n}_{k}, \kappa_{k}) = [Q(s_{k}, \overline{n}_{k}, \kappa_{k})]^{\kappa_{k}}, \qquad Q(s_{k}, \overline{n}_{k}, \kappa_{k})= \kappa_{k}/[\overline{n}_{k}(1 - s_{k}) + \kappa_{k}],
\label{IDIV1}
\end{equation}
the probability distribution associated with Eq. (\ref{DR16}) is infinitely divisible. 
The property spelled out by Eq. (\ref{IDIV1}) has an interesting 
physical interpretation: it means that the multiplicity distributions of the relic gravitons 
can be phrased as a compound Poisson process \cite{KAS,KAR}. Also the reverse 
is true: since the Poisson processes are infinitely divisible, also the class of distribution 
of Eqs. (\ref{DR15}) and (\ref{DR16}) must be infinitely divisible. 

To clarify the last statement, let us first stress that the class of distributions of Eqs. (\ref{DR15}) and (\ref{DR16}) have been deduced from a quantum mechanical production mechanism. The same class of probability distributions could also be obtained from a 
random sum of random variables \cite{KAS,KAR}.  Let us then consider the sum of ${\mathcal N}$ mutually independent random variables
\begin{equation}
S_{{\mathcal N}} = X_{1} +X_{2} +\,....+\, X_{\mathcal N}.
\label{TH1}
\end{equation}
When ${\mathcal N}$ is fixed  the probability generating function  of the sum $S_{{\mathcal N}}$ is simply given by the product of the generating functions of the various distributions (see e.g. \cite{KAR}). This is the property already exploited in Eq. (\ref{IDIV1}).
It can happen however that  ${\mathcal N}$ is itself a random variable and the corresponding pgf is obtained by compounding the probability distribution of the sum with the one of the random variables themselves. If the $X_{i}$ of Eq. (\ref{TH1}) are identically and independently distributed random variables with probability distribution $p_{n}$, their 
pgf is defined as $P(s) = \sum_{n} s^{n} p_{n}$, in full analogy with the quantities 
already introduced before (see Eqs. (\ref{BE9})--(\ref{M14}) and (\ref{M14a})).
In case ${\mathcal N}$ is itself a random variable, it must be characterized by a certain discrete probability distribution (be it $g_{k}$) whose associated pgf will be given, by definition, as $G(s)$. In the theory of discrete random processes, the generating function of a random number of random variables is given by \cite{KAR}
\begin{equation}
C(s) = G[P(s)], \qquad P(s) = \sum_{n=0}^{\infty} s^{n} p_{n}, \qquad G(s) =  \sum_{k=0}^{\infty} s^{k} g_{k},
\label{TH0}
\end{equation}
where $C(s)$ is  referred to as the compound distribution. To close the circle it remains to prove that Eqs. (\ref{DR15}) and (\ref{DR16}) can arise as the probability distribution of the random sum of  independent and identically distributed random variables. Let then consider the case when ${\mathcal N}$ 
is itself randomly distributed as a Poisson distribution with multiplicity 
$\overline{N}_{\mathrm{c}}$. We are here considering the case of a random number 
of Poissonian sources characterized by a multiplicity $\overline{N}_{c}$; each of these 
variables will be characterized by a generating function $P(s)$. The generating function of the random sum of random variables 
will then be given by Eq. (\ref{TH0})  as ${\mathcal S}(s_{k}) = e^{\overline{N}_{c}[P(s_{k}) -1]}$.
We are interested in the situation where the compound distribution 
${\mathcal S}(s_{k})$ is given by Eq. (\ref{DR16}). The request that the pgf of the 
random sum of random variables is given by Eq. (\ref{DR16}) fixes the $P(s)$ which is given by 
\begin{equation}
P(s_{k},\,\overline{N}_{c},\,\overline{n}_{k},\,\kappa_{k} )= 1 - \frac{\kappa_{k}}{\overline{N}_{c}}\ln{\biggl[\frac{\overline{n}_{k}}{\kappa_{k}}( 1- s_{k}) +1\biggr]},
\label{TH5}
\end{equation}
where $\overline{n}$ and $\kappa$ are the parameters appearing in Eqs. (\ref{DR15}) and (\ref{DR16}). Equation (\ref{TH5}) is the pgf of a logarithmic distribution and it ultimately demonstrates that the multiplicity distributions of the relic gravitons are infinitely divisible since they can be obtained as a compound Poisson process. There is a direct connection between the multiplicity distributions of the relic gravitons, their 
infinite divisibility and the irreducible representations 
of the $SU(1,1)$. Indeed, the square modulus of the Wigner matrix 
elements associated with the positive discrete series of the 
irreducible representations of the $SU(1,1)$ group 
encompass the class of infinitely divisible distributions 
discussed in this section. The analysis presented in appendix 
\ref{APPA} clarifies this connection that is also 
useful in the light of the explicit form of the Hamiltonian 
operator of our problem.
 
\subsection{Second-order correlation effects and multiplicity distributions}
The correlation properties of the produced gravitons can also be assessed through 
the degree of second-order coherence. A heuristic argument stipulates that the degree of second-order coherence should be directly related to the variance of the multiplicity distribution. Indeed, the quantum optical analogy between gravitons and optical photons \cite{HBTG1} would suggest   that we may neglect the momentum dependence and just consider the degree of second-order coherence of a single mode of the field \cite{MD1,MD2}. In this case the degree of second-order temporal coherence is given by:
\begin{equation}
g^{(2)}(\Delta\tau) = \frac{\langle \widehat{a}^{\dagger}(\tau )  \widehat{a}^{\dagger}(\tau+ \Delta\tau) \, \widehat{a}(\tau + \Delta\tau)\, \widehat{a}(\tau)\rangle}{\langle \widehat{a}^{\dagger}(\tau) \, \widehat{a}(\tau)\rangle \langle \widehat{a}^{\dagger}(\tau+ \Delta \tau) \, \widehat{a}(\tau + \Delta\tau)\rangle},
\label{CORR1}
\end{equation}
where the quantum mechanical expectation values are computed on a specific state. The operator 
appearing in the numerator at the right hand side of Eq. (\ref{CORR1}) is Hermitian and it represents, for a single 
mode, the correlation of the intensities of the radiation field appearing in the context of  
Hanbury Brown-Twiss interferometry (HBT) \cite{HBT1,HBT2,HBT3}. In the zero-delay limit 
(i.e. $\Delta \tau \to 0$) the degree of second-order coherence of Eq. (\ref{CORR1}) can be written as $g^{(2)}(0) \equiv g^{(2)}$ and it is directly related to the dispersion $D^2$ as\footnote{The result of Eq. (\ref{CORR2}) follows by recalling that $\langle \widehat{a}^{\dagger}\,\widehat{a}^{\dagger} \,\widehat{a}\,\widehat{a}\rangle = \langle\widehat{n}^2 \rangle - \langle \widehat{n} \rangle$ where $\widehat{n} = \widehat{a}^{\dagger}\,\widehat{a}$. }:
\begin{eqnarray}
g^{(2)} &=& \frac{\langle \widehat{a}^{\dagger} \,\,\widehat{a}^{\dagger} \,\,\widehat{a}\,\, \widehat{a}\rangle}{\langle \widehat{a}^{\dagger}\, \widehat{a}\rangle^2 }
\nonumber\\
&=& 1 + \frac{D^2 - \overline{n}}{\overline{n}^2}, \qquad D^2 =  \langle\widehat{n}^2\rangle - \langle\widehat{n}\rangle^2 \equiv  \langle\widehat{n}^2\rangle - \overline{n}^2.
\label{CORR2}
\end{eqnarray}
Equations (\ref{CORR1})--(\ref{CORR2}) relate the degree of second-order coherence 
with the dispersion and with the average multiplicity of a specific quantum states.
In a heuristic perspective we may now replace the quantum averages with 
statistical averages computed in terms of the multiplicity distribution (\ref{DR15}) 
valid for a single mode of the field. If we then use  $p_{n}(\overline{n}, \kappa)$
for the estimate of $D^2$ we obtain
\begin{equation}
\frac{D^2}{\overline{n}^2} = \frac{1}{\kappa} + \frac{1}{\overline{n}},
\label{CORR3}
\end{equation}
which is exactly of Eq. (\ref{DR18}) written in the case of a single mode of the field.
Once Eq. (\ref{CORR3}) is inserted into Eq. (\ref{CORR2}) we obtain 
\begin{equation}
g^{(2)} = 1 + \frac{1}{\kappa}.
\label{CORR4}
\end{equation}
In the limit $\kappa \to \infty$ the degree of second-order coherence 
goes to $1$ and this is consistent with the absence of correlation since 
the original $p_{n}(\overline{n}, \kappa)$ turns into a Poisson distribution. 
This limit is consistent with the occurrence that $D^2= \langle\hat{n} \rangle $ exactly 
when $g^{(2)}\to 1$, as it happens for a standard coherent state characterized by a Poisson multiplicity distribution. In a more direct perspective, for a single mode coherent state we would have $\widehat{a} | \alpha \rangle = \alpha |\alpha\rangle$; from Eq. (\ref{CORR2}) we would have that $g^{(2)}= 1$. In the Glauber theory of optical coherence it is possible to introduce the degrees of higher-order coherence \cite{GL1,GL2,GL3}. By 
definition a coherent state is coherent to all orders (i.e. $g^{(1)}= g^{(2)}= .\,.\,.\,.= g^{(n-1)} = g^{(n)} =1$).
For $\kappa \to 1$ the case of the Bose-Einstein distribution is recovered since Eq. (\ref{CORR4}) implies 
$g^{(2)} \to 2$: this is the situation of a chaotic mixture where the role of correlations is dominant \cite{MD1,MD2}. 
The results described by Eq. (\ref{CORR4}) suggest that  $g^{(2)}>1$ and this means, in a quantum optical language, that the gravitons are always  bunched with a super-Poissonian statistics. It is interesting that the case 
 $g^{(2)} <1$ (associated with antibunched photons and sub-Poissonian field statistics) \cite{MD1,MD2} is never realized since $\kappa > 0$. While Eqs. (\ref{CORR2})--(\ref{CORR3}) provide   
a physical analogy for the correlation properties 
 of the relic gravitons, the states leading to the result (\ref{CORR4}) 
 would rigorously follow from a density matrix 
with Pascal weights $p_{n}(\overline{n},\kappa)$: 
 \begin{equation}
 \widehat{\rho} = \sum_{n=0}^{\infty} \, p_{n}(\overline{n},\kappa) 
 |n \rangle \langle n|. 
 \label{CORR3a}
 \end{equation}
The  argument based on Eqs. (\ref{CORR3})--(\ref{CORR4}) relies on the 
single-mode approximation and it neglects completely the momentum dependence.
In what follows we shall swiftly address both issues also by referring to 
previous analyses that qualitatively confirm the conclusions of the heuristic arguments 
of the present subsection.

\subsection{Two-particle inclusive density}
In quantum optics the degrees of second-order coherence are determined by correlating the intensities of the radiation emitted by a source \cite{HBT1,HBT2,HBT3}. This strategy is customarily used, with complementary purposes, both in quantum optics and
in high-energy physics \cite{HBTH1,HBTH1a,HBTH2,HBTH3,HBTH4} where the interference of the intensities has been employed to determine the hadron fireball dimensions roughly corresponding to the linear size of the interaction region in proton-proton collisions. The correct framework where to address the second-order interference 
effects is the quantum theory of optical coherence that relates the coherence properties of a state (or of a source) to the minimization of the indetermination relations \cite{GL1,GL2,GL3}. The theory of optical coherence 
is usually formulated in terms of light fields that have a vector structure but it is possible to extend 
the Glauber theory to the tensor case \cite{HBTG1} (see also \cite{HBTG2,HBTG3}).  
If cosmic gravitons are detected by independent interferometers the correlated outputs are employed to estimate the degrees of second-order coherence. If we avoid the polarization dependence may introduce 
 the single-particle and the two particles (inclusive) density \cite{AU6}
\begin{equation}
\rho_{1}(\vec{k}) = \langle \widehat{A}^{\,\dagger}(\vec{k})\,\,\widehat{A}(\vec{k})\rangle, \qquad \rho_{2}(\vec{k}_{1}, \vec{k}_{2}) = \langle \widehat{A}^{\,\dagger}(\vec{k}_{1})\,\,\widehat{A}^{\,\dagger}(\vec{k}_{2})\,\,\widehat{A}(\vec{k}_{2})\,\,\widehat{A}(\vec{k}_{1})\rangle,
\label{HH1a}
\end{equation}
where $\widehat{A}(\vec{k},\tau)$ (and its Hermitian conjugate) are just a set of creation and annihilation operators that are non-zero inside the volume of the particle source. Barring 
for the technical details we would  have that the normalized 
second-order correlation function reads \cite{AU6}:
 \begin{equation}
 C_{2}(\vec{k}_{1}, \vec{k}_{2}) = \frac{\rho_{2}(\vec{k}_{1}, \vec{k}_{2})}{\rho_{1}(\vec{k}_{1})\,\,\rho_{1}(\vec{k}_{2})} \to 3 + {\mathcal O}\biggl(\frac{1}{\sqrt{\overline{n}(k_{1})\, \overline{n}(k_{2})}}\biggr).
 \label{HH3a}
 \end{equation}
Equation (\ref{HH3a}) is in fact the analog of Eq. (\ref{CORR3}) 
in the case where the momenta of the species are taken 
into account. It turns out that the value of $C_{2}(\vec{k}_{1}, \vec{k}_{2})$ is  always enhanced in comparison with so-called Poissonian limit so that the statistics of the relic gravitons is always super-Poissonian and  super-chaotic. Indeed in the limit of a large number of graviton pairs $C_{2}(\vec{k}_{1}, \vec{k}_{2}) \to 3$ whereas $C_{2}(\vec{k}_{1}, \vec{k}_{2}) \to 2$ in the case of a chaotic mixture. This result is slightly refined by taking into account the polarisation structure of the correlators \cite{HBTG2,HBTG3}. In this case $C_{2}(\vec{k}_{1}, \vec{k}_{2}) \leq 3$ but the statistics always remains super-Poissonian.  

The result of Eq. (\ref{HH3a}) can also be obtained in a more 
direct way from the field operators $\hat{\mu}_{ij}(x)$ that consist of a positive and of a negative frequency part, i.e.  $\hat{\mu}_{ij}(x) = \hat{\mu}_{ij}^{(+)}(x) + \hat{\mu}_{ij}^{(-)}(x)$, with $\hat{\mu}_{ij}^{(+)}(x)= \hat{\mu}_{ij}^{(-)\,\dagger}(x)$.  The field operators describing the positive and negative frequency parts can be expressed as:
\begin{eqnarray}
\hat{\mu}^{(-)}(\vec{x}, \tau) &=& \frac{\sqrt{2} \ell_{P}}{(2\pi)^{3/2}} \sum_{\alpha} \int \frac{d^{3} k}{\sqrt{2 k}} e^{(\alpha)}_{ij} \, \hat{a}_{- \vec{k}\, \alpha}^{\dagger}(\tau) \, e^{- i \vec{k}\cdot\vec{x}} \to  \frac{\sqrt{2} \ell_{P}}{(2\pi)^{3/2}} \int \frac{d^{3} k}{\sqrt{2 k}} \, \hat{a}_{- \vec{k}\, \alpha}^{\dagger}(\tau) \, e^{- i \vec{k}\cdot\vec{x}} ,
\label{degA1}\\
\hat{\mu}^{(+)}(\vec{x}, \tau) &=& \frac{\sqrt{2} \ell_{P}}{(2\pi)^{3/2}} \sum_{\alpha}
\int \frac{d^{3} k}{\sqrt{2 k}} e^{(\alpha)}_{ij} \, 
\hat{a}_{\vec{k}\, \alpha}(\tau) e^{- i \vec{k}\cdot\vec{x}}\to \frac{\sqrt{2} \ell_{P}}{(2\pi)^{3/2}} 
\int \frac{d^{3} k}{\sqrt{2 k}} \, 
\hat{a}_{\vec{k}\, \alpha}(\tau) e^{- i \vec{k}\cdot\vec{x}}.
\label{degA2}
\end{eqnarray}
Instead of considering full polarization structure as discussed in Ref. \cite{HBTG3}, for the sake 
of simplicity we can focus on the single-polarization approximation where the first-order Glauber correlator reads
\begin{equation}
{\mathcal S}^{(1)}(x_{1},\, x_{2}) = \langle  \hat{\mu}^{(-)}(x_{1}) \, \hat{\mu}^{(+)}(x_{2}) \rangle.
\label{SSS1}
\end{equation} 
Equation (\ref{SSS1}) describes the correlation of the amplitudes (i.e. Young interferometry)
Similarly we can also write the second-order Glauber correlator
\begin{equation}
{\mathcal S}^{(2)}(x_{1}, x_{2}) = \langle  \hat{\mu}^{(-)}(x_{1}) \, \hat{\mu}^{(-)}(x_{2})\hat{\mu}^{(+)}(x_{2}) \hat{\mu}^{(+)}(x_{1}) \rangle
\label{SSS2}
\end{equation}
accounting for the correlation of the intensities at two 
separate space-time points (i.e. HBT interferometry). This choice corresponds to the interferometric strategy pioneered by Hanbury Brown and Twiss (HBT) as opposed to the standard Young-type experiments where only amplitudes (rather than intensities) are allowed to interfere.
The applications of the HBT  ideas range from stellar astronomy   to subatomic 
physics. The interference of the intensities has been used to determine the hadron 
fireball dimensions corresponding to the linear size of the 
interaction region in proton-proton collisions. To disambiguate the possible origin of large-scale  curvature perturbations and of relic gravitons, probably the only hope is the analysis of the degree of second-order coherence, as we shall argue. The intensity must be Hermitian and this requirement ultimately imposes the form of the operator 
inside the average of Eq. (\ref{SSS2}). From Eqs. (\ref{SSS1})--(\ref{SSS2}) the corresponding degrees of quantum coherence are defined as 
\begin{eqnarray}
g^{(1)}(x_{1},\, x_{2}) = \frac{{\mathcal S}^{(1)}(x_1,\, x_{2})}{\sqrt{{\mathcal S}^{(1)}(x_1)} \,\, \sqrt{{\mathcal S}^{(1)}(x_2)}},
\qquad
g^{(2)}(x_{1},\, x_{2}) = \frac{{\mathcal S}^{(2)}(x_1,\, x_{2})}{{\mathcal S}^{(1)}(x_1) \,\, {\mathcal S}^{(1)}(x_2)}.
\label{SSS4}
\end{eqnarray}
The degrees of coherence follow then from the estimates of ${\mathcal S}^{(1)}(x_1,\, x_{2})$ and ${\mathcal S}^{(2)}(x_1,\, x_{2})$. If we now insert Eqs. (\ref{DR2})--(\ref{DR3}) into Eq. (\ref{SSS1}) we obtain 
\begin{equation}
{\mathcal S}^{(1)}(\vec{x}_{1},\, \vec{x}_{2}; \tau_{1},\, \tau_{2}) = \frac{1}{4\pi^2}\int k d k \,j_{0}(k r) \,  v_{k}^{*}(\tau_{1}, \tau_{i}) v_{k}(\tau_{2}, \tau_{i}) \kappa_{k},
\label{SSS5}
\end{equation}
where $r= |\vec{x}_{1} - \vec{x}_{2}|$ and $j_{0}(k r) = \sin{k r}/(k r)$ is the zeroth-order spherical Bessel function \cite{abr1,abr2}. In Eq. (\ref{SSS5}) we are implicitly considering, for convenience, the case $\eta \to \tau$. We can also insert  Eqs. (\ref{DR2})--(\ref{DR3}) into Eq. (\ref{SSS2}) and obtain:
 \begin{eqnarray}
 {\mathcal S}^{(2)}(x_{1}, x_{2}) &=& \frac{1}{4(2\pi)^6} \int \frac{d^{3} k}{k} \, \int \frac{d^{3} p}{p} 
 \kappa_{k}\kappa_{p}
\nonumber\\
&\times& \biggl\{ |v_{k}(\tau_{1},\tau_{i})|^2 \,\, |v_{p}(\tau_{2}, \tau_{i})|^2 
 +\biggl[ v_{k}^{*}(\tau_{1},\tau_{i}) 
 v_{p}^{*}(\tau_{2},\tau_{i}) v_{k}(\tau_{2},\tau_{i})v_{p}(\tau_{1},\tau_{i}) 
 \nonumber\\
 &+& v_{k}^{*}(\tau_{1},\tau_{i}) 
 u_{k}^{*}(\tau_{2},\tau_{i}) u_{p}(\tau_{2},\tau_{i})v_{p}(\tau_{1},\tau_{i})\biggr] 
 e^{- i (\vec{k} - \vec{p})\cdot\vec{r}}\biggr\}.
 \label{SSS6}
\end{eqnarray}
The analog of the zero-delay limit of Eqs. (\ref{CORR1})--(\ref{CORR2}) is represented by $r \to 0$ and 
$|k\tau_{1} |\ll 1$ and $|p \tau_{2}| \ll 1$ \cite{HBTG3}. It follows in this limit that ${\mathcal S}^{(2)}(\tau_{1}, \tau_{2}) \simeq 3 \,{\mathcal S}^{(1)}(\tau_{1}) {\mathcal S}^{(1)}(\tau_{2})$; this means, according to Eq. (\ref{SSS4}) that $g^{(2)}( \tau_{1}, \tau_{2}) \to 3 $. As Eq. (\ref{HH3a}), also the result obtained here relies on the single polarization approximation. In the observations of the relic gravitons, however, we might 
want to explore also a more inclusive perspective where all the polarizations are treated together. In this case, as 
suggested in Refs. \cite{HBTG2,HBTG3} the degree of second-order coherence can be a bit reduced 
by always remaining sharply super-Poissonian. 

\renewcommand{\theequation}{5.\arabic{equation}}
\setcounter{equation}{0}
\section{The high frequency gravitons}
\label{sec5}
The multiplicity distributions of the relic gravitons are a general consequence 
of the Hamiltonian of Eqs. (\ref{QA1})--(\ref{QA2}). Although the results discussed so far do not depend on the specific  choices of $a_{1}(\tau)$ and $a_{2}(\tau)$ (see Eq. (\ref{AC2}) and discussion thereafter), the potentially different profiles of  $a_{1}(\tau)$ and $a_{2}(\tau)$ may affect  the values of the averaged multiplicities, of the dispersions and of the other factorial moments of the distribution. 
The shapes of $a_{1}(\tau)$ and $a_{2}(\tau)$ also 
determine the frequency domains of the spectrum and the overall 
consistency of the scenario. The goal of this section is to analyze 
 the distributions for different ranges of the comoving frequency. For the sake of concreteness the attention will now be focussed on the minimal framework where the scale factor $a_{1}(\tau)= a_{2}(\tau) = a(\tau)$ is  a function of the conformal time coordinate $\tau$ and not of $\eta$ (since, in this case, $\eta\equiv \tau$). Broadly speaking this choice corresponds to the concordance scenario; the forthcoming discussion must also comply with all the most relevant phenomenological bounds eventually applicable to the diffuse backgrounds of gravitational radiation\footnote{If $a_{1}(\tau) \neq a_{2}(\tau)$ the time coordinate and the underlying cosmological evolution will be different (see, for instance, \cite{SZ3} and \cite{MG3}). These differences will however affect the averaged multiplicities and the variances but not the general form of the multiplicity distribution.}. 

\subsection{The averaged multiplicity and the maximal frequency in the standard lore}
In the standard lore the relic gravitons are produced during an inflationary phase followed by a stage dominated by radiation extending down to the matter-radiation equality.  This conventional timeline leads to a quasi-flat plateau in the spectral energy density \cite{FL1,FL2,FL3}  (see also \cite{AU5} for an extended review).The averaged multiplicities of the produced gravitons depend on the comoving three-momentum $k \leq \overline{k}_{\mathrm{max}}$ where $\overline{k}_{\mathrm{max}}$ denotes the maximal $k$-mode beyond which the spectrum is exponentially suppressed. Although different 
post-inflationary evolutions modify  the quantitative expression of $\overline{k}_{\mathrm{max}}$, there always exist a maximal frequency that cannot exceed the THz range
\cite{AU6}. The explicit computation of $\overline{n}_{k}$ requires the evolution of $u_{k,\,\alpha}$ and $v_{k,\,\alpha}$; for this purpose we first rewrite  Eq. (\ref{DNT8}) in the conformal time coordinate and recall that, in the standard situation of Eqs. (\ref{AC1})--(\ref{AC1b}), $\eta \to \tau$ and $b(\eta) \to a(\tau)$.  The difference of $u_{k,\,\alpha}$ and $v_{k,\,\alpha}^{\ast}$ obeys then the following decoupled equation:
\begin{equation}
(u_{k,\,\alpha} -v^{\ast}_{k,\,\alpha})^{\prime\prime} + \biggl[ k^2 - \bigl( {\mathcal H}^2 + {\mathcal H}^{\prime}\bigr) \biggr] (u_{k,\,\alpha} -v^{\ast}_{k,\,\alpha}) =0, 
\label{FFF1}
\end{equation}
where ${\mathcal H} = a^{\prime}/a$ and the prime denotes, throughout this section, a derivation with respect to the conformal time coordinate $\tau$. The term $({\mathcal H}^2 + {\mathcal H}^{\prime})$ of Eq. (\ref{FFF1}) 
can also be rewritten as\footnote{In the previous section the prime has been also employed to indicate a derivation of the pgf with respect to its own argument $s$. Since  these quantities never appear in the same context the two notations cannot be confused. Finally, the overdot denotes here a derivation with respect to the cosmic time coordinate $t$ (recall that $a(\tau) \, d\tau = d\,t$).} 
\begin{equation}
{\mathcal H}^2 + {\mathcal H}^{\prime} =  a^2 H^2 (2 - \epsilon), \qquad \epsilon = - \dot{H}/H^2,
\label{FFF1a}
\end{equation}
and $\epsilon$ is the standard slow-roll parameters. During a stage of accelerated expansion $\epsilon \ll 1$ but as soon as the inflationary stage terminates $\epsilon = - \dot{H}/H^2 = 3\dot{\varphi}^2/(\dot{\varphi}^2 + 2\,V)\to 1$ where $\varphi$ denotes now the inflaton field with a potential $V(\varphi)$; in practice the end of the inflationary stage occurs  when $\dot{\varphi}^2$ and $V$ coincide within an order of magnitude. Once Eq. (\ref{FFF1}) 
is solved either exactly or approximately (see e.g. the discussion of appendix \ref{APPB}) the orthogonal combination follows by explicit derivation with respect to the conformal time coordinate:
\begin{equation}
u_{k,\,\alpha} + v^{\ast}_{k,\,\alpha} = [(u_{k,\,\alpha} -v^{\ast}_{k,\,\alpha})^{\prime} - {\mathcal H} (u_{k,\,\alpha} -v^{\ast}_{k,\,\alpha})]/k,
\label{FFF1a2}
\end{equation}
as it can be explicitly deduced always from Eq. (\ref{DNT8}).
According to Eqs. (\ref{FFF1a})--(\ref{FFF1a2}) the maximal wavenumber of the spectrum is proportional to $a\,H$ evaluated at the end of the inflationary stage so that the production of gravitons effectively takes place for $k$-modes that do not exceed $k_{\mathrm{max}}$, i.e.
\begin{equation}
 k^2 < k_{\mathrm{max}}^2 \simeq a_{\mathrm{max}}^2 \, H_{\mathrm{max}}^2 = a_{1}^2 \, H_{1}^2.
\label{FFF1b}
\end{equation}
The last equality in Eq. (\ref{FFF1b}) implies that $H_{\mathrm{max}}$ is comparable with the expansion rate at the onset 
of the decelerated stage of expansion\footnote{As already mentioned in the first part of section \ref{sec2}, the convention that $a_{0} = 1$ is adopted throughout; this 
means that at the present conformal time $\tau_{0}$ the comoving and the physical frequencies of the relic gravitons coincide. }.  Since the decrease of the Hubble rate during inflation
is very small (and typically ${\mathcal O}(\epsilon)$) in  Eq. (\ref{FFF1b}) $H_{i}$,  $H_{\mathrm{max}}$ and $H_{k}$ coincide within an order of magnitude; we remind that $H_{k}$ denotes the expansion 
rate when the typical wavelength $2\pi/k$ crosses
the Hubble radius. For the inflationary expansion rate we are then going to adopt the following estimates  
\begin{equation}
H_{k}/M_{P} \simeq H_{\mathrm{max}}/M_{P} \simeq \sqrt{\pi \, r_{T} \, {\mathcal A}_{{\mathcal R}}}/4,
\label{FFF1ba}
\end{equation}
where ${\mathcal A}_{\mathcal R} = {\mathcal O}(2.41) \times 
10^{-9}$ denotes the amplitude of curvature inhomogeneities 
deduced from large-scale experiments \cite{LL0,LL1,LL2,LL3,LL4,LL5}; note that in Eq. (\ref{FFF1ba})  we used that $\epsilon \simeq r_{T}/16$, as implied by the consistency relations (see e.g. \cite{CL1,CL2}).
The total redshift during the radiation stage (i.e. between $a_{1}$ to $a_{eq}$) determines the value of the maximal frequency $\overline{\nu}_{\mathrm{max}}$ in the standard lore
\begin{equation}
\overline{\nu}_{\mathrm{max}} = \frac{ (2 \Omega_{R0})^{1/4}}{2 \pi} \,\, \biggl(\frac{g_{s,\, eq}}{g_{s, \, r}}\biggr)^{1/3} \,\biggl(\frac{g_{\rho,\, r}}{g_{\rho, \, eq}}\biggr)^{1/4} \,\, \sqrt{H_{\mathrm{max}} \, H_{0}}.
\label{FFF1c}
\end{equation}
In Eq. (\ref{FFF1c}) $g_{\rho}$ is the number of relativistic degrees of freedom in the plasma while $g_{s}$ denotes the effective number of relativistic degrees of freedom appearing in the entropy density; in the conventional situation $g_{s,\, r}= g_{\rho,\, r} = 106.75$ and $g_{s,\, eq}= g_{\rho,\, eq} = 3.94$. Always in Eq. (\ref{FFF1c}) $\Omega_{R\, 0}$ indicates the critical fraction of radiation and 
we remind that, in the spatially flat case, the redshift to equality is chiefly determined as $a_{0}/a_{eq} = \Omega_{M\,0}/\Omega_{R\,0}$ where $\Omega_{M\,0}$ is the critical fraction of non-relativistic matter. Taking into account 
the typical numerical values of the different parameters and the 
result of Eq. (\ref{FFF1ba}), Eq. (\ref{FFF1c}) can be written in more explicit terms
\begin{equation}
\overline{\nu}_{\mathrm{max}} = 271.93 \, {\mathcal C}(g_{s}, g_{\rho},\tau_{r},\tau_{eq}) \, \biggl(\frac{{\mathcal A}_{{\mathcal R}}}{2.41\times 10^{-9}}\biggr)^{1/4}\,\,
\biggl(\frac{r_{T}}{0.06}\biggr)^{1/4} \,\, \biggl(\frac{h_{0}^2 \, \Omega_{R\,0}}{4.15\times 10^{-5}}\biggr)^{1/4} \,\,\mathrm{MHz}.
\label{FFF1h}
\end{equation}
 We remark that the impact of ${\mathcal C}(g_{s}, g_{\rho},\tau_{r},\tau_{eq})$  on $\overline{\nu}_{\mathrm{max}}$ is not essential:
for typical values of the late-time parameters (i.e. $g_{\rho, \, r} = g_{s,\, r} = 106.75$ 
and  $g_{\rho, \, eq} = g_{s,\, eq} = 3.94$) we have ${\mathcal C}(g_{s}, g_{\rho},\tau_{r}, \tau_{eq}) =0.7596$ and  $\overline{\nu}_{\mathrm{max}}$ gets shifted from $271.93$ MHz to $206.53$ MHz. 
For $\nu < \overline{\nu}_{\mathrm{max}}$ the averaged multiplicity at the present time is given by\footnote{In this section it is 
practical to consider directly the dependence of the averaged multiplicity upon the comoving frequency $\nu$ and this 
is why we prefer to write $\overline{n}_{\nu}$ rather than $\overline{n}_{k}$; similarly we shall write $v_{\nu}$ instead of $v_{k}$ and so on. We remind that, in the present units, $ k = 2\pi \nu$.}:
\begin{equation}
\overline{n}_{\nu} = \kappa_{\nu}\, \bigl|v_{\nu}\bigr|^2 = \kappa_{\nu} \,\biggl(\frac{\nu}{\overline{\nu}_{\mathrm{max}}}\biggr)^{\overline{m}_{T} -4}, \qquad \nu< \nu_{\mathrm{max}}.
\label{FFF1d}
\end{equation}
 To deduce Eq. (\ref{FFF1d}), Eqs. (\ref{FFF1a}) and (\ref{FFF1a2}) must be solved explicitly. For the sake of completeness the derivation of the averaged multiplicity has been illustrated in appendix \ref{APPB} within the perspective 
 provided by the Wentzel–Kramers–Brillouin (WKB) approximation.
The spectral index $\overline{m}_{T}$ appearing in Eq. (\ref{FFF1d}) depends both on the inflationary dynamics and on the post-inflationary evolution; if we adopt the same timeline leading to Eq. (\ref{FFF1c}) the value of $\overline{m}_{T}$ is given by \cite{FL1,FL2,FL3} (see also \cite{AU5,KK3} and the last part of appendix \ref{APPB}):
\begin{equation}
\overline{m}_{T} = - 2\, \epsilon/(1 - \epsilon) \simeq - 2 \epsilon + {\mathcal O}(\epsilon^2) = - r_{T}/8 + {\mathcal O}(r_{T}^2).
\label{FFF1e}
\end{equation}
As in Eq. (\ref{FFF1h}), also in the last equality of Eq. (\ref{FFF1e}) the consistency relations have been employed; moreover, since   $r_{T}\ll 1$ we have that $\overline{m}_{T} \ll 1$. This means, overall, that in the concordance 
scenario the averaged multiplicity of  Eq. (\ref{FFF1d}) approximately scales as $\nu^{-4}$. 
When the comoving frequency gets of the order of $\overline{\nu}_{\mathrm{max}}$ the averaged multiplicity gets $\overline{n}_{\nu} = {\mathcal O}(1)$ and for  $\nu > \overline{\nu}_{\mathrm{max}}$ 
the pair production process described by Eqs. (\ref{DR2})--(\ref{DR3}) and (\ref{FFF1})--(\ref{FFF1a2}) is exponentially suppressed. As observed long 
ago \cite{BIRREL,PTOMS,PARKTH} (see also \cite{MD5,MD6}) the ratio of the square moduli of the mixing coefficients must scale as:
\begin{equation}
|v_{\nu}|^2/|u_{\nu}|^2 = |v_{\nu}|^2/(1 + |v_{\nu}|^2) = e^{-\gamma (\nu/\overline{\nu}_{\mathrm{max}})}.
\label{FFF1f}
\end{equation}
The value of $\gamma$ (typically between $1$ and $3$) can be estimated both 
analytically and numerically \cite{TLCDM1,TLCDM2,TLCDM3} and it is controlled by the smoothness of the transition between the inflationary and the post-inflationary stages.  Equation (\ref{FFF1f}) applies in all the situations where the evolution of the expansion rate is continuous across the various transitions and it implies that the production of particles falls off faster than any inverse power of $\nu$ (or of $k$) \cite{BIRREL}. Since  Eqs. (\ref{FFF1d})--(\ref{FFF1f}) are complementary limits of a single expression, the mean number of pairs can be written in the following interpolating form:
\begin{equation}
\overline{n}_{\nu} = \kappa_{\nu} \,\gamma\, \frac{(\nu/\overline{\nu}_{\mathrm{max}})^{\overline{m}_{T} - 3}}{e^{\gamma\,(\nu/\overline{\nu}_{\mathrm{max}}) } -1}.
\label{FFF1g}
\end{equation}
The averaged multiplicity introduced in Eq. (\ref{FFF1g}) correctly interpolates between the regime $\nu< \nu_{\mathrm{max}}$ (where $\overline{n}_{\nu}$ scales as in Eq. (\ref{FFF1d})) and $\nu> \nu_{\mathrm{max}}$ 
(where $|v_{\nu}|^2$ is exponentially suppressed as in Eq. (\ref{FFF1f})).  From the actual values of $\overline{\nu}_{\mathrm{max}}$ and from the observation that $\overline{m}_{T} \ll 1$ it follows that $\overline{n}_{\nu}/\kappa_{\nu} \gg 1$ when $\nu \ll \overline{\nu}_{\mathrm{max}}$. 
Let us finally consider, for the sake of illustration, some numerical estimates. In the kHz range (roughly corresponding to the highest frequency probed by wide-band detectors) Eqs. (\ref{FFF1h})--(\ref{FFF1d}) approximately give:
\begin{equation}
\overline{n}_{\nu}/\kappa_{\nu} = {\mathcal O}(10^{20}) (\nu/\mathrm{kHz})^{-4}.
\label{FFFi}
\end{equation}
To obtain Eq. (\ref{FFFi}) we estimate $\overline{\nu}_{\mathrm{max}} = {\mathcal O}(300) \, \mathrm{MHz}$ in the limit $\overline{m}_{T} \ll 1$.  Equation (\ref{FFFi}) implies that when $\nu \ll \overline{\nu}_{\mathrm{max}}$ the multiplicity distribution falls in the regime 
$\overline{n}_{\nu} \gg \kappa_{\nu}$. When $\nu = {\mathcal O}(\nu_{\mathrm{max}})$ we have instead that $\overline{n}_{\nu}$ and $\kappa_{\nu}$ are of the same order. Finally for $\nu \gg \overline{\nu}_{\mathrm{max}}$ the average multiplicity gets suppressed as if $\kappa_{\nu} \to \infty$.

\subsection{Absolute bounds on the maximal frequency}
Equations (\ref{FFF1c})--(\ref{FFF1h}) provide an estimate of $\overline{\nu}_{\mathrm{max}}$
when the post-inflationary evolution is dominated by radiation.  
For a different timeline the maximal frequency $\nu_{\mathrm{max}}$ 
does not necessarily coincide with Eqs. (\ref{FFF1c})--(\ref{FFF1h}) and 
the simplest illustrative possibility is to consider a single expanding stage between the end of inflation and the onset of the radiation epoch at a putative scale $H_{r}$ (see also Eqs. 
(\ref{APPB12})--(\ref{APPB13}) and discussion therein). In this situation the connection between $\nu_{\mathrm{max}}$ and $\overline{\nu}_{\mathrm{max}}$ is given by:
\begin{equation}
\nu_{\mathrm{max}} = (H_{r}/H_{1})^{\frac{\delta -1}{2(\delta +1)}} \, \overline{\nu}_{\mathrm{max}},
\label{GGG1}
\end{equation}
where $H_{1}$ and $H_{r}$ denote, respectively, the Hubble rates at the end of the inflationary stage and at the onset of the epoch dominated by radiation. 
The values of $\delta$ appearing in Eq. (\ref{GGG1}) control the  
timeline between $H_{1}$ and $H_{r}< H_{1}$ and when $\delta > 1$ the expansion rate is {\em faster} than radiation; the maximal frequency of Eq. (\ref{GGG1}) gets then {\em smaller} than ${\mathcal O}(300)$ MHz. In the opposite situation (i.e. for an expansion rate {\em slower} than radiation)  we have instead $\delta < 1$ and $\nu_{\mathrm{max}} < \overline{\nu}_{\mathrm{max}}$. Finally when $\delta \to 1$ we have $\nu_{\mathrm{max}} \to \overline{\nu}_{\mathrm{max}}$ corresponding to a radiation stage extending between $H_{1}$ and the equality time (see also (\ref{APPB12})--(\ref{APPB13}) and the related considerations).

Depending on the hierarchy between $H_{r}$ and $H_{1}$ it is possible to conceive situations where\footnote{For instance a lower bound on $H_{r}$ is obtained by requiring $H_{r} > H_{\mathrm{bbn}}$ where $H_{\mathrm{bbn}}$ indicates the 
Hubble rate at the time of big-bang nucleosynthesis (BBN). Since the plasma 
must be already dominated by radiation for typical temperatures $T_{\mathrm{bbn}} = {\mathcal O}(\mathrm{MeV})$ we should also impose $H_{r} > 10^{-44}\, M_{P}$. 
By taking the extreme values $H_{r} = {\mathcal O}(10^{-44})\, M_{P}$ and 
$H_{1} = {\mathcal O}(10^{-6})\, M_{P}$ we would obtain $\nu_{\mathrm{max}} = {\mathcal O}(100) \, \mathrm{THz}$. This value exceeds however 
the quantum bound that we are going to discuss.} 
$\nu_{\mathrm{max}} = {\mathcal O}(\mathrm{GHz})$ or even $\nu_{\mathrm{max}} > {\mathcal O}(\mathrm{THz})$. 
Although from the classical viewpoint the maximal frequency of Eq. (\ref{GGG1}) depends both on the inflationary dynamics and on the decelerated evolution, in a quantum mechanical perspective it is possible to demonstrate that there always exist an absolute bound stipulating that $\nu_{\mathrm{max}}$ cannot exceed frequencies ${\mathcal O}(\mathrm{THz})$ \cite{AU6}. If we look at Eq. (\ref{FFF1g}) we can argue that as $\nu \to \overline{\nu}_{\mathrm{max}}$ the average multiplicity becomes ${\mathcal O}(1)$. With this observation in mind we can estimate the spectral energy density of the relic gravitons 
in critical units and write\footnote{We might want to restore the $\hbar$ dependence in $\Omega_{gw}(\nu,\tau_{0})$. We must then consider that the energy of a single graviton is given by $\hbar\, \omega$ where 
$\omega = k c$; another $\hbar$ comes from the definition of Planck mass. This $\Omega_{gw}(\nu,\tau_{0}) \propto \hbar^2$ as suggested in \cite{AS10}. This occurrence shows that, in practice, the diffuse backgrounds of relic gravitons have a quantum mechanical origin. In spite of this observation, as mentioned at the beginning of section \ref{sec2} we use throughout units $\hbar = c = k_{B} =1$.} 
\begin{equation}
\Omega_{gw}(\nu, \tau_{0}) = \frac{1}{\rho_{c}} \frac{ d \langle \rho_{gw} \rangle}{d \ln{\nu}} = \frac{128 \pi^3}{3} \biggl(\frac{\nu}{\sqrt{H_{0} \, M_{P}}}\biggr)^{4} \, \overline{n}_{\nu}.
\label{GGG2}
\end{equation}
Equation (\ref{GGG2}) can be always referred to a putative $\nu_{\mathrm{max}}$ beyond which the ratio of the mixing coefficients is exponentially suppressed as established in Eq. (\ref{FFF1f}). According to  Eq. (\ref{GGG2})  the maximal frequency of the spectrum corresponds in fact to the production of a single pair of gravitons with opposite three-momenta. For this reason we have, in practice, 
\begin{equation}
\Omega_{gw}(\nu, \tau_{0}) =   \frac{128 \pi^3}{3} \frac{\nu_{\mathrm{max}}^{4}}{H_{0}^2 \, M_{P}^2}
 \biggl(\frac{\nu}{\nu_{\mathrm{max}}}\biggr)^{4}  \, \overline{n}_{\nu_{\mathrm{max}}},
 \label{GGG3}
 \end{equation} 
 where $\overline{n}_{\nu_{\mathrm{max}}} = {\mathcal O}(1)$. Thanks to Eq. (\ref{GGG3}) 
 we can deduce the absolute upper bound on the maximal frequency 
 of cosmic gravitons which is approximately given by \cite{AU6}
\begin{equation}
\nu_{\mathrm{max}} < {\mathcal O}(10^{-2}) \, \sqrt{H_{0} \, M_{P}} < {\mathcal O}(\mathrm{THz}),
\label{GGG4}
\end{equation}
where we estimated  $H_{0} = {\mathcal O}(10^{-18}) \mathrm{aHz}= {\mathcal O}(\mathrm{aHz})$ and $M_{P} = {\mathcal O}(10^{43}) \mathrm{Hz}$.
If we would simply require $\Omega_{gw}(\nu_{\mathrm{max}}, \tau_{0}) < 1$ 
(as suggested by the critical density bound) the factor 
 ${\mathcal O}(10^{-2})$ would be absent. This factor is however essential 
 to comply with the big-bang nucleosynthesis (BBN) bound setting a limit on the 
 concentration of massless species at the time of the synthesis of light nuclei.
Indeed, the argument leading to Eq. (\ref{GGG4}) can be refined 
if the averaged multiplicity of the produced gravitons is directly inserted into the BBN bound \cite{BBN1,BBN2,BBN3}
\begin{equation}
h_{0}^2 \, \int_{\nu_{bbn}}^{\nu_{\mathrm{max}}} \,\Omega_{gw}(\nu,\tau_{0}) \,\,d\ln{\nu} < 5.61\times 10^{-6} \biggl(\frac{h_{0}^2 \,\Omega_{\gamma0}}{2.47 \times 10^{-5}}\biggr) \, \Delta N_{\nu},
\label{GGG5}
\end{equation}
where $\Omega_{\gamma\,0}$ is the (present) critical fraction of CMB photons.  Equation (\ref{GGG5}) sets a constraint  on the extra-relativistic species possibly present at the time of nucleosynthesis. Since this limit is relevant in the context of neutrino physics, the bound is often expressed in via\footnote{The actual bounds on $\Delta N_{\nu}$ range from $\Delta N_{\nu} \leq 0.2$ to $\Delta N_{\nu} \leq 1$ so that the integrated spectral density in Eq. (\ref{GGG5}) must range, at most, between  $10^{-6}$ and $10^{-5}$. }  $\Delta N_{\nu}$ (i.e. the contribution of supplementary neutrino species). 
In the general case (not necessarily coinciding with the conventional lore) $\overline{n}_{\nu}$ is expressed as
\begin{equation}
\overline{n}_{\nu} = \kappa_{\nu} \,\gamma\, \frac{(\nu/\nu_{\mathrm{max}})^{m_{T} - 3}}{e^{\gamma\,(\nu/\nu_{\mathrm{max}}) } -1}.
\label{GGG6}
\end{equation}
Equation (\ref{GGG6}) generalizes Eq. (\ref{FFF1g}) since now $\nu_{\mathrm{max}}$ 
is given by Eq. (\ref{GGG1}) while $m_{T}$ becomes:
\begin{equation}
m_{T} = \frac{2 - 4 \epsilon}{1 - \epsilon} - 2 \delta = \frac{32 - 4 r_{T}}{16 - r_{T}} - 2 \delta.
\label{GGG7}
\end{equation}
In Eq. (\ref{GGG7}) the second equality follows by enforcing the consistency relations. In the limit $r_{T} \ll 1$ we can expand the first term in Eq. (\ref{GGG7}) and obtain $m_{T} = 2 ( 1 - \delta) - r_{T}/8 + {\mathcal O}(r_{T}^2)$. When $\delta \to 1$ 
we have instead $m_{T} \to \overline{m}_{T} = - r_{T}/8 + {\mathcal O}(r_{T}^2)$ which is the result already obtained in Eq. (\ref{FFF1e}). 
\begin{figure}[!ht]
\centering
\includegraphics[height=8.2cm]{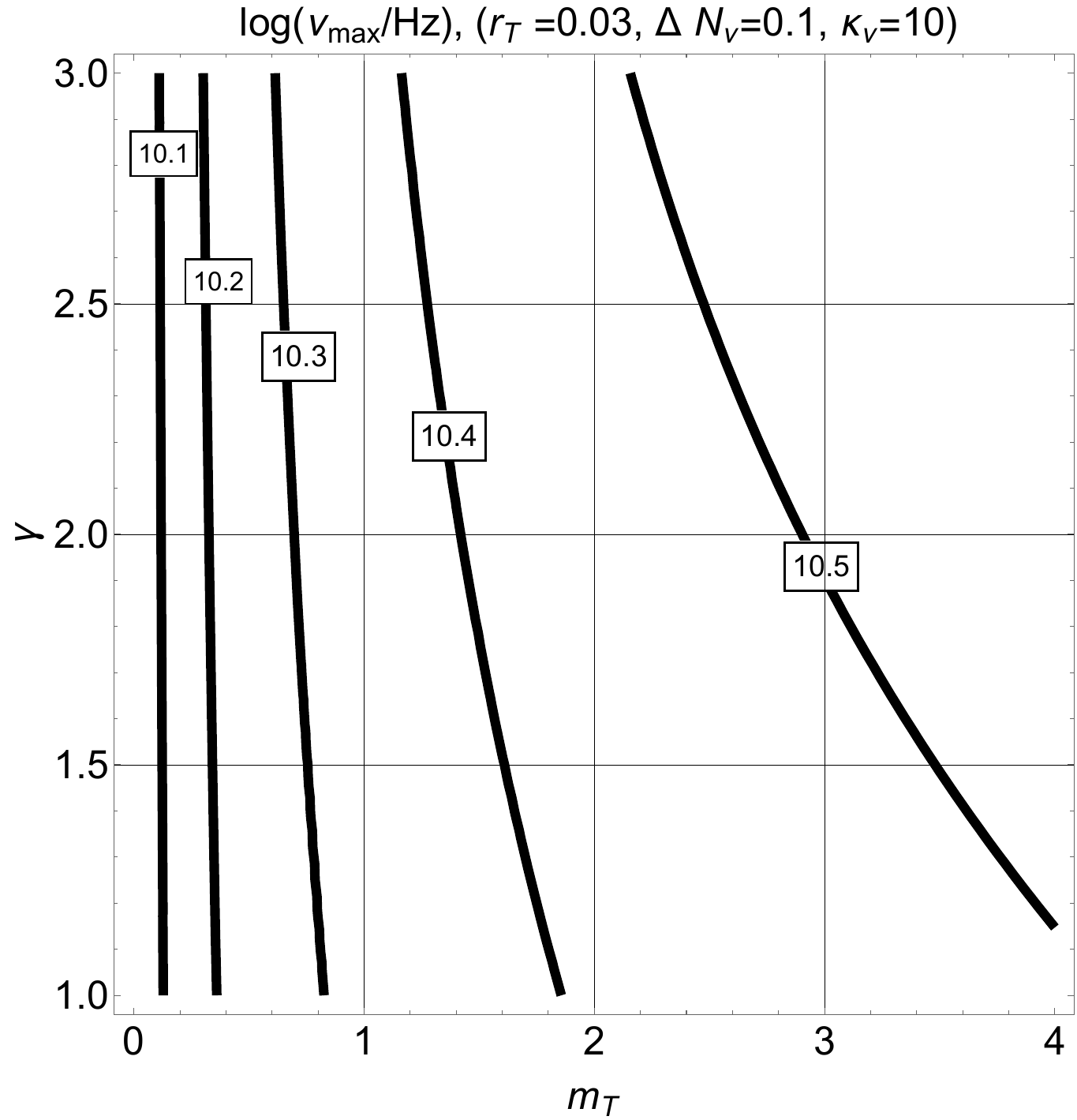}
\includegraphics[height=8.2cm]{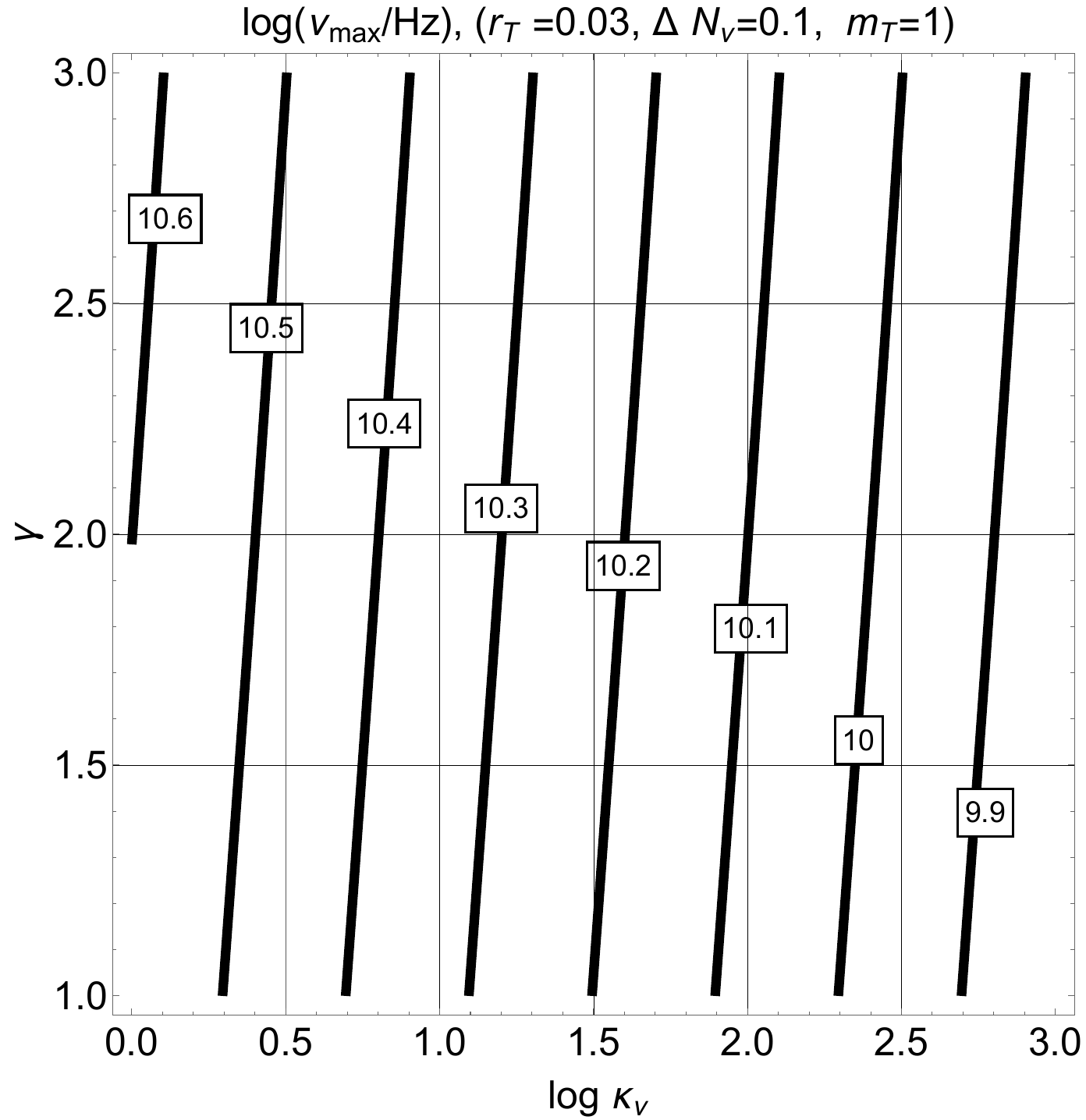}
\includegraphics[height=8.2cm]{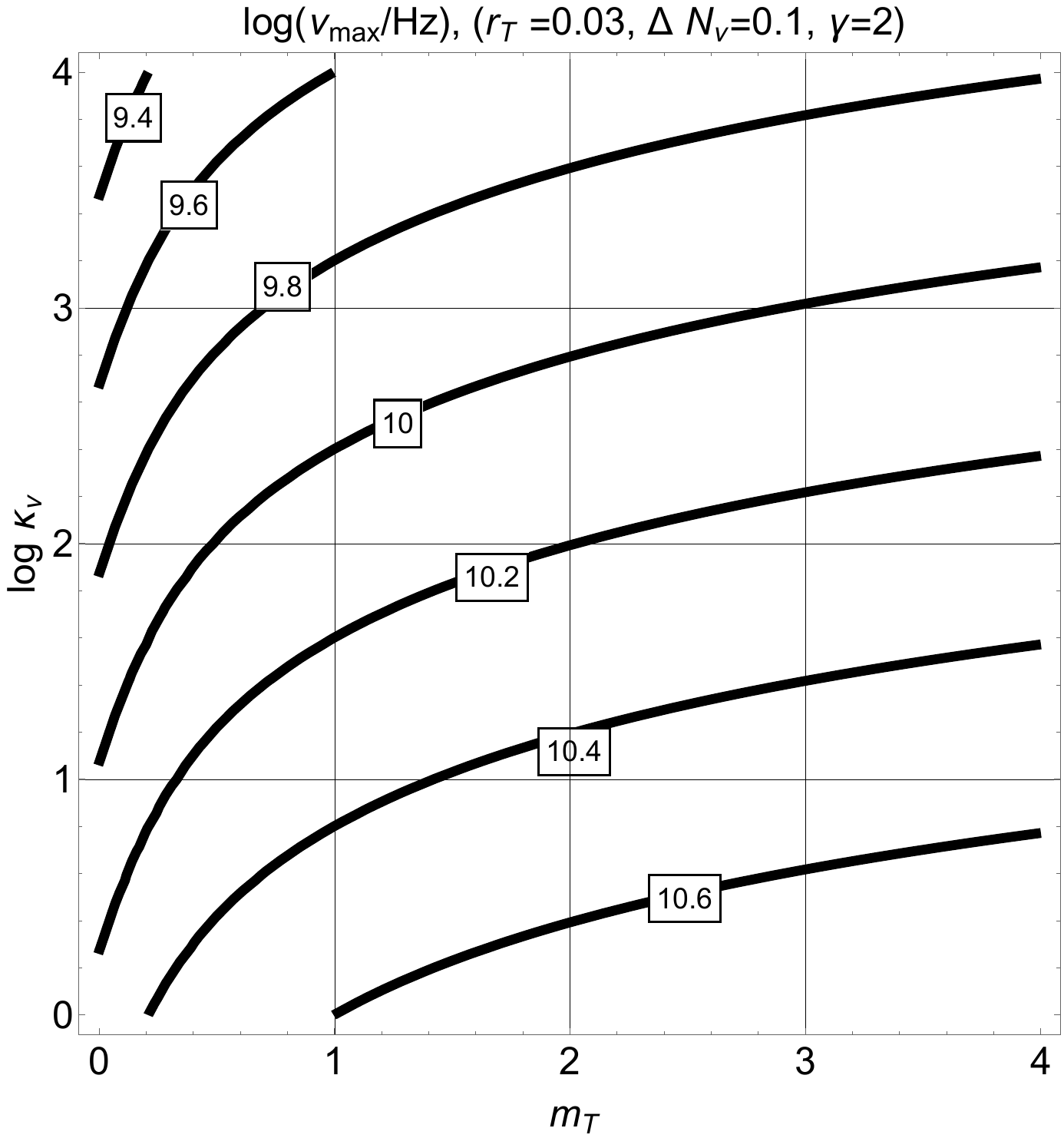}
\includegraphics[height=8.2cm]{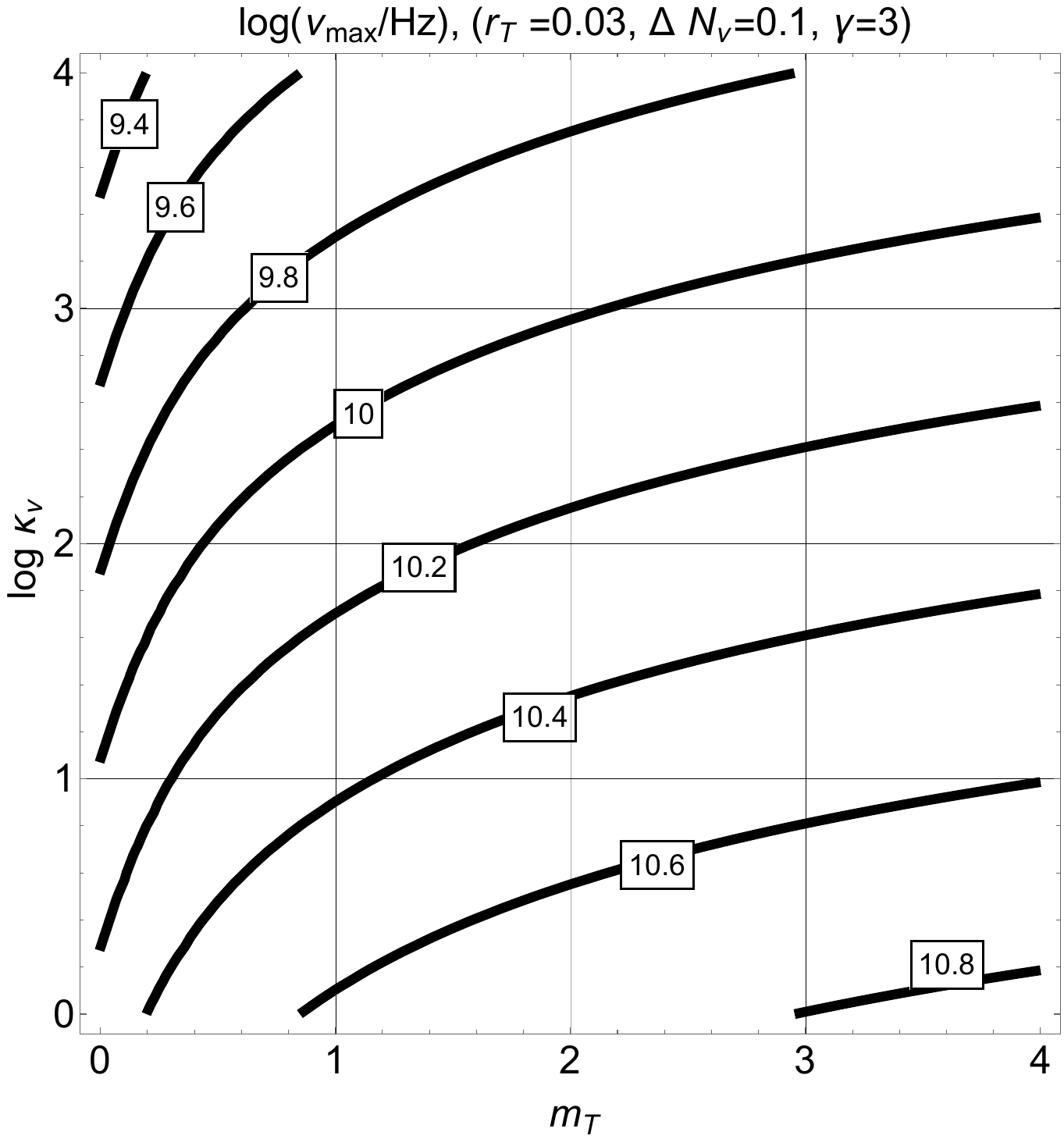}
\caption[a]{The different labels in both plots indicate the common logarithms of $(\nu_{\mathrm{max}}/\mathrm{Hz})$ determined from Eqs. (\ref{GGG11})--(\ref{GGG12}) and (\ref{GGG13}). In all the four plots the minimal value of $m_{T}$ corresponds to the one of the concordance scenario supplemented by the consistency conditions (i.e. $m_{T} \simeq - r_{T}/8$) whereas the value of $r_{T}$ is consistent with the current determinations (i.e. $r_{T} < 0.035$ ).  The two upper plots illustrate the parameter space in the $(\gamma,\, m_{T})$ and in the $(\gamma, \log{\kappa_{\nu}})$ planes. The two lower plots involve instead the $(\log{\kappa_{\nu}}, \,m_{T})$ plane for two different values of $\gamma$.}
\label{FIGURE1}      
\end{figure}
If we now insert Eq. (\ref{GGG6}) into the spectral energy density (\ref{GGG2}) we obtain 
\begin{equation}
\Omega_{gw}(\nu,\tau_{0}) = \frac{128 \,\pi^3}{3} \gamma \kappa_{\nu} \biggl( \frac{\nu_{\mathrm{max}}}{\sqrt{H_{0}\, M_{P}}}\biggr)^{4} \frac{(\nu/\nu_{\mathrm{max}})^{m_{T} +1}}{e^{\gamma\,(\nu/\nu_{\mathrm{max}}) } -1}.
\label{GGG8}
\end{equation}
The result of Eq. (\ref{GGG8}) can be also presented in a more eloquent 
manner that is particularly suitable for the applications, namely 
\begin{equation}
h_{0}^2\, \Omega_{gw}(\nu,\tau_{0}) = 3.66 \times 10^{-37} \biggl(\frac{\nu_{\mathrm{max}}}{\mathrm{kHz}}\biggr)^4 \overline{\Omega}_{gw}(\nu/\nu_{\mathrm{max}},\, m_{T},\, \gamma,\, \kappa_{\nu}),
\label{GGG9}
\end{equation}
where $\overline{\Omega}_{gw}(\nu/\nu_{\mathrm{max}},\, m_{T},\, \gamma,\, \kappa_{\nu})$ 
indicates the following scale-invariant and dimensionless combination:
\begin{equation}
\overline{\Omega}_{gw}(\nu/\nu_{\mathrm{max}},\,m_{T},\, \gamma,\, \kappa_{\nu}) = 
\gamma \kappa_{\nu} \frac{x^{m_{T}+1}}{e^{\gamma x}-1}, \qquad x = \nu/\nu_{\mathrm{max}}.
\label{GGG10}
\end{equation}
Equations (\ref{GGG9})--(\ref{GGG10}) can now be inserted into Eq. (\ref{GGG5}) 
with the result that the BBN bound translates into the condition 
\begin{equation}
3.66 \times 10^{-37} \biggl(\frac{\nu_{\mathrm{max}}}{\mathrm{kHz}}\biggr)^4\, 
Q(m_{T}, \, \gamma, \, \kappa_{\nu}) \leq 5.61 \, \Delta N_{\nu} \,\biggl(\frac{h_{0}^2 \,\Omega_{\gamma0}}{2.47 \times 10^{-5}}\biggr),
\label{GGG11}
\end{equation}
that constrains both $\nu_{\mathrm{max}}$ and $Q(m_{T}, \, \gamma, \, \kappa_{\nu})$
\begin{equation}
Q(m_{T}, \, \gamma, \, \kappa_{\nu}) = \kappa_{\nu} \, \gamma \, \int_{\nu_{\mathrm{bbn}}/\nu_{\mathrm{max}}}^{\infty} \, \frac{x^{m_{T}}}{e^{\gamma x} -1} \, d x.
\label{GGG12}
\end{equation}
The integral of Eq. (\ref{GGG12}) is convergent in the limit $x \to \infty$ (i.e. $\nu\to \infty$) whereas the lower 
limit\footnote{The lower limit of integration leads to a potential divergence 
for $m_{T}\to 0$ but this never occurs since $m_{T}$ never goes to zero. In the standard lore $m_{T} \simeq - r_{T}/8 \ll 1$
but the exact scale-invariant case is never reached. When $m_{T} >0$ we can ignore 
the lower bound of integration in $Q(m_{T}, \, \gamma, \, \kappa_{\nu})$. } goes in practice to $0$ since 
$\nu_{\mathrm{bbn}} = {\mathcal O}(10^{-2})$ nHz and 
$\nu_{\mathrm{max}} \gg  \nu_{\mathrm{bbn}}$. Equation (\ref{GGG11}) implies that 
\begin{equation}
\nu_{\mathrm{max}} \leq  {\mathcal O}(10^{11}) \,\, \biggl[\frac{\Delta N_{\nu}}{Q(m_{T}, \gamma)}\biggr]^{1/4} \, \mathrm{Hz} < \mathrm{THz},
\label{GGG13}
\end{equation}
and $[\Delta N_{\nu}/Q(m_{T}, \gamma)]< {\mathcal O}(10)$. Instead of using the approximation (\ref{GGG13}) the condition 
of Eq. (\ref{GGG11}) can be analyzed numerically and the result is summarized in Fig. \ref{FIGURE1} for the different 
portions of the parameter space. In the four plots of Fig. \ref{FIGURE1} the labels appearing on the various curves indicate 
the common logarithm of $\nu_{\mathrm{max}}$ expressed in Hz (i.e. $\log{(\nu_{\mathrm{max}}/\mathrm{Hz})}$, as reminded on top of each 
of the four plots of Fig. \ref{FIGURE1}). We consider values of $m_{T}$ between $-r_{T}/8$ and $4$ while the values of $\gamma$ typically vary between $1$ and $3$. The values of $\kappa_{\nu}$ are typically fixed by backreaction considerations \cite{AS2} and a simple argument suggests 
that at the onset of the dynamical evolution the total energy content of the gravitons would roughly be of the order 
of $\kappa_{\nu}\, H^4$; this quantity should always undershoot $3 \, H^2 \, \overline{M}_{P}^2$. Therefore we should have, broadly speaking,
that $\kappa_{\nu} < {\mathcal O}(10^{10})$; in a conservative perspective we therefore considered $1\leq \kappa_{\nu} < {\mathcal O}(10^{4})$.
As we can clearly see from the different plots of Fig. \ref{FIGURE1} the maximal frequency is always smaller than the THz.

\subsection{Multiplicity distributions in different frequency intervals}
\begin{figure}[!ht]
\centering
\includegraphics[height=6.5cm]{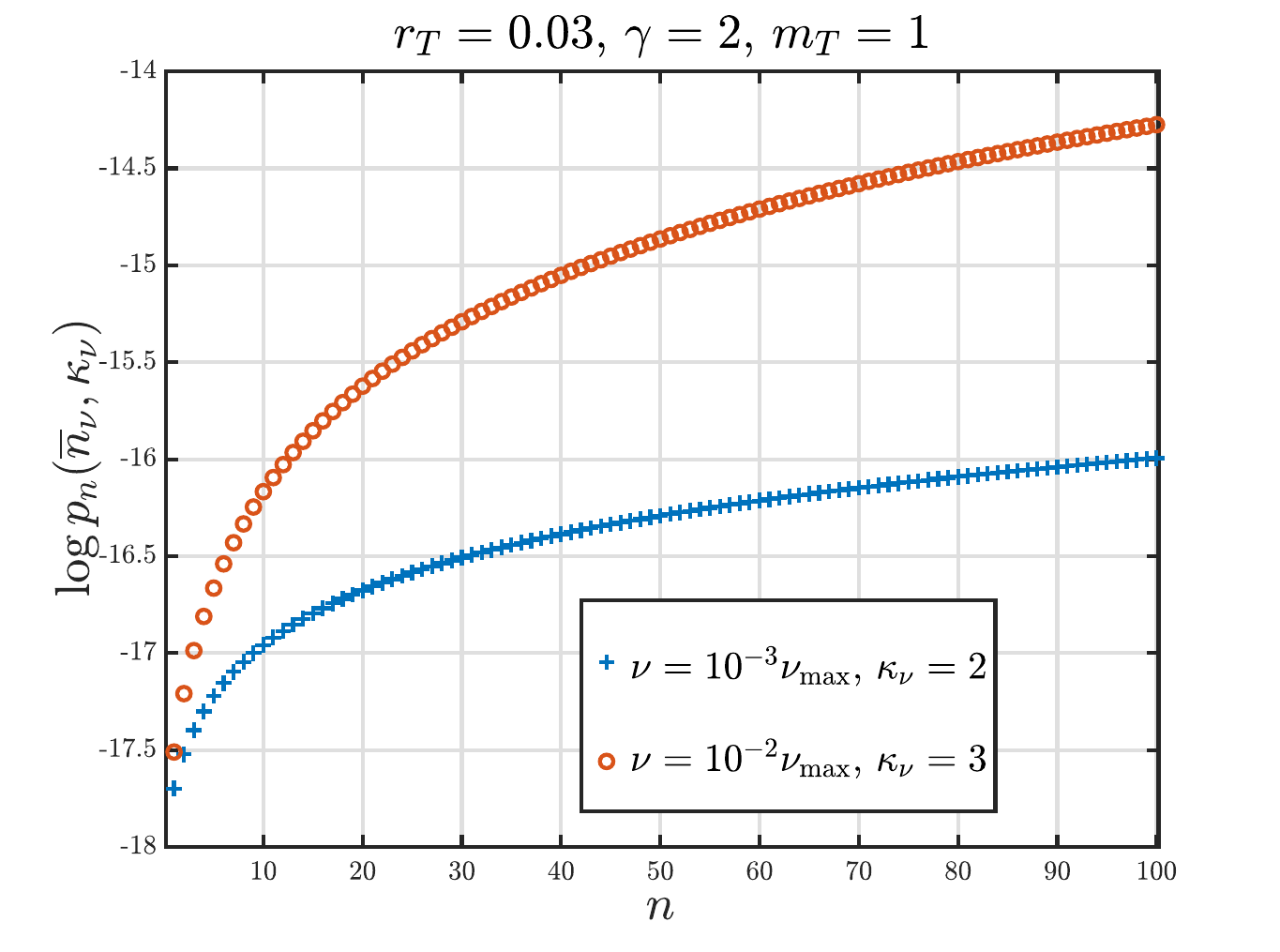}
\includegraphics[height=6.5cm]{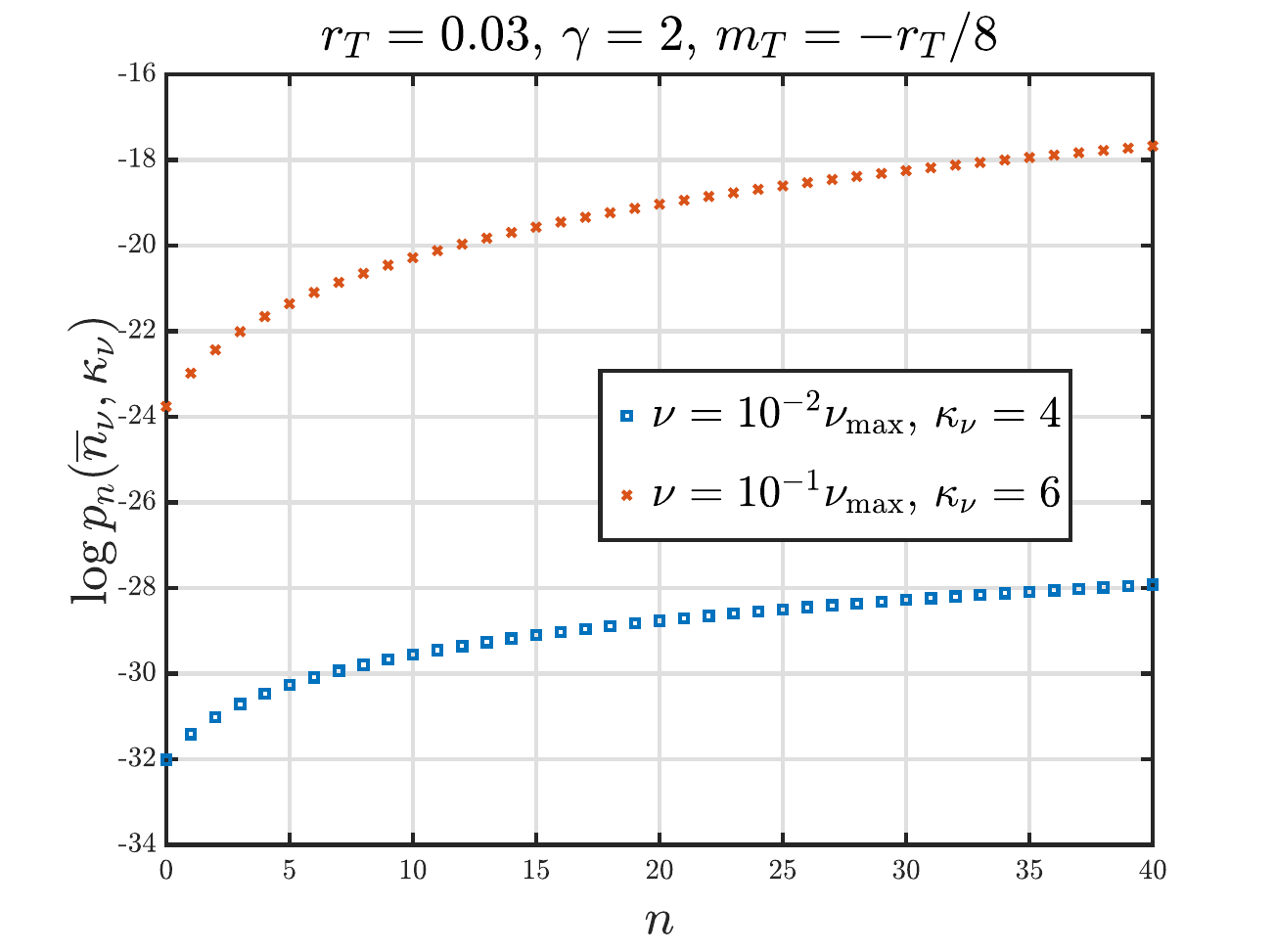}
\caption[a]{In both plots the common logarithm of $p_{n}(\overline{n}_{\nu}, \kappa_{\nu})$ is illustrated. As the legends of the plots clarify we are in the regime $\nu < \nu_{\mathrm{max}}$. The shape of the multiplicity distribution 
suggests that we are already in the asymptotic regime characterized by a Gamma distribution (see Eq. (\ref{av14}) and discussion therein).
In the left plot we have chosen $m_{T} = 1$ while the plot at the right we have posited $m_{T} = - r_{T}/8$ 
as it happens in the case of the concordance scenario. In this and in the following plots the large-scale bound  
on the tensor to scalar ratio (i.e. $r_{T} < 0.035$) has been always enforced by choosing $r_{T} =0.03$.}
\label{FIGURE2}      
\end{figure}
Some of the relevant technical features of the distributions have been already scrutinized in section \ref{sec4} but their different physical limits also depend on the frequency domain that determines the averaged multiplicity (see, e.g.   Eqs. (\ref{FFF1g})--(\ref{FFF1f}) and (\ref{GGG6})). In the concordance scenario $\overline{n}_{\nu}$ can be ${\mathcal O}(10^{20})$ already in the kHz region. When the post-inflationary expansion rate is slower than radiation $\overline{n}_{\nu}$ may get even larger but as soon as $\nu = {\mathcal O}(\nu_{\mathrm{max}})$ the averaged multiplicity becomes ${\mathcal O}(1)$ and it is finally suppressed when $\nu\gg \nu_{\mathrm{max}}$. The properties of the multiplicity distribution are determined by the interplay between $\overline{n}_{\nu}$ and $\kappa_{\nu}$: besides the first regime $\overline{n}_{\nu} \gg \kappa_{\nu}$ there are two supplementary physical cases corresponding to  $\overline{n}_{\nu} = {\mathcal O}(\kappa_{\nu})$ and to $\kappa_{\nu} \gg \overline{n}_{\nu}$. In what follows we shall refer to the multiplicity distribution (\ref{DR15}) written in the form\footnote{Strictly speaking the multiplicity 
$n$ could have been indicated by $n_{\nu}$ but we avoided this style for the sake of conciseness.} 
\begin{eqnarray}
p_{n}(\overline{n}_{\nu}, \kappa_{\nu}) = \frac{\Gamma(\kappa_{\nu} + n)}{\Gamma(\kappa_{\nu}) \Gamma(n+1)}   \biggl(\frac{\overline{n}_{\nu}}{\overline{n}_{\nu} + \kappa_{\nu}}\biggr)^{n} \biggl(\frac{\kappa_{\nu}}{\overline{n}_{\nu}+ \kappa_{\nu}}\biggr)^{\kappa_{\nu}},
\label{MULT1}
\end{eqnarray}
where the averaged multiplicity $\overline{n}_{\nu}$ is now given by Eq. (\ref{FFF1g}).  The purpose of the remaining part of this section is to illustrate 
the shapes of the multiplicity distributions and their modifications depending 
upon the frequency interval.

\subsubsection{The region $\overline{n}_{\nu} \gg \kappa_{\nu}$}
When $\nu < \nu_{\mathrm{max}}$  the averaged multiplicity gets always larger  
than $\kappa_{\nu}$ as it follows from the interpolating expressions of $\overline{n}_{\nu}$  discussed in 
Eqs. (\ref{FFF1g}) and (\ref{GGG6})--(\ref{GGG7}). Thus, for $\nu < \nu_{\mathrm{max}}$ and $\overline{n}_{\nu} \gg \kappa_{\nu}$ the results of Eq. (\ref{av14}) would suggest that the multiplicity distributions of the relic gravitons could be approximated by a Gamma distribution. To illustrate this physical sitaution 
we  first consider $\nu = {\mathcal O}(10^{-2})\, \nu_{\mathrm{max}}$ 
and $\kappa_{\nu} = {\mathcal O}(1)$.  In the two plots of Fig. \ref{FIGURE2} Eq. (\ref{MULT1}) is illustrated for increasing values of $\kappa_{\nu}$. The frequency is selected in the range where $\overline{n}_{\nu}$ always 
exceeds $\kappa_{\nu}$: for instance we have from Eq. (\ref{FFF1g}) that when $m_{T} ={\mathcal O}(1)$ the ratio $\overline{n}_{\nu}/\kappa_{\nu}$ is ${\mathcal O}(10^{8})$ for $\nu = {\mathcal O}(10^{-2})\, \nu_{\mathrm{max}}$.
In the left plot of Fig. \ref{FIGURE2} we have set $m_{T} = 1$ while in the plot at the right $m_{T} = - r_{T}/8$ as it happens for the concordance paradigm. In both plots we can see the characteristic shape of the Gamma distribution arising  from
of $p_{n}(\overline{n}_{\nu},\kappa_{\nu})$ in the limit $\overline{n}_{\nu} > \kappa_{\nu}$. 
\begin{figure}[!ht]
\centering
\includegraphics[height=6.2cm]{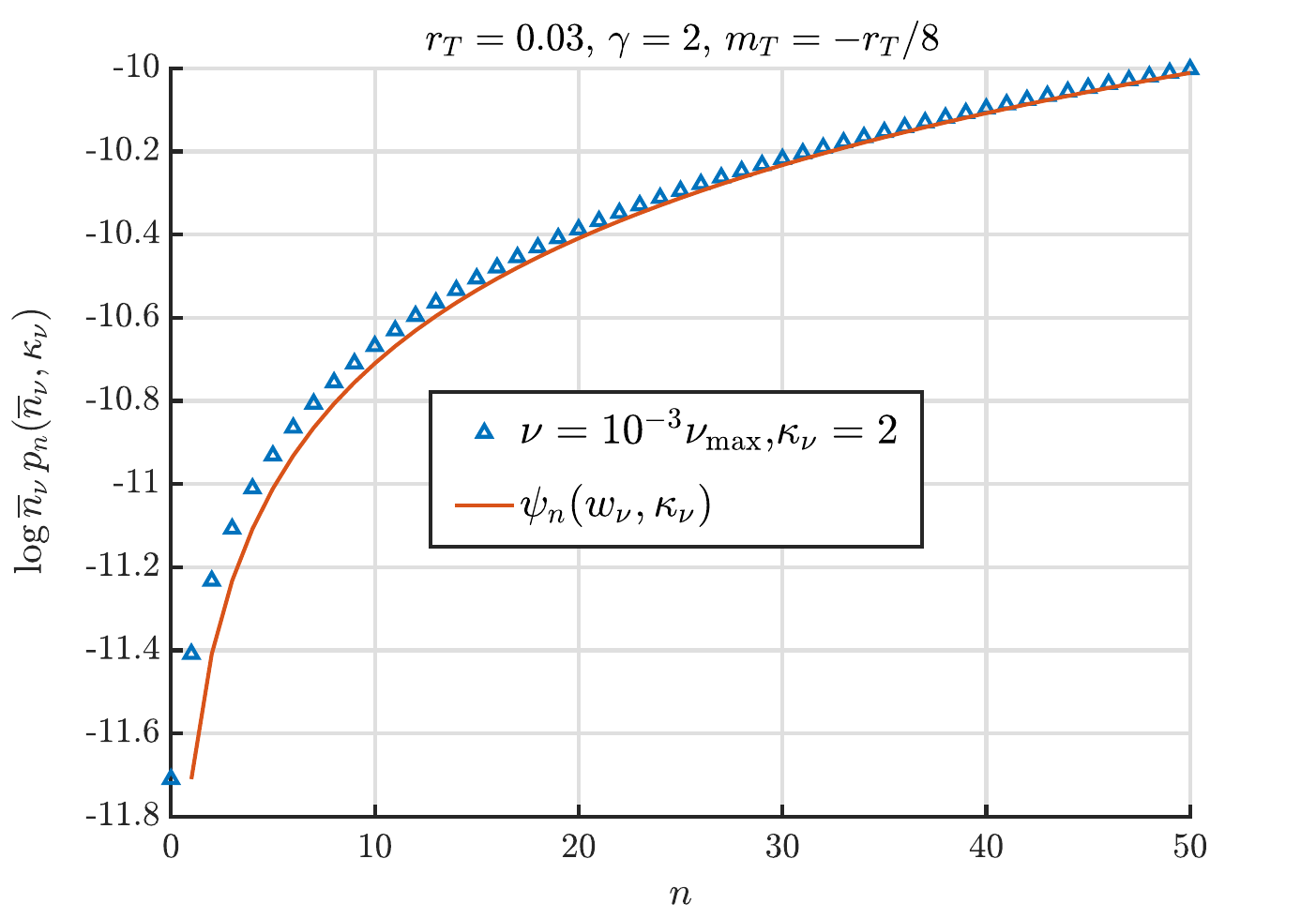}
\includegraphics[height=6.2cm]{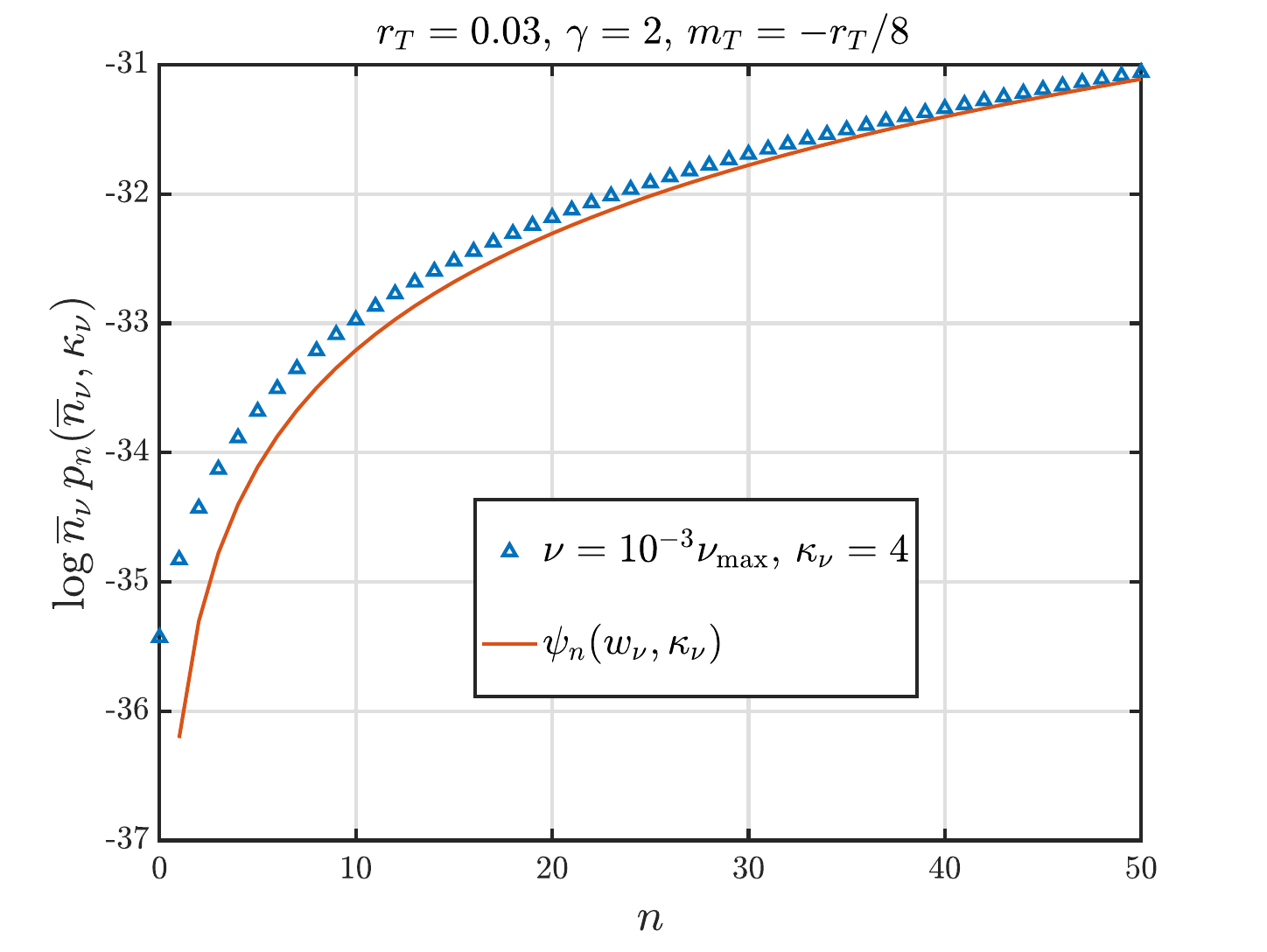}
\caption[a]{In both plots the common logarithm of $\overline{n}_{\nu} \,\,p_{n}(\overline{n}_{\nu}, \kappa_{\nu})$ is illustrated. With 
the full line we denote the approximation of the original distribution in the limit $\overline{n}_{\nu}/\kappa_{\nu}\gg 1$. For the 
same averaged multiplicity the asymptotic form of the distribution (i.e. $\psi_{n}(w_{\nu}, \kappa_{\nu}$)) is reached faster 
when $\kappa_{\nu}$ is comparatively smaller. We note that in both plots we have chosen $m_{T} = \overline{m}_{T} = - r_{T}/8$.}
\label{FIGURE3}      
\end{figure}
To characterize this limit  even more explicitly we plot 
directly the distributions in KNO variables (see  Eq (\ref{av14}) and discussion therein). In other words we directly plot $\overline{n}_{\nu} \, p_{n}(\kappa_{\nu})$ and show that the obtained expression agrees well 
with the correctly normalized Gamma distribution that we rewrite as 
\begin{equation}
\lim_{\overline{n}_{\nu}/\kappa_{\nu}\gg 1} \overline{n}_{\nu} \,\,p_{n}(\overline{n}_{\nu}, \kappa_{\nu}) \to \psi_{\kappa_{\nu}}(w_{\nu}, \kappa_{\nu}), 
\label{MULT2}
\end{equation}
where $\psi_{n}(w_{\nu}, \kappa_{\nu})$ is the asymptotic form of the distribution, i.e.
\begin{equation}
\psi_{n}(w_{\nu}, \kappa_{\nu})= \frac{\kappa_{\nu}^{\kappa_{\nu}}}{\Gamma(\kappa_{\nu})} \,\,w_{\nu}^{\kappa_{\nu}-1}\,\, e^{-\kappa_{\nu} w_{\nu}}.
\label{MULT3} 
\end{equation} 
In Eq. (\ref{MULT3}) $w_{\nu} = n/\overline{n}_{\nu}$ and when $\overline{n}_{\nu} \gg \kappa_{\nu}$ [e.g. $\nu = {\mathcal O}(10^{-6}) \nu_{\mathrm{max}}$, $m_{T} = {\mathcal O}(1)$ and $\kappa_{\nu} = {\mathcal O}(1)$] the exact form of $\overline{n}_{\nu} \, p_{n}(\overline{n}_{\nu}, \kappa_{\nu})$ cannot be graphically distinguished from the Gamma distribution of Eq. (\ref{MULT2}). This is why in Fig. \ref{FIGURE3} we selected an appropriate range of parameters where $\overline{n}_{\nu}$ is not so large and $\kappa_{\nu}$ is not so small although the KNO limit $\overline{n}_{\nu} > \kappa_{\nu}$ is always realized.
\begin{figure}[!ht]
\centering
\includegraphics[height=6.3cm]{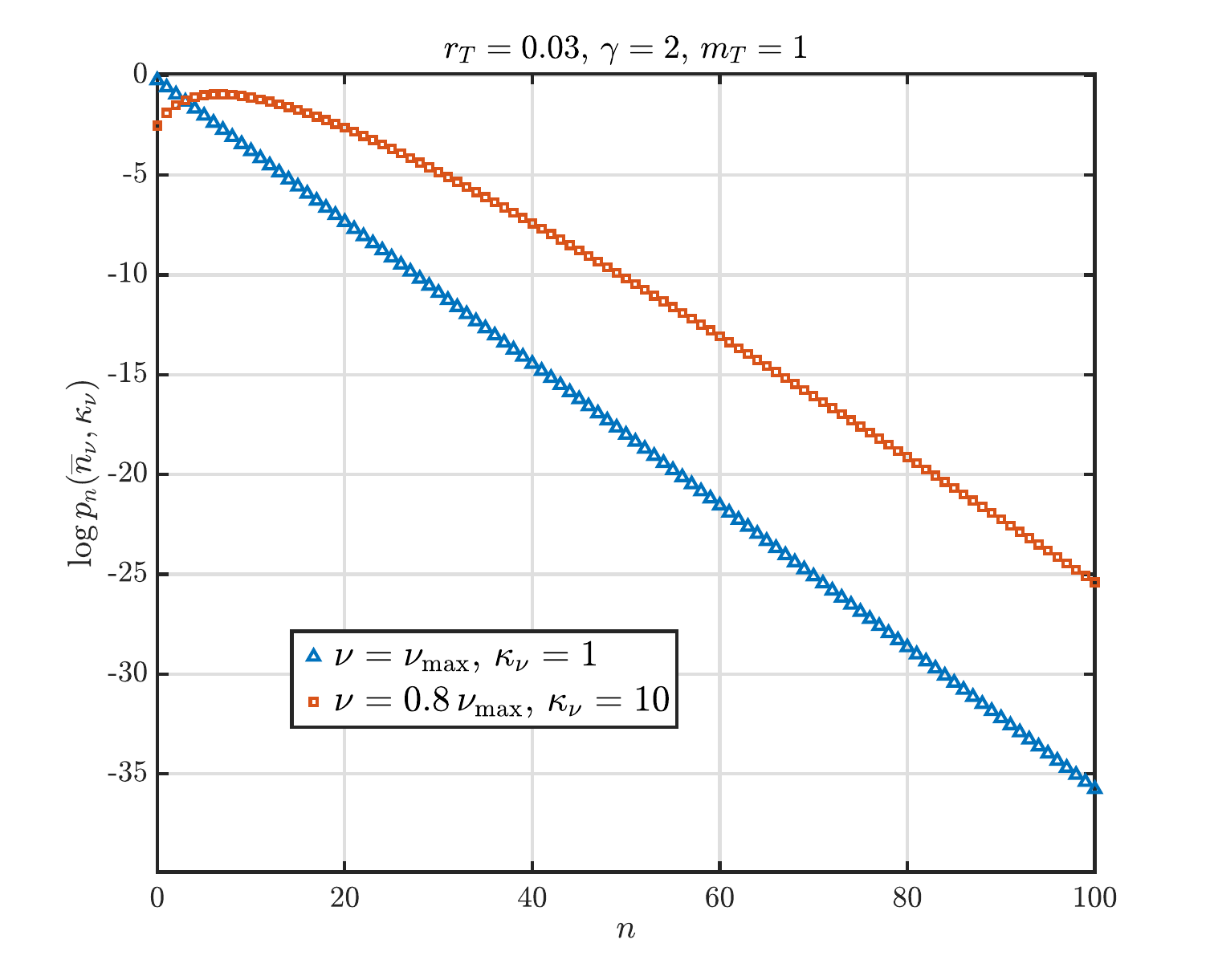}
\includegraphics[height=6.3cm]{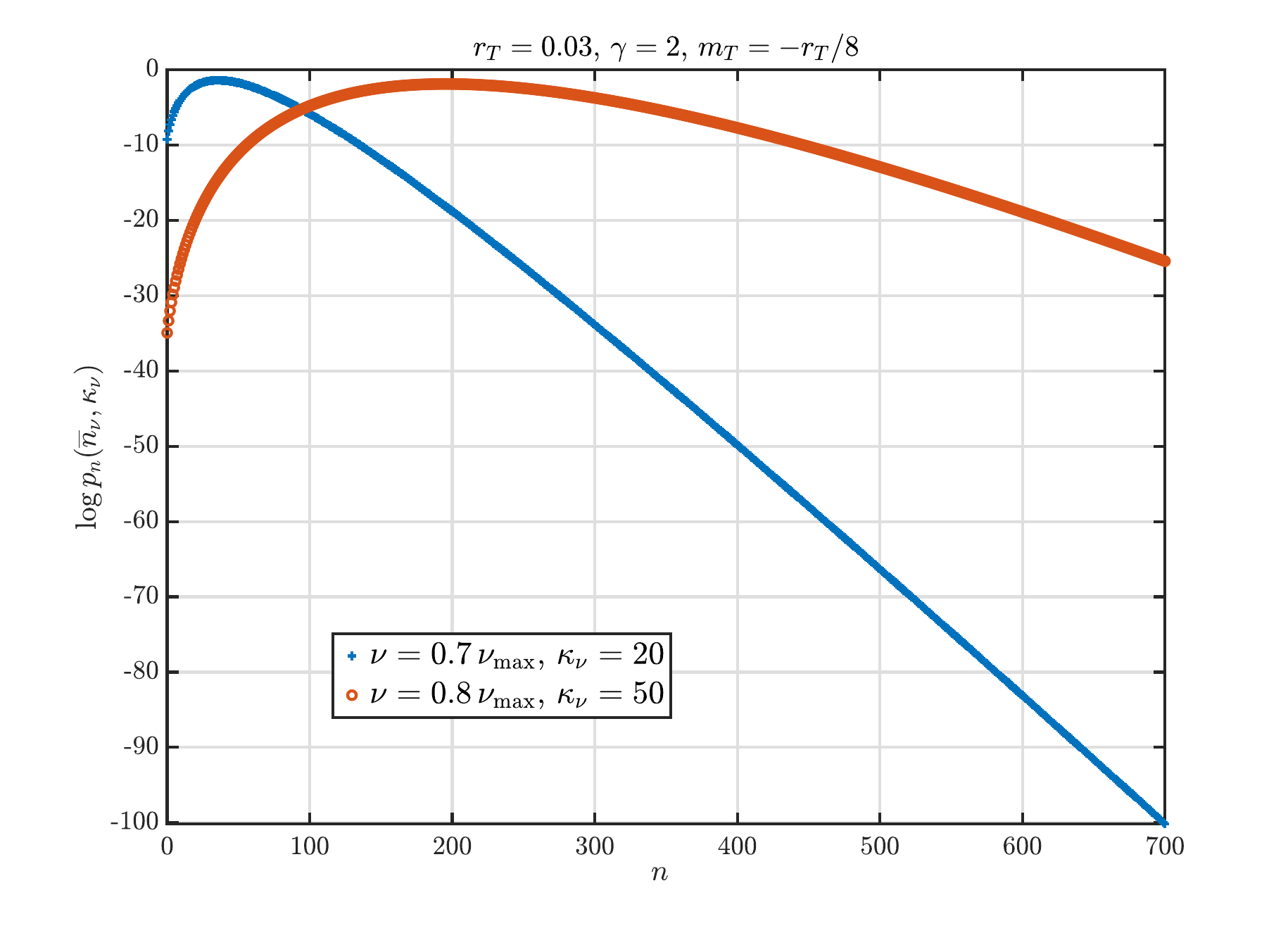}
\caption[a]{In both plots we report the common logarithm of $p_{n}(\overline{n}_{\nu}, \kappa_{\nu})$ when $\overline{n}_{\nu} = {\mathcal O}(1)$ and $\kappa_{\nu} < 100$. In the left plot the case $\kappa \to 1$ corresponds to the Bose-Einstein distribution. In the right plot when 
$ \kappa_{\nu} \to 50$ the multiplicity distribution becomes progressively more symmetric as it happens for the Poisson distribution. }
\label{FIGURE4}      
\end{figure}
The comparison between the two plots of Fig. \ref{FIGURE3}  demonstrates, as expected, that for $\kappa_{\nu} = 2$ the KNO limit is reached earlier than in the case $\kappa_{\nu} =4$. Moreover the shapes illustrated in Figs. \ref{FIGURE2} and \ref{FIGURE3} remain valid when the averaged multiplicity are larger of even much larger.
As we already stressed it can easily happen that $\overline{n}_{\nu} = {\mathcal O}(10^{20})$: in all these cases the limit defined by Eqs. (\ref{MULT2})--(\ref{MULT3}) is, in practice, an excellent approximation for all the phenomenologically acceptable values of  $\kappa_{\nu}$.

\subsubsection{The region $\overline{n}_{\nu} = {\mathcal O}(\kappa_{\nu})$}
When the typical frequency of the gravitons is close to the maximal frequency we enter a regime 
where, in practice,  $\overline{n}_{\nu} = {\mathcal O}(\kappa_{\nu})$. Since $\kappa_{\nu}$ and $\overline{n}_{\nu}$ are comparable, in this range the limits explored in Fig. \ref{FIGURE3} are not realized and the characteristic features on the Bose-Einstein distribution appear more clearly. In the left plot of Fig. \ref{FIGURE4} we illustrate the situation where $\overline{n}_{\nu} > 1$ 
and $\kappa_{\nu} = 1$; this happens when $\nu$ and $\nu_{\mathrm{max}}$ are very close. We remind that for $\kappa_{\nu} \to 1$ the results of sections \ref{sec3} and \ref{sec4} imply that the multiplicity distribution is of Bose-Einstein kind. In the left plot of Fig. \ref{FIGURE4} we also consider the case $\kappa_{\nu} = 10$. In the right plot of Fig. \ref{FIGURE4} the value of $\kappa_{\nu}$ increases 
and we see that for $\kappa_{\nu} \to 50$ the multiplicity distribution becomes progressively more symmetric.
In this limit the $p_{n}(\overline{n}_{\nu},\,\kappa_{\nu})$ effectively becomes a Poisson distribution whose specific  features are discussed in the following subsection.
\subsubsection{The region $\kappa_{\nu} \gg \overline{n}_{\nu}$}
\begin{figure}[!ht]
\centering
\includegraphics[height=6.3cm]{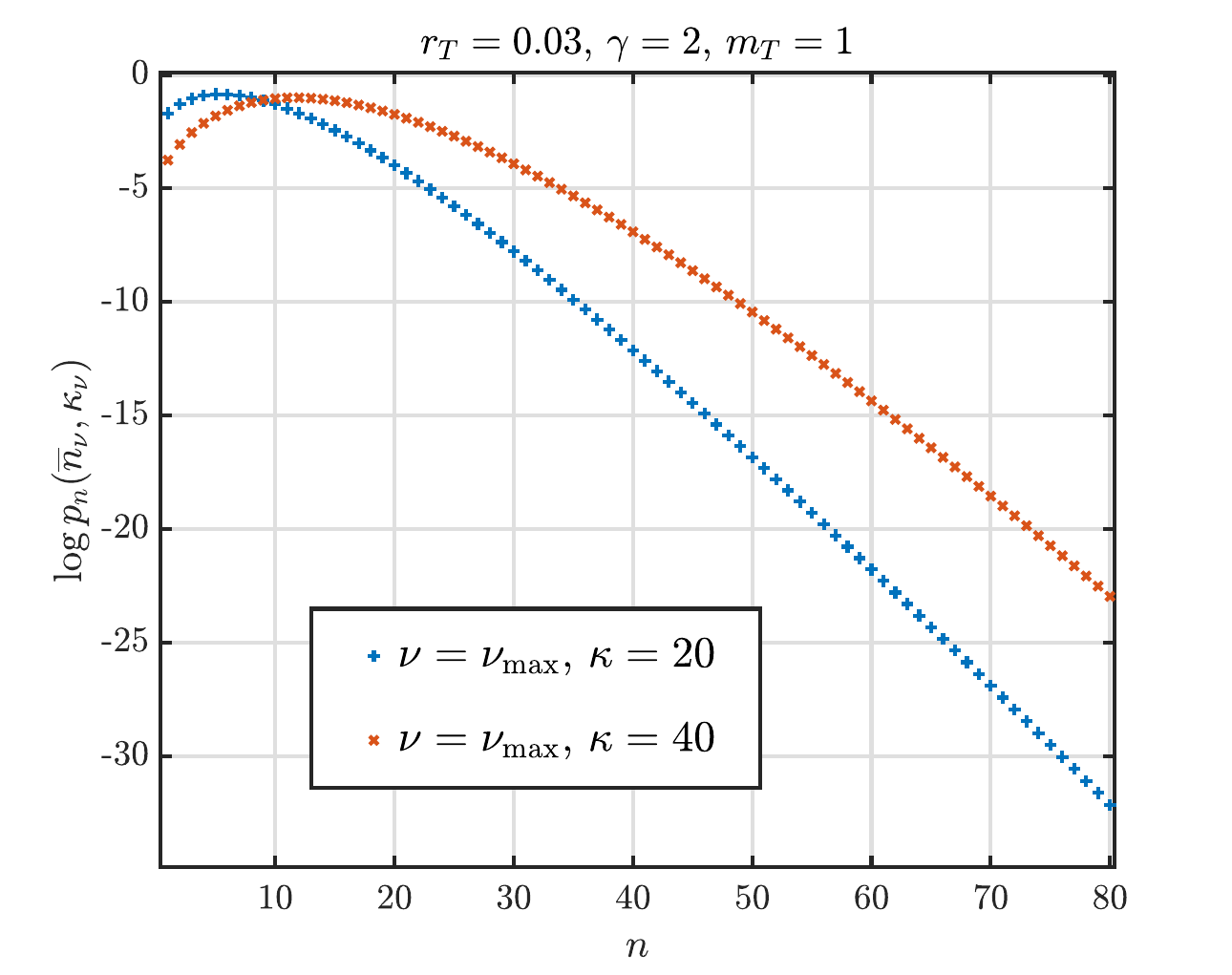}
\includegraphics[height=6.3cm]{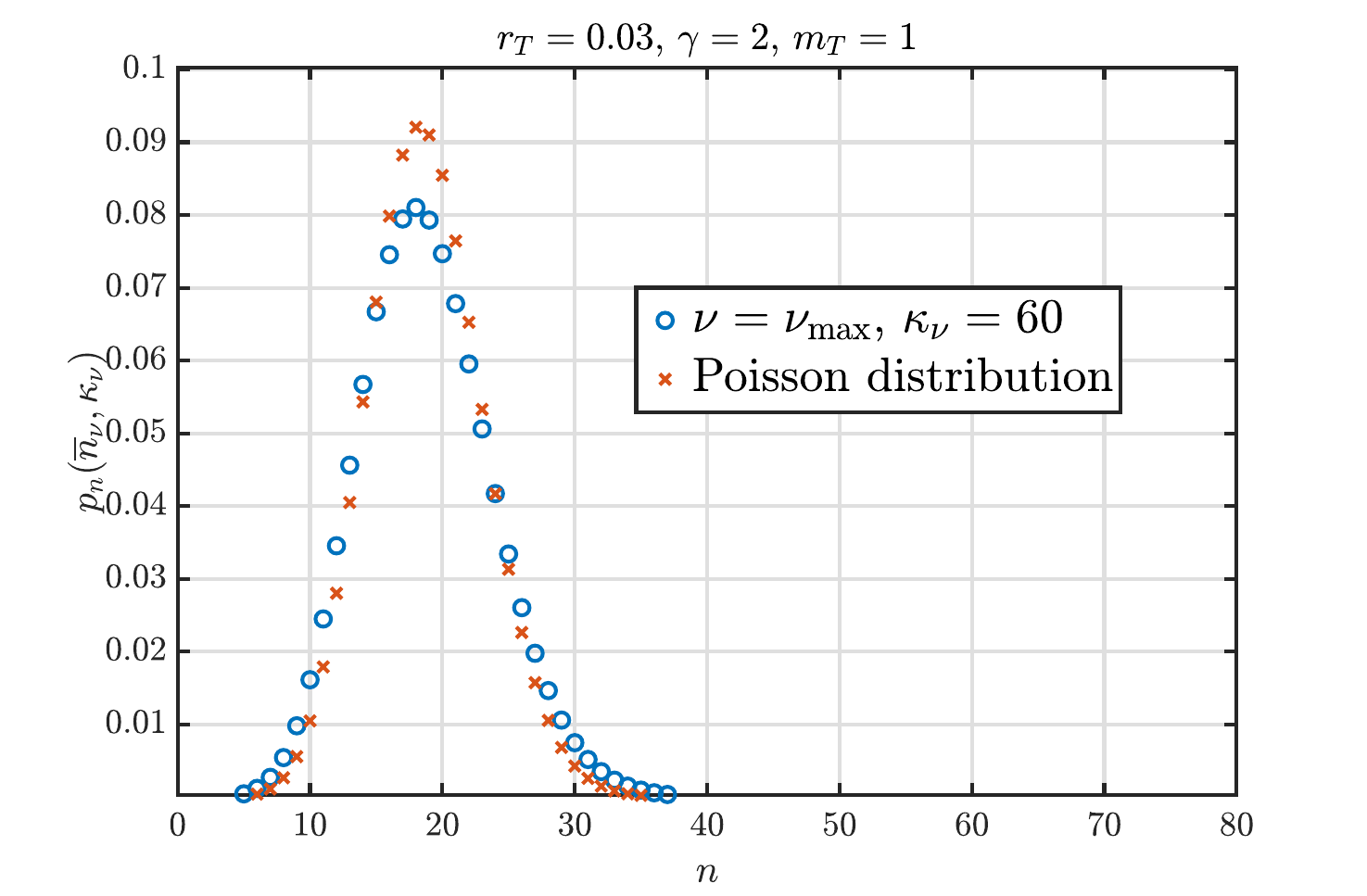}
\caption[a]{In the left plot we illustrate the common logarithm of $p_{n}(\overline{n}_{\nu}, \kappa_{\nu})$ for different sets of parameters. 
In Poisson limit it is also significant to plot directly the multiplicity distribution (and not simply its common logarithm) since,  in this way, the symmetry properties of the distributions are more evident. In both plots 
the averaged multiplicity is ${\mathcal O}(1)$ and $\kappa_{\nu} \gg \overline{n}_{\nu}$. In the right plot the results obtained from $p_{n}(\overline{n}_{\nu}, \kappa_{\nu})$ are compared with the canonical form of the Poisson distribution obtained in terms of the same averaged multiplicity.}
\label{FIGURE5}      
\end{figure}
We already established from Eqs. (\ref{DR18}) and (\ref{DR20}) that in the limit 
$\kappa_{\nu} \to \infty$   the gravitons are produced independently and the dispersion of the distribution coincides with its mean value (i.e. $D_{\nu}^2 \to \overline{n}_{\nu}$) as 
required in the context of a Poisson process. This general observation is now more concretely 
illustrated in Fig. \ref{FIGURE5} for the regime 
$\kappa_{\nu} \gg \overline{n}_{\nu}$. In the left plot we 
considered the situation where $\overline{n}_{\nu} = {\mathcal O}(1)$ while 
the values of $\kappa_{\nu}$ have been selected to be, respectively, $20$ and $40$. 
The condition $\overline{n}_{\nu} = {\mathcal O}(1)$ is realized for typical 
frequencies close to $\nu_{\mathrm{max}}$, as assumed in Fig. \ref{FIGURE5}.
For the two cases of the left plot we clearly see that the distribution becomes progressively 
more symmetric as $\kappa_{\nu}$ passes from $20$ to $40$. To appreciate 
the symmetry properties it is useful to present the results on a linear scale (as opposed to the logarithmic scale). For this reason in the right plot of Fig. \ref{FIGURE5} the $p_{n}(\overline{n}_{\nu}, \kappa_{\nu})$ a linear scale has been employed; more specifically the blobs correspond to $p_{n}(\overline{n}_{\nu},\kappa_{\nu})$
with $\kappa_{\nu} = 60$ while the crosses are computed from a putative Poisson 
distribution having exactly the same averaged multiplicity. From the right plot 
of Fig. \ref{FIGURE5} we can see that the limit $D_{\nu}^2 \to \overline{n}_{\nu}$ is 
approximately verified even if the hierarchy between $\overline{n}_{\nu}$ and $\kappa_{\nu}$ 
only involves an order of magnitude. 

\subsection{Multiplicity distributions and post-inflationary evolution}
In the examples presented so far the averaged multiplicity follows from 
the ratio between $\nu$ and $\nu_{\mathrm{max}}$: after
imposing the bounds on the maximal frequency given in Eq. (\ref{GGG13})  (and illustrated in Fig. \ref{FIGURE1}) the value of $\overline{n}_{\nu}$ for a given spectral slope $m_{T}$
can be directly obtained. In the last part of this discussion we are going to study the multiplicity distributions directly in terms of the parameters that define the inflationary and the post-inflationary evolutions. 
\begin{figure}[!ht]
\centering
\includegraphics[height=6.5cm]{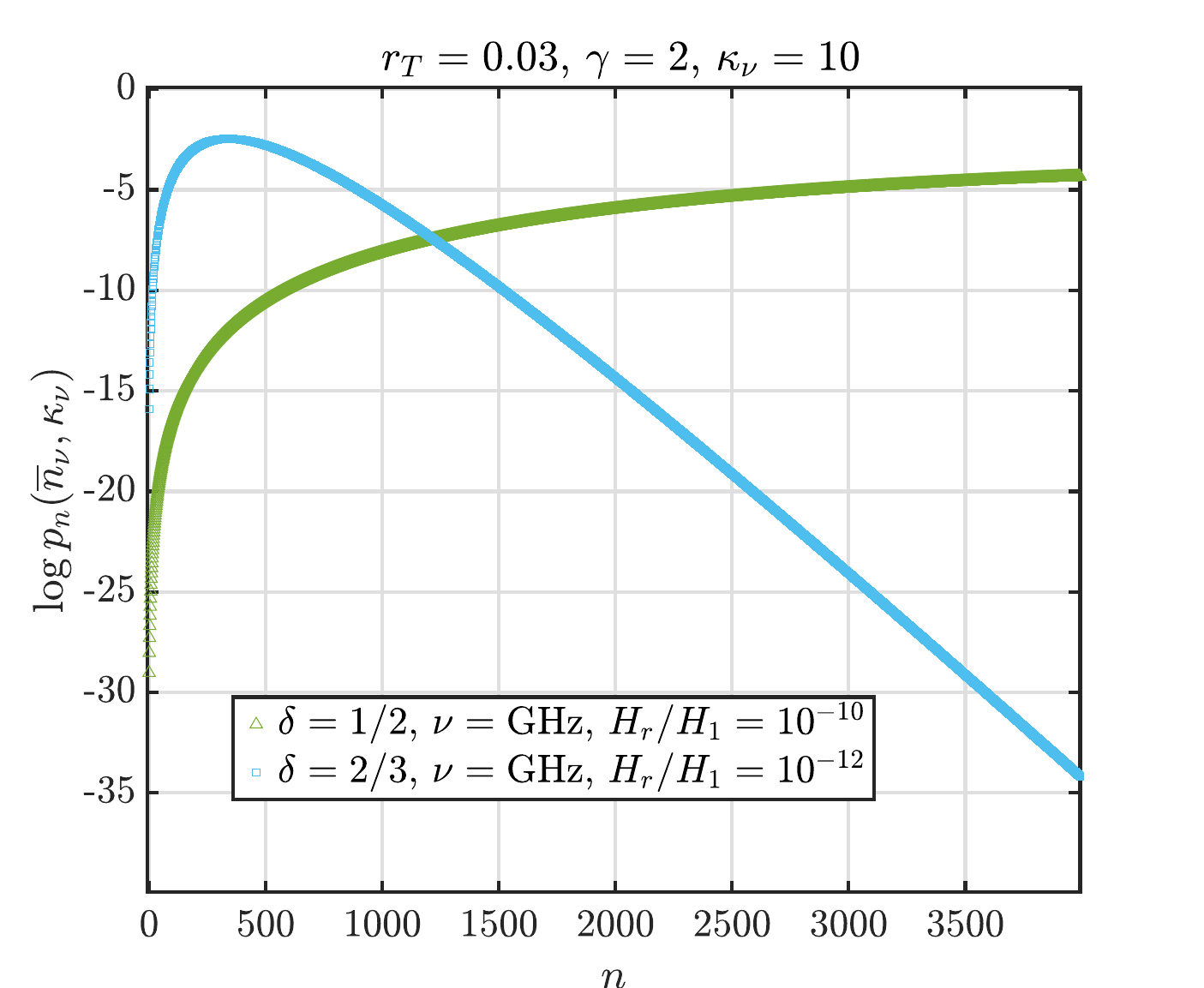}
\includegraphics[height=6.5cm]{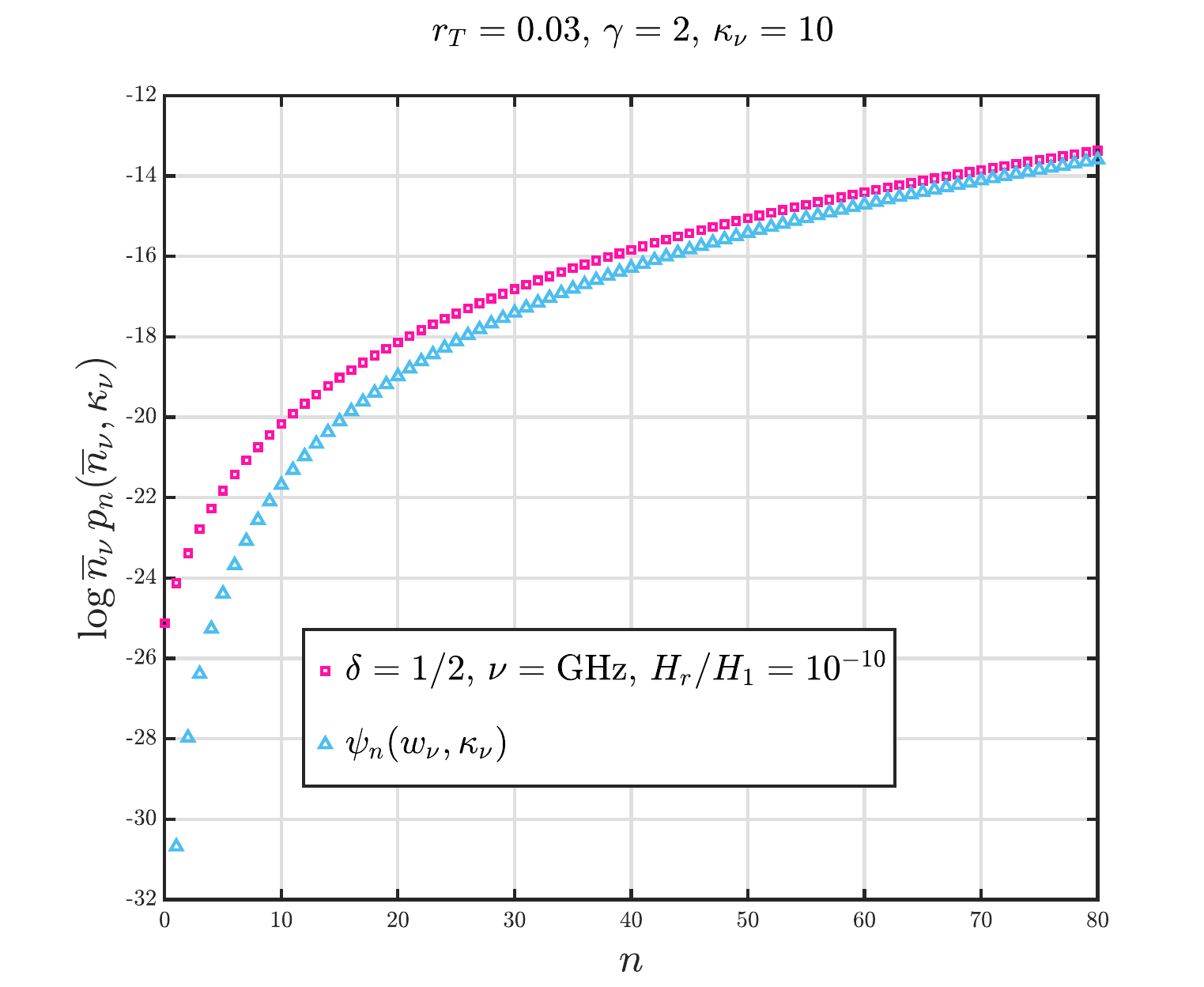}
\caption[a]{In both plots the multiplicity distributions have been illustrated in terms of the underlying post-inflationary evolution. We have considered the case where the post-inflationary expansion rate is slower than radiation (i.e. $\delta <1$) while $H_{r}/H_{1}$ 
varies between $10^{-10}$ and $10^{-12}$. In the right plot the result obtained 
from $p_{n}(\overline{n}_{\nu}, \kappa_{\nu})$ is compared with the Gamma distribution 
in  valid for the same set of parameters.}
\label{FIGURE6}      
\end{figure}
The quantum bound of Eq. (\ref{GGG13}) implies that $\nu_{\mathrm{max}} < {\mathcal O}(\mathrm{THz})$. From a purely classical perspective, however, the value of $\nu_{\mathrm{max}}$ should depend on the timeline of the expansion rate as described by Eqs. (\ref{GGG1}) and (\ref{APPB13}) in the case of a single stage of post-inflationary evolution. When this second viewpoint is adopted the  
dependence of $\nu_{\mathrm{max}}$ upon the post-inflationary timeline is specified 
by selecting $\delta$ and $H_{r}$. If $\delta < 1$ (as discussed in Fig. \ref{FIGURE6}) the post-inflationary evolution is slower than radiation; $H_{r}$ determines instead 
 the expansion rate when the radiation-dominated evolution is recovered.
Although more complicated cases can be envisaged (i.e. various post-inflationary stages with different expansion rates), for the present illustrative purposes it seems sufficient to consider a single post-inflationary 
stage;  more general possibilities can be however found in Ref. \cite{KK3}.

The value of $\delta$ does not only determine $\nu_{\mathrm{max}}$ 
but also the frequency dependence of $\overline{n}_{\nu}$ for $\nu< \nu_{\mathrm{max}}$.
Recalling in fact Eqs. (\ref{GGG7}) and (\ref{APPB11}) we have that in the limit $r_{T} \ll 1$ 
$m_{T} \simeq (2 - 2 \delta)$. This also means that $\overline{n}_{\nu} \simeq \kappa_{\nu} 
(\nu/\nu_{\mathrm{max}})^{-2 - 2 \delta}$. For a post-inflationary regime dominated 
by radiation we then recover the result of Eq. (\ref{FFF1d}) and $\overline{n}_{\nu}$ scales 
as $\nu^{-4}$. However when $\delta < 1$ the spectral energy density is comparatively larger 
at high frequencies, as first discussed with slightly different notations, in Ref. \cite{FL5}. 
\begin{figure}[!ht]
\centering
\includegraphics[height=6.3cm]{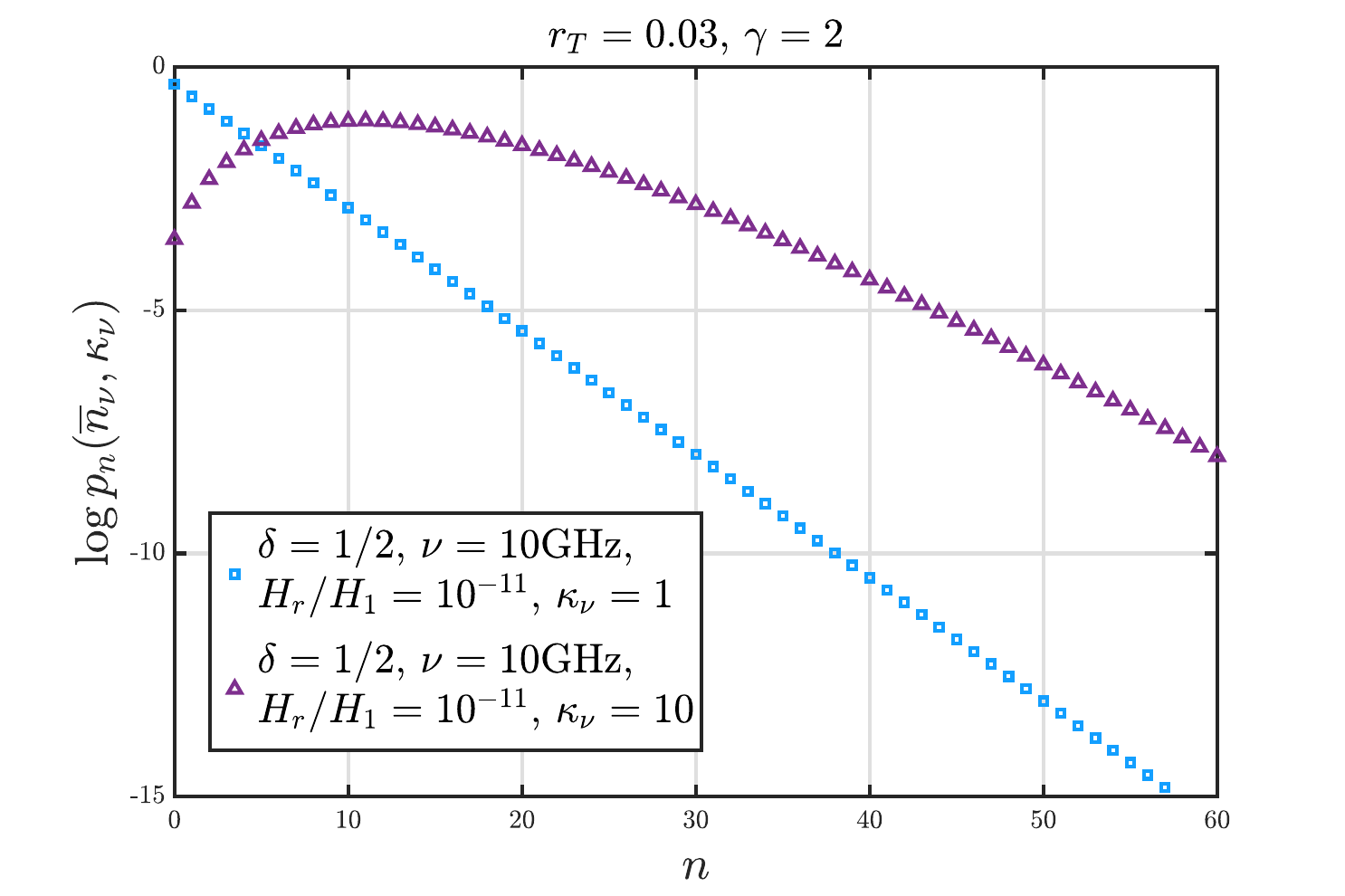}
\includegraphics[height=6.3cm]{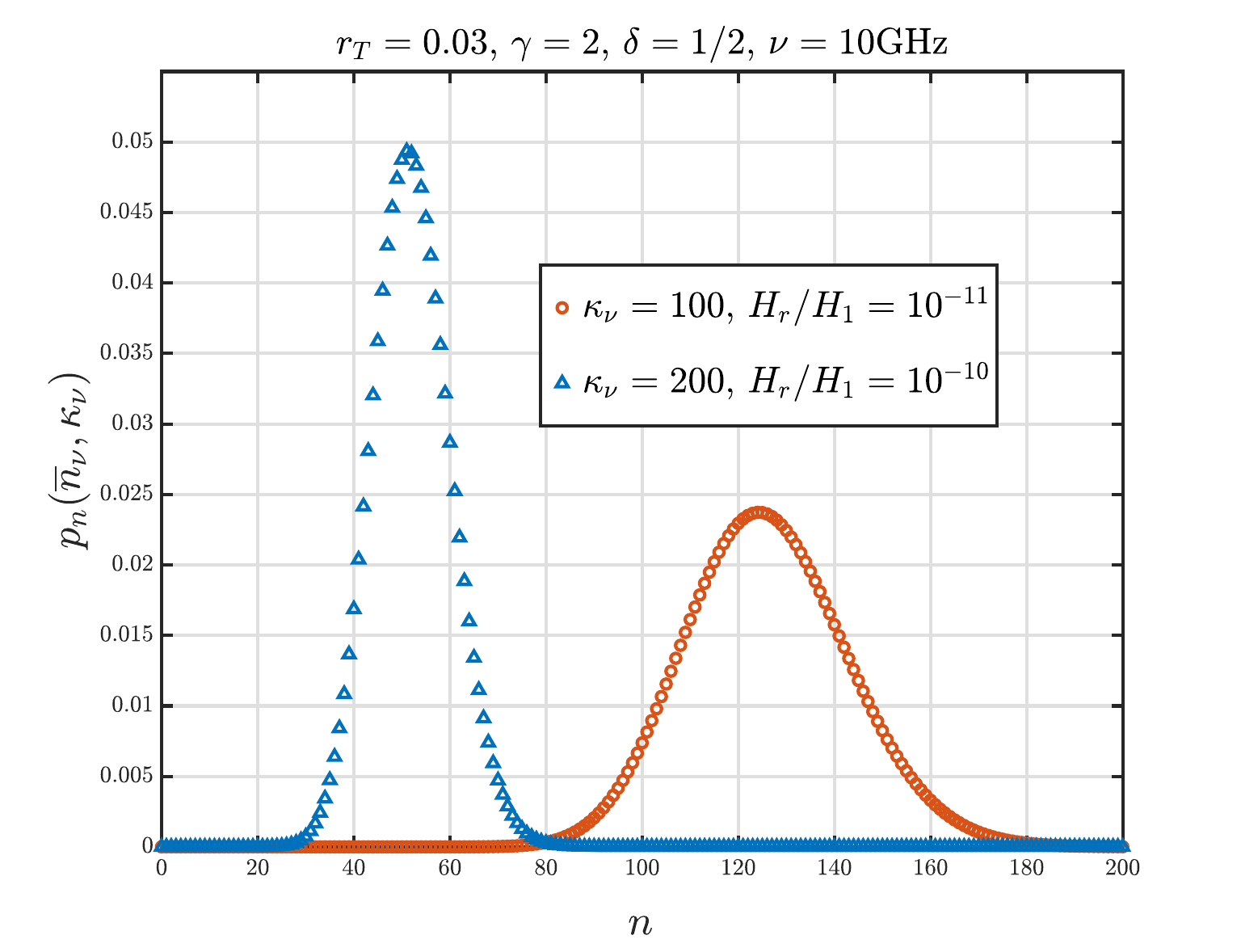}
\caption[a]{The Poisson limit is illustrated always in the case when the post-inflationary evolution is slower than radiation. In the left plot the case $\kappa_{\nu} = 1$ is compared with $\kappa_{\nu} = 10$.  In the right plot we report instead the choices $\kappa_{\nu} =100$ and $\kappa_{\nu}= 200$ always 
in the case of a post-inflationary expansion rate slower than radiation. Note the linear 
scale in the right plot and opposed to the logarithmic scale of the right plot.}
\label{FIGURE7}      
\end{figure}
In the left plot of Fig. \ref{FIGURE6} we illustrate the multiplicity distributions for $\kappa_{\nu} =10$; in the same plot the post-inflationary expansion rate is slower than radiation (i.e. $\delta < 1$) while $H_{r}/H_{1} = 10^{-10}$ for the upper curve with the triangles and $H_{r}/H_{1} =10^{-12}$ for the curve with the squares (see the legend of the left plot). The difference 
between the parameters of the two cases modifies the physical regime: the curve with the triangles  already tends to a Gamma distribution, while for the curve with the squares the 
Pascal  distribution is still far from the KNO limit. In the right plot of Fig. \ref{FIGURE6} 
the Gamma distribution in the KNO limit is compared with the result directly 
obtained from $p_{n}(\overline{n}_{\nu}, \kappa_{\nu})$.  In Fig. \ref{FIGURE7} the expansion rate is slower than radiation (i.e.  $\delta=1/2$) and $H_{r}/H_{1} = 10^{-11}$. In the left plot the cases $\kappa_{\nu} =1$ and $\kappa_{\nu} =10$ are compared (see the legend of the plot)
While for $\kappa_{\nu} \to 1$ the distribution is clearly of Bose-Einstein kind, in the case $\kappa_{\nu} \to 10$ the patterns of the Poisson limit become already evident. In the right plot of Fig. \ref{FIGURE7} we then 
compute the distributions for $\kappa_{\nu} = 100$ and $\kappa_{\nu} = 200$. The symmetry 
of the curves (typical of the Poisson distribution) is better appreciated if the results are plotted 
on a linear scale, as opposed to the logarithmic scale employed, for instance, in the left plot of Fig. \ref{FIGURE7}.

\begin{figure}[!ht]
\centering
\includegraphics[height=8cm]{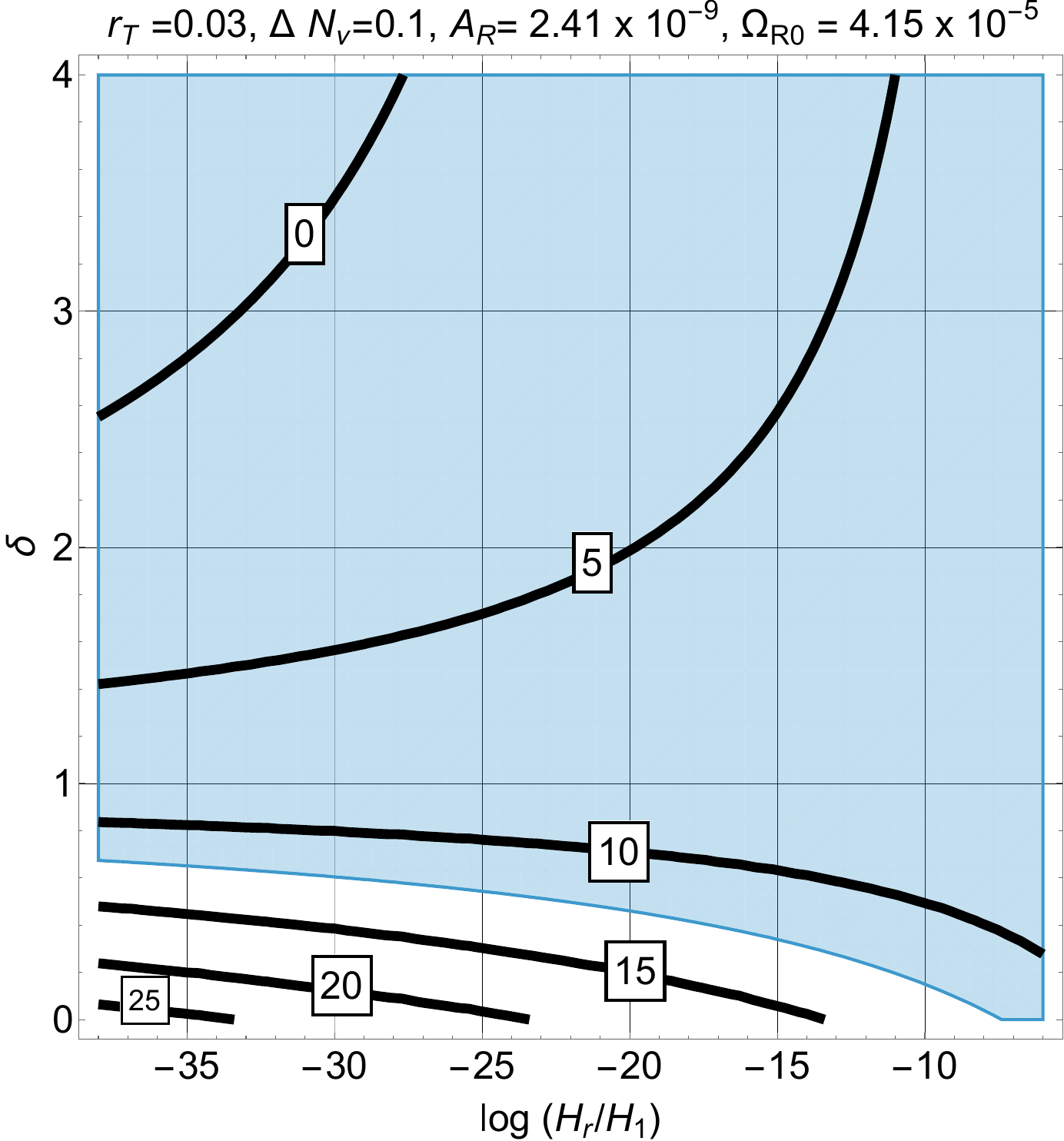}
\caption[a]{The shaded area is obtained by requiring that 
the maximal frequency determined from Eq. (\ref{GGG1}) never exceeds the THz region. 
The labels appearing on the various curves correspond to the common logarithm 
of $\nu_{\mathrm{max}}$ expressed in Hz. All the parameters of the examples 
discussed in this subsection (see Figs. \ref{FIGURE6} and \ref{FIGURE7}) are extracted 
from the allowed region of the parameter space illustrated here. }
\label{FIGURE8}      
\end{figure}

The parameters of Figs. \ref{FIGURE7} and \ref{FIGURE8} have been selected 
by enforcing the quantum bound on the maximal frequency of the spectrum. We already 
pointed out that the classical determination of $\nu_{\mathrm{max}}$ of Eq. (\ref{GGG1}) 
may overshoot the quantum bound of Eq. (\ref{GGG13}). The values of $\delta$ and $H_{r}/H_{1}$ must then comply with a condition ensuring that $\nu_{\mathrm{max}}$ (obtained 
from Eq. (\ref{GGG1})) does not exceed the THz domain. This condition leads to a 
specific region in the plane defined by $\delta$ (the post-inflationary expansion rate) and 
$H_{r}/H_{1}$. In Fig. \ref{FIGURE8} we illustrate, for the sake of completeness, this plane always for of a single post-inflationary stage expanding at different rates. 

\subsection{High frequency sensitivities}
\begin{figure}[!ht]
\centering
\includegraphics[height=8cm]{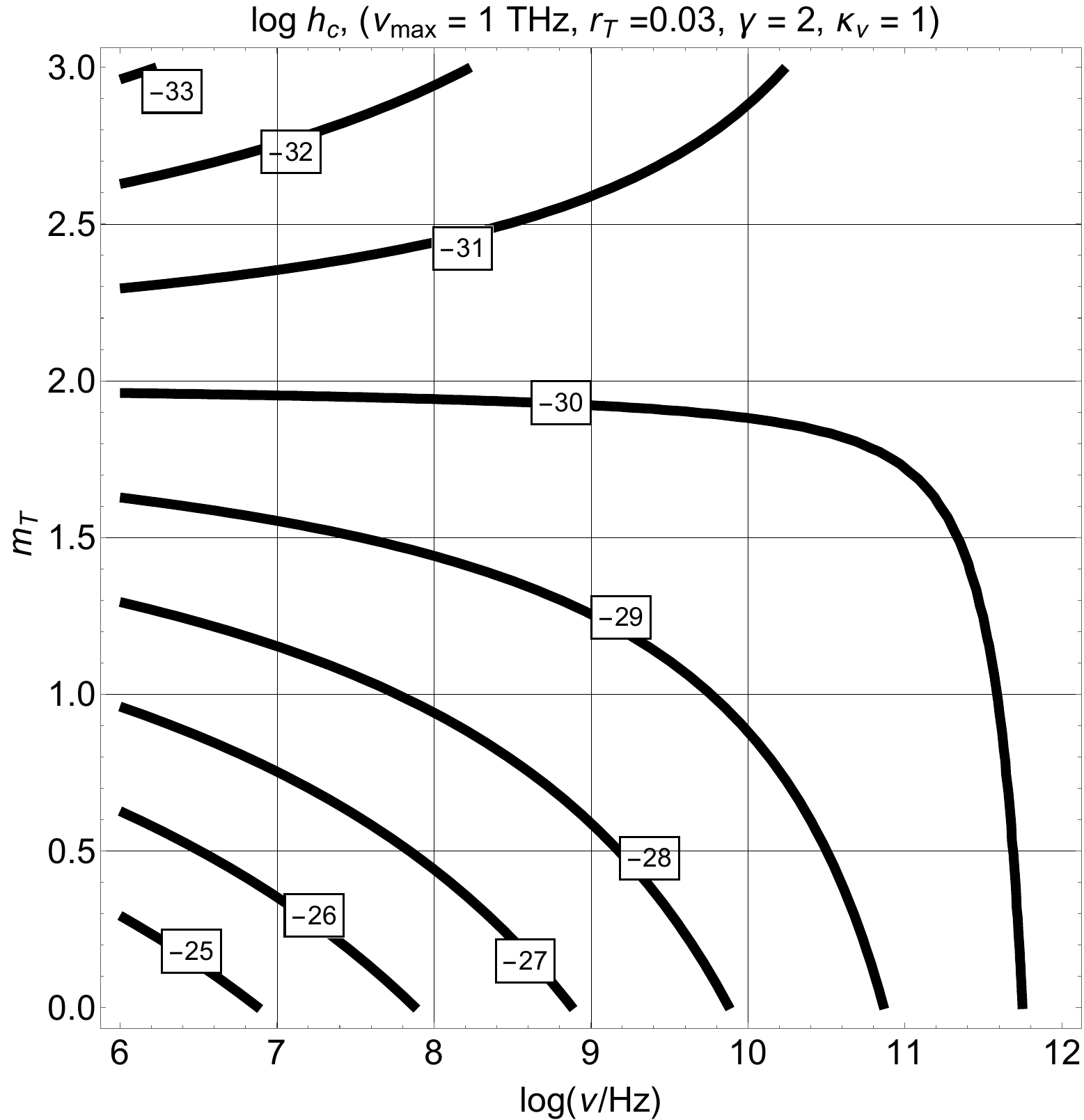}
\includegraphics[height=8cm]{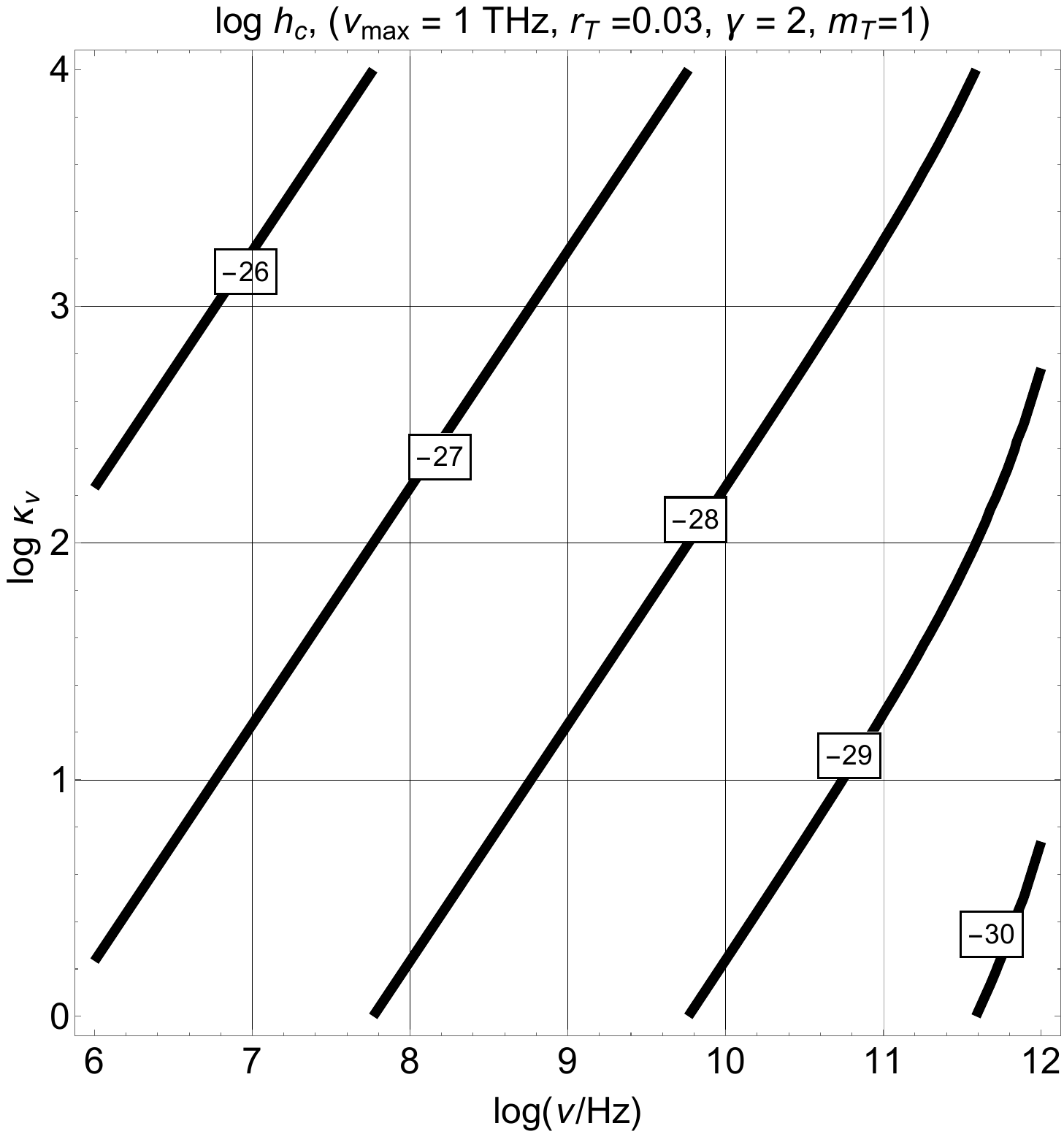}
\includegraphics[height=8cm]{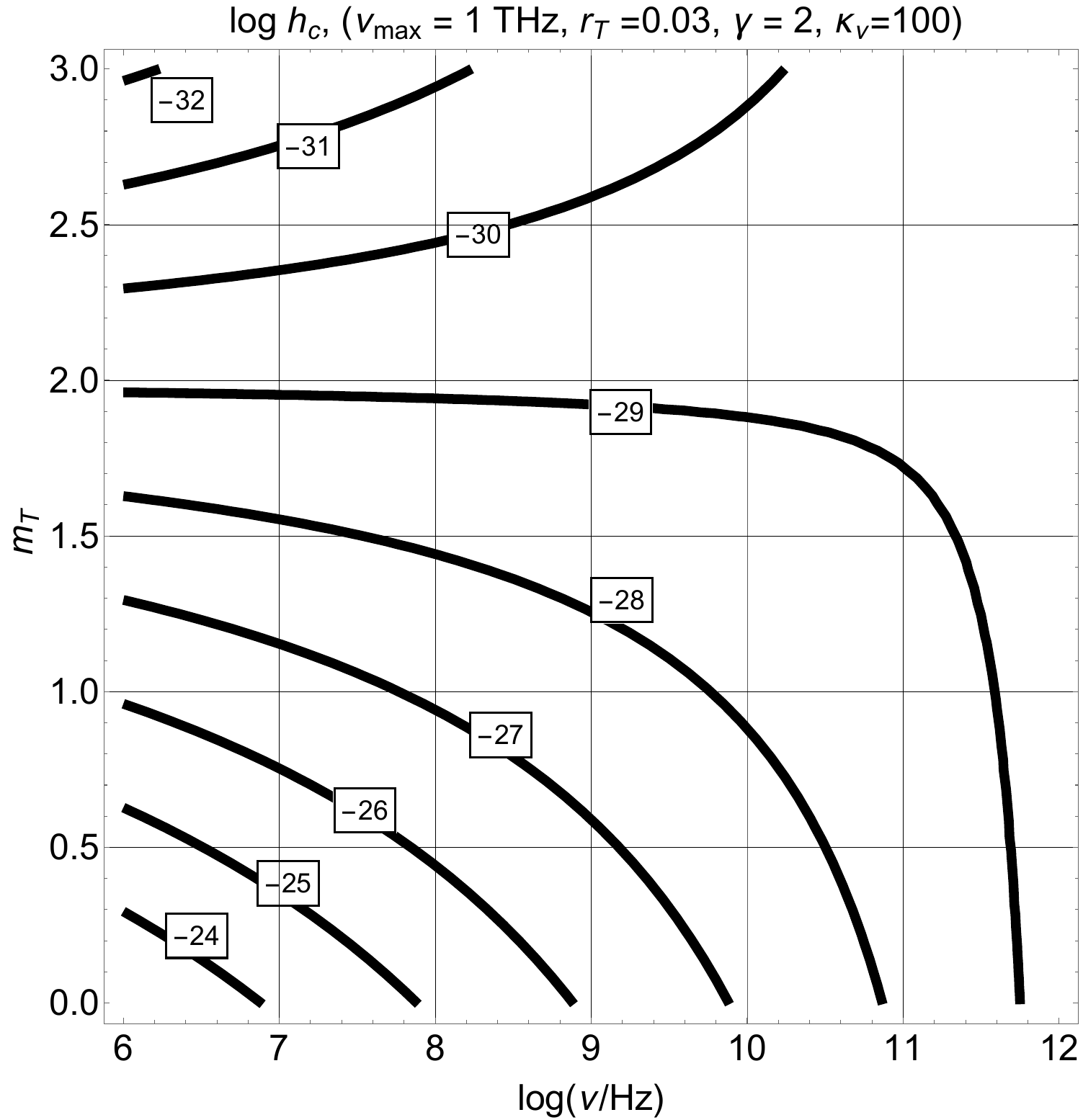}
\includegraphics[height=8cm]{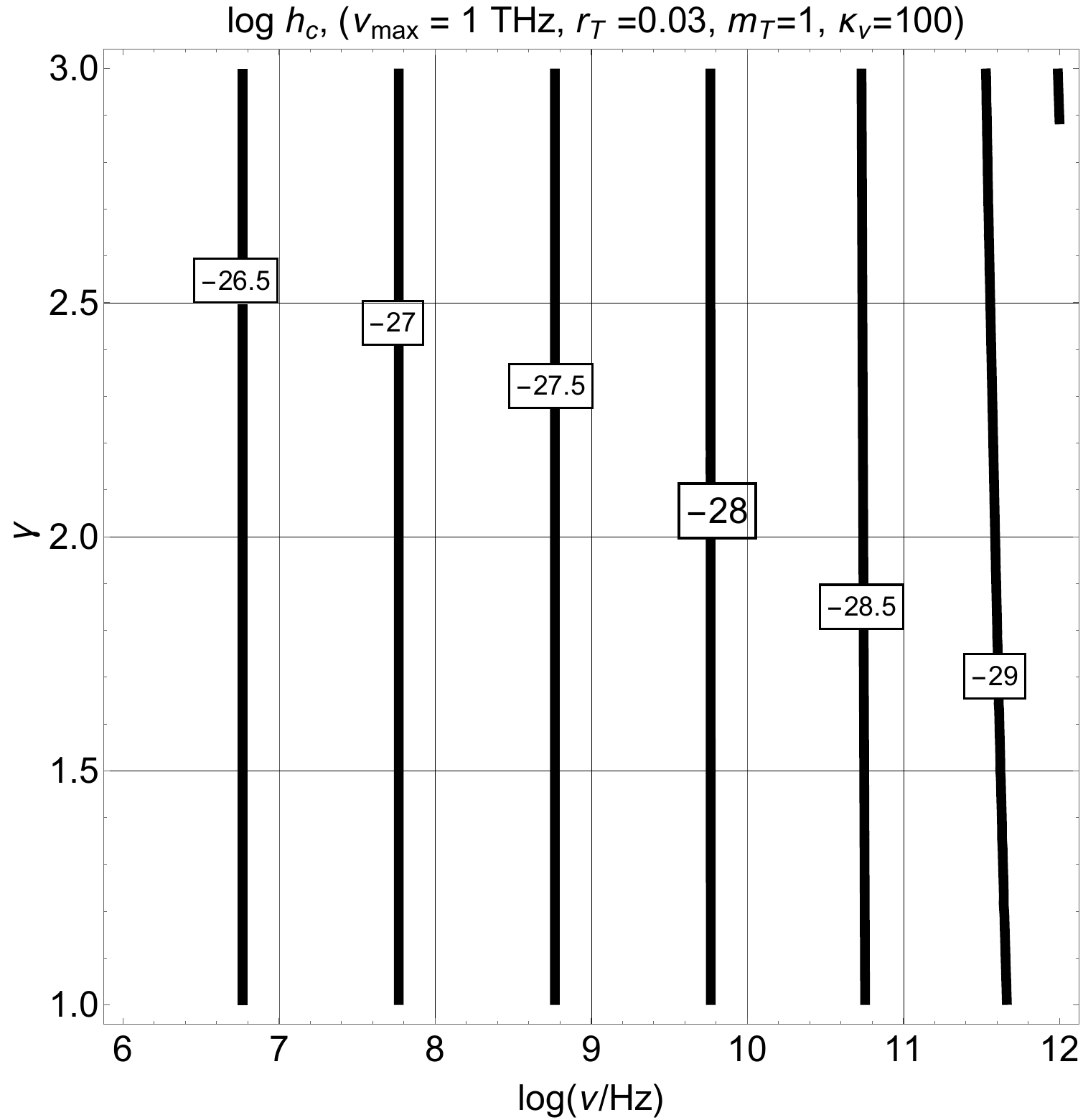}
\caption[a]{The different labels in the four plots of this figure correspond to the common logarithms 
of the chirp amplitude. In all the plots $\nu_{\mathrm{max}}$ is fixed 
to its maximal value compatible with the phenomenological limits, i.e. $\nu_{\mathrm{max}}= {\mathcal O}(\mathrm{THz})$. In the two upper plots the variation of the chirp amplitude 
is illustrated in the planes $[m_{T},\, \log{(\nu/\mathrm{Hz})}]$ and $[\log{\kappa_{\nu}},\, \log{(\nu/\mathrm{Hz})}]$. In the two lower plots $\kappa_{\nu}$ is fixed to ${\mathcal O}(100)$ and the values 
of $h_{c}$ are examined in the planes  $[m_{T},\, \log{(\nu/\mathrm{Hz})}]$ and $[\gamma\, \log{(\nu/\mathrm{Hz})}]$. In all the plots the minimal value of $m_{T}$ corresponds to the one of the concordance scenario supplemented by the consistency conditions (i.e. $m_{T} \simeq - r_{T}/8$) whereas the value of $r_{T}$ is consistent with the current determinations (i.e. $r_{T} < 0.035$); see also \cite{MGINV} for the possibility of invisible gravitons in the aHz range.}
\label{FIGUREX}      
\end{figure}

The statistical properties of the relic gravitons are encoded in their multiplicity distributions whose 
specific shapes are also affected by the range of the comoving frequencies. At the moment the regions that are observationally accessible include the aHz domain (where the CMB is the largest electromagnetic detector of relic gravitons \cite{LL0,LL1,LL2,LL3,LL4,LL5}),  the nHz range (where the pulsar timing arrays are currently providing essential upper bounds and, hopefully, 
more direct observations \cite{NANO1,NANO2,PPTA1,PPTA2})  and, last but not least, the audio band (i.e. between few Hz and $10$ kHz  where the interferometric observations 
are now scrutinizing the astrophysical sources and, indirectly, the relic gravitons \cite{AU1,AU2,AU3,AU4,AU5}). For the present ends, however, the high frequency and ultra-high frequency ranges are also extremely relevant as suggested long ago \cite{FL5} (see also \cite{QQ1,QQ3,QQ4}). Indeed, the spectral energy density is potentially larger in the MHz and GHz bands and the region of the maximal frequency (in the THz domain) can provide essential information on the production mechanism \cite{AU6}.  In particular the MHz and GHz detectors are essential to probe the statistical properties of the relic gravitons and especially their quantumness \cite{CAV1,CAV2,CAV3,CAV4,CAV5,CAV6,CAV7,WL1,WL2}. The high frequency detectors may also provide a unique information on the post-inflationary expansion rates and, more generally, on the early stages of the evolution of the primeval plasma \cite{MGrev2}. Although the sensitivity requirements in separated frequency domains are potentially very different, 
the same targets of the audio band are often adopted in the MHz or GHz domains and, within this logic, the minimal sensitivities are generically assigned as 
\begin{equation}
h_{c}^{(\mathrm{min})}(\nu,\tau_{0}) \simeq \sqrt{S^{(\mathrm{min})}_{h}(\nu,\tau_{0}) \, \mathrm{Hz}}  \simeq 10^{-21}, \qquad \nu \geq {\mathcal O}(\mathrm{MHz}).
\label{SENS1}
\end{equation}
In Eq. (\ref{SENS1}) $h_{c}(\nu,\tau_{0})$ and $S_{h}(\nu,\tau_{0})$ denote, respectively, 
the chirp  and the spectral amplitudes; sometimes $S_{h}(\nu,\tau_{0})$ is also called power spectral density. The descriptions based upon $h_{c}(\nu,\tau_{0})$ and $S_{h}(\nu,\tau_{0})$ are used interchangeably but these quantities have physically different properties especially in the case of the relic gravitons\footnote{Since, as already discussed, the random backgrounds of cosmological origin are non-stationary \cite{AS9}
The autocorrelation function is not invariant under a shift of the time coordinate. In this situation 
 Fourier transform of the autocorrelation function is not associated with a well defined spectral amplitude as the Wiener-Khintchine theorem would imply in the case of a stationary process \cite{STOC1,WK1,WK2}. For this reason, even if the chirp amplitude is most suitable quantity for assigning the sensitivities in what follows we are also going to consider the spectral amplitudes since this is the practice (which is, as we said, only partially correct in the case of the relic signals).}.  When all the relevant wavelengths are shorter than the Hubble radius 
the chirp amplitude has a simple relation with the spectral energy density given by $\Omega_{gw}(\nu,\, \tau_{0}) = 2\, \pi^2 \, \nu^2 h_{c}^2(\nu,\tau_{0})/( 3 \, H_{0}^2)$. This is why $h_{c}(\nu,\tau_{0})$ is dimensionless and it can be directly expressed through the averaged multiplicity as:
\begin{equation}
h_{c}(\nu,\tau_{0}) = \sqrt{ 64\, \pi} (\nu/M_{P})\, \sqrt{\overline{n}_{\nu}}.
\label{SENS2}
\end{equation}
Recalling the interpolating expression of $\overline{n}_{\nu}$ (see Eq. (\ref{GGG6}) and discussion therein) the chirp amplitude must be exponentially 
suppressed for $ \nu \gg \nu_{\mathrm{max}}$, as suggested by Eq. (\ref{SENS2}). In the opposite limit the chirp amplitude scales instead as:
\begin{equation}
h_{c}(\nu, \tau_{0}) \to \sqrt{64 \pi \kappa_{\nu}} (\nu_{\mathrm{max}}/M_{P}) \, (\nu/\nu_{\mathrm{max}})^{-1 + m_{T}/2}.
\label{SENS3}
\end{equation}
Equations (\ref{SENS2})--(\ref{SENS3}) are incompatible with Eq. (\ref{SENS1}) and with the absolute bounds on $\nu_{\mathrm{max}}$ deduced in Eq. (\ref{GGG13}) (see also Fig. \ref{FIGURE1} and the related discussion). If we evaluate Eq. (\ref{SENS3}) for 
$\nu= {\mathcal O}(\nu_{\mathrm{max}})$  we could discover, under the assumption that the minimal sensitivities around the maximal frequency range
correspond to Eq. (\ref{SENS1}), that $\nu_{\mathrm{max}}$ should be, at least, ${\mathcal O}(10^{8}) \, \mathrm{THz}$. Such a range of values of $\nu_{\mathrm{max}}$ is in sharp contradiction with all the bounds of Fig. \ref{FIGURE1} and it would suggest, in particular, a violation of the BBN limits \cite{BBN1,BBN2,BBN3}. In Fig. \ref{FIGUREX} the values appearing in the labels of the various plots correspond to the common logarithms of the chirp amplitudes.
\begin{figure}[!ht]
\centering
\includegraphics[height=8cm]{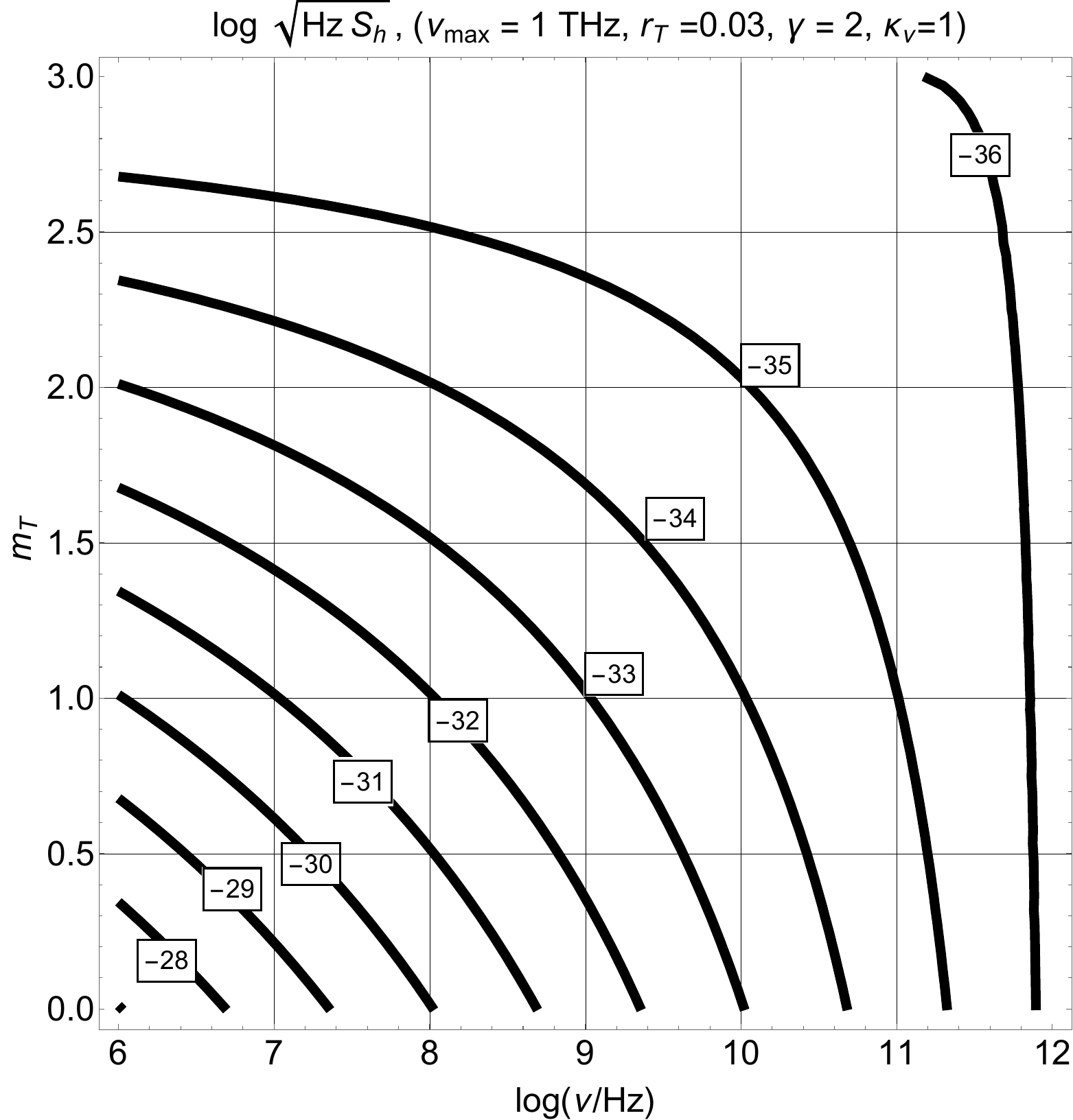}
\includegraphics[height=8cm]{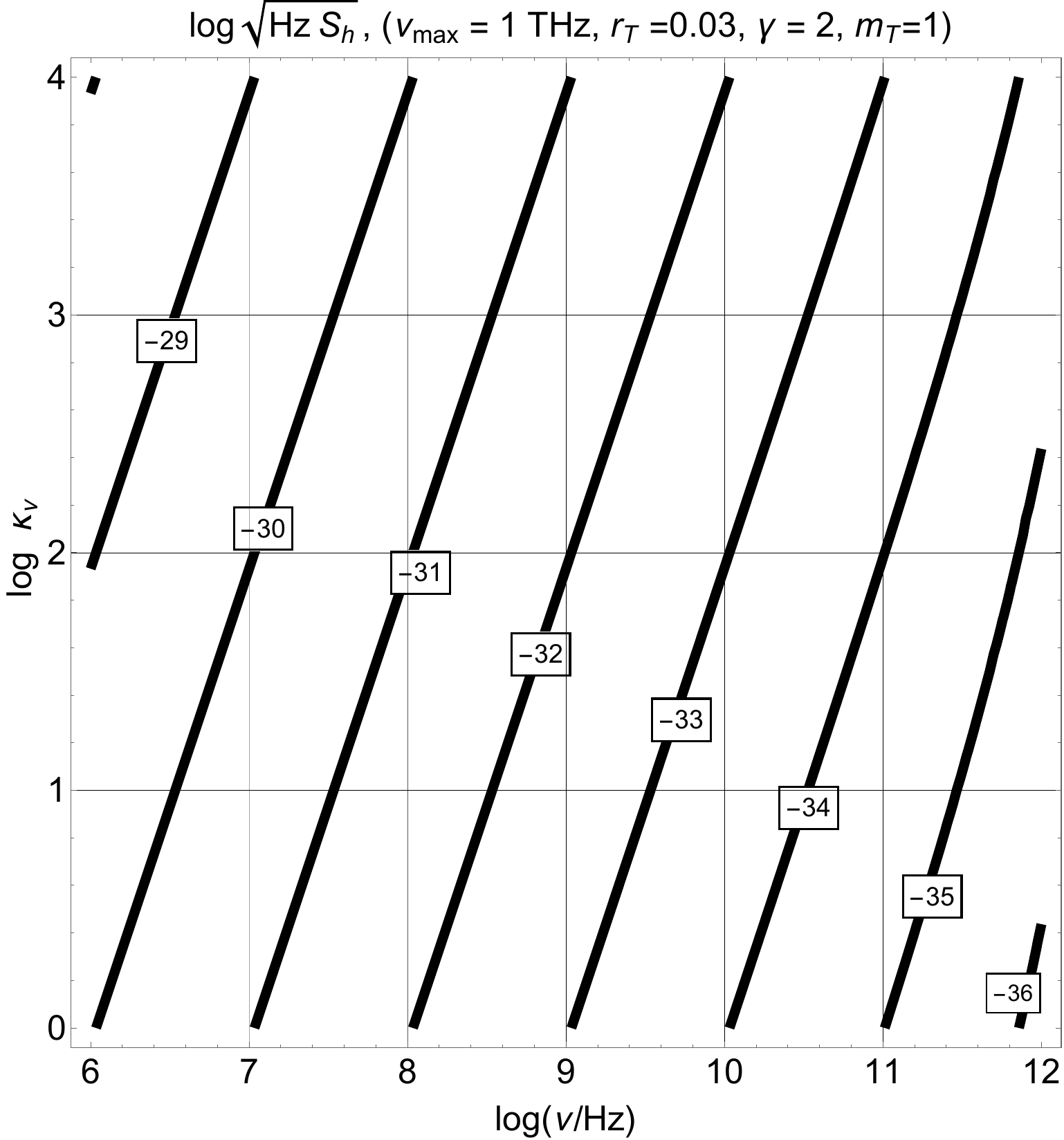}
\includegraphics[height=8cm]{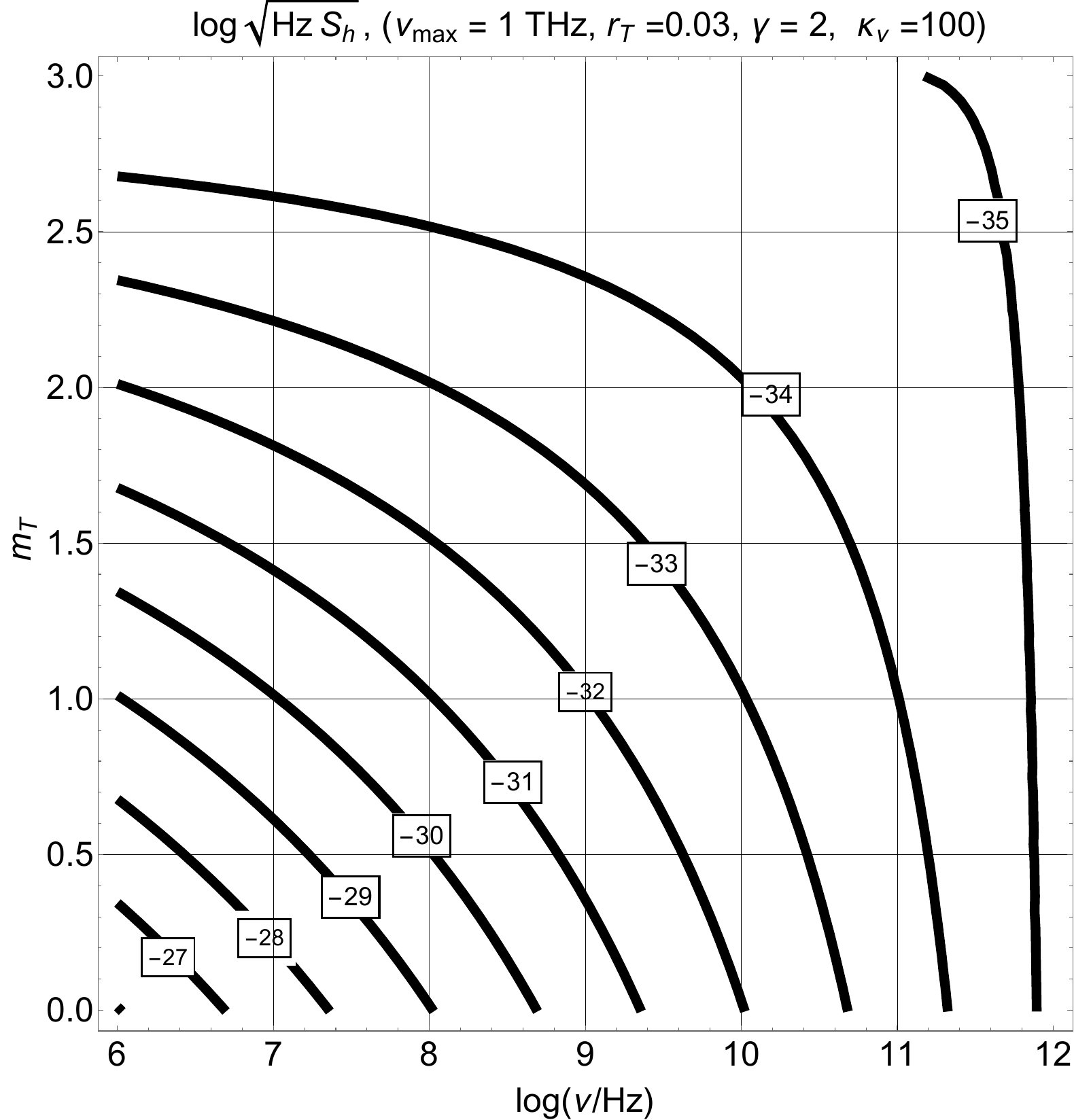}
\includegraphics[height=8cm]{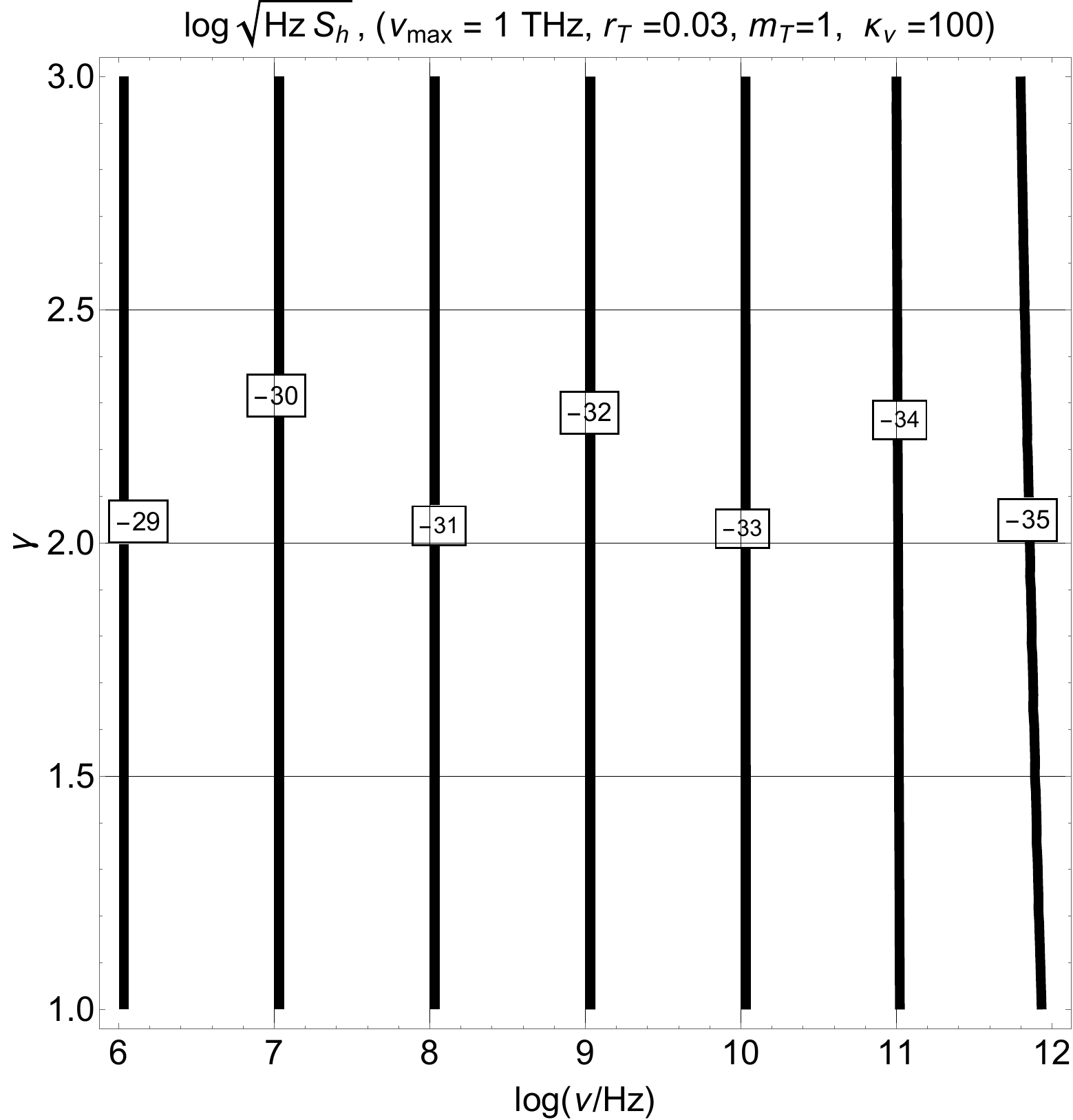}
\caption[a]{The different labels in all the plots correspond to the common logarithms of $\sqrt{S_{h}}$.
Since $S_{h}$ is measured in Hz, the labels give actually the values of $\log{\sqrt{S_{h}\, \mathrm{Hz}}}$, as stressed on top of each plot. The plots illustrated in this figure 
are the counterpart of the ones of Fig. \ref{FIGUREX} with the difference 
that we are now concerned with the spectral amplitude.  As in Fig. \ref{FIGUREX} $\nu_{\mathrm{max}}$ has been fixed 
to $\nu_{\mathrm{max}}= {\mathcal O}(\mathrm{THz})$. In the two upper plots we discuss the variation of the spectral amplitude  
in the planes $[m_{T},\, \log{(\nu/\mathrm{Hz})}]$ and $[\log{\kappa_{\nu}},\, \log{(\nu/\mathrm{Hz})}]$. In the two lower plots $\kappa_{\nu}$ has been fixed to be ${\mathcal O}(100)$.}
\label{FIGUREY}      
\end{figure}

We see from the results of Fig. \ref{FIGUREX} that depending on the parameters the 
minimal chirp amplitude ranges between $10^{-27}$ and $10^{-33}$. These values
have a non-trivial dependence on the frequency, on the spectral index $m_{T}$ and also on $\kappa_{\nu}$.  Once the chirp amplitude is normalized directly in the THz region the bounds on $\nu_{\mathrm{max}}$ derived in Eq. (\ref{GGG13}) can be translated into limits on $h_{c}^{(\mathrm{min})}$:
 \begin{equation}
h^{(\mathrm{min})}_{c}(\nu, \tau_{0}) < 8.13 \kappa_{\nu} \times 10^{-32} \biggl(\frac{\nu}{0.1\, \mathrm{THz}}\biggr)^{-1 +{m}_{T}/2}.
\label{SENS4}
\end{equation}
The results of Fig. \ref{FIGUREX} and of Eq. (\ref{SENS4}) suggest that the minimal sensitivities 
of Eq. (\ref{SENS1}) for a frequency domain encompassing the MHz and GHz regions 
are irrelevant for the potential detection of the high frequency gravitons.  
Equation (\ref{SENS4}) also clarifies why 
$h_{c}^{(\mathrm{min})}$ must be at least ${\mathcal O}(10^{-32})$ (or smaller) for a potential detection of cosmic gravitons in the THz domain.  We also note that for $m_{T} > 2$ the largest signal occurs at the largest frequency, for $m_{T} \leq 2$ the frequencies smaller than the THz are observationally convenient. If we consider, for instance, the case $m_{T} \to 1$ (which is, incidentally, typical of a post-inflationary stiff phase when we neglect here all the possible logarithmic enhancements) we would have that the chirp amplitude at in the MHz range could be ${\mathcal O}(10^{-28})$. When $m_{T} \to 2$ we would have instead that $h_{c}(\nu,\tau_{0})$ is the same at higher and smaller frequencies. For $m_{T} \to 3$ the chirp amplitude at lower frequencies gets even smaller. There is therefore a mutual compensation between the optimal frequency, the features of the signal and the 
noises (especially the thermal one) indicating that the highest possible frequency 
(close to $\nu_{\mathrm{max}}$) is not always the most convenient. Also this aspect should be 
taken into account if the goal is really an accurate assessment of the required sensitivities of high frequency instruments like the ones suggested in Refs. \cite{CAV1,CAV2,CAV3,CAV4,CAV5,CAV6,CAV7,WL1,WL2}.
\begin{figure}[!ht]
\centering
\includegraphics[height=6.2cm]{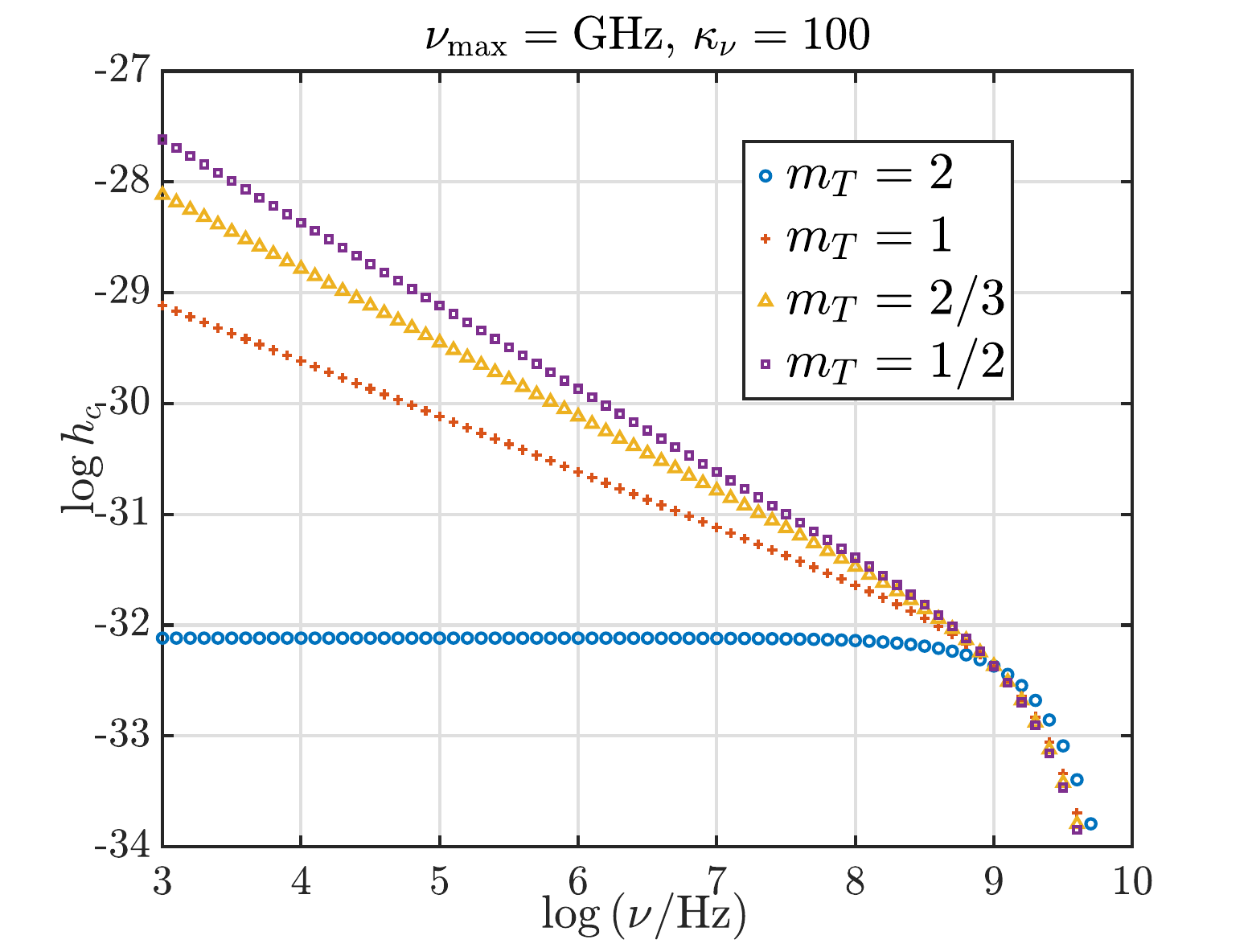}
\includegraphics[height=6.2cm]{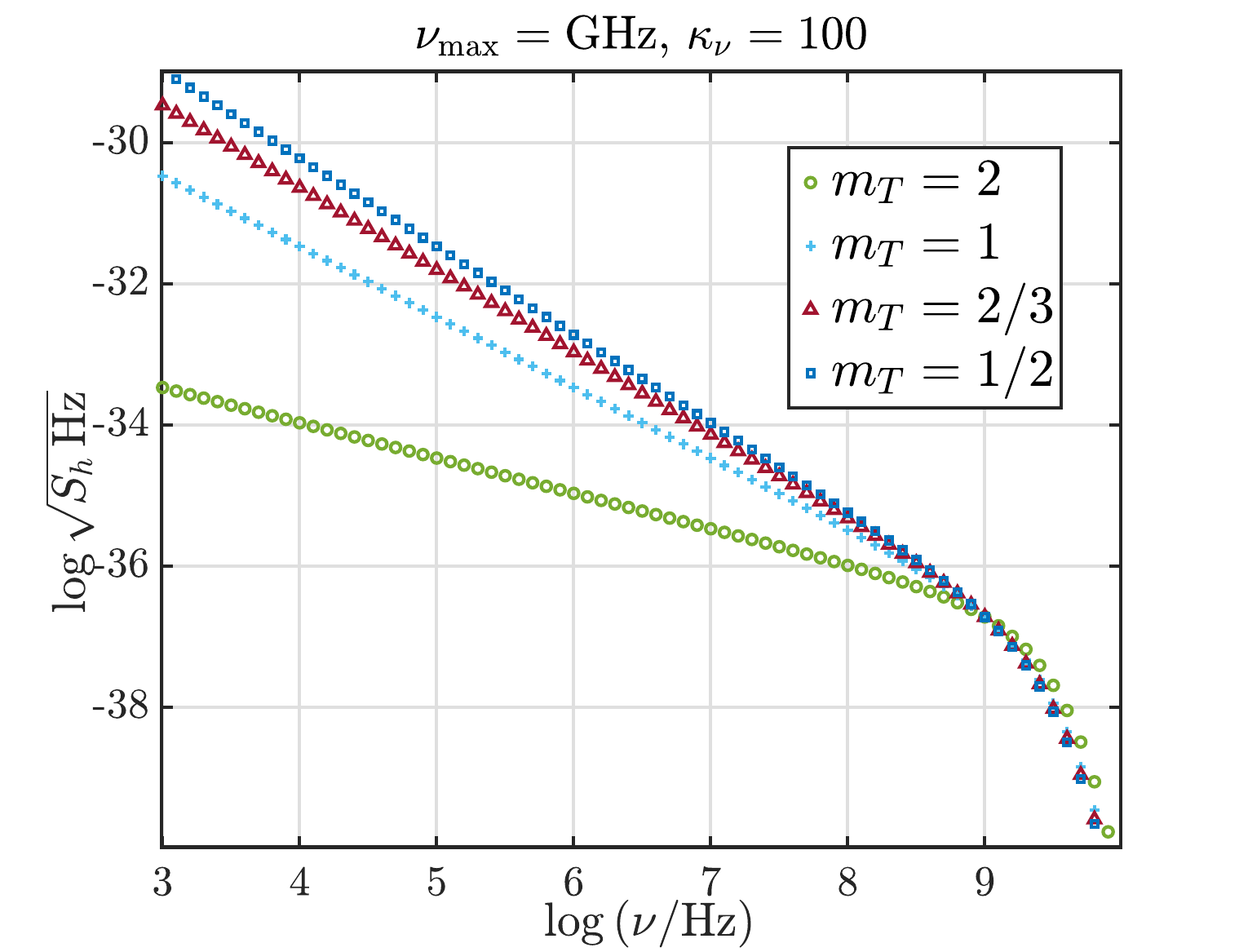}
\caption[a]{In the left plot we illustrate the common logarithm of the chirp amplitude as a function of the comoving frequency. In the right plot we instead illustrate the common logarithm 
of $\sqrt{S_{h} \, \mathrm{Hz}}$. In both plots all the remaining parameters haven been fixed in the same way. As suggested by Figs. \ref{FIGUREX} and \ref{FIGUREY} the minimal detectable spectral amplitude must always be $2$ or $3$ orders of magnitude smaller than the chirp amplitude.}
\label{FIGUREZ}      
\end{figure}

So far we discussed the conditions implied by Eq. (\ref{SENS1}) in term of the chirp amplitude $h_{c}(\nu, \tau_{0})$
and similar conclusions would follow if we would consider the spectral amplitude $\sqrt{S_{h}(\nu,\tau_{0})}$. The equivalence of the two descriptions assumes however that the diffuse backgrounds of relic gravitons are in fact equivalent to a stationary (and homogeneous) random process: only in this case the spectral amplitude and the autocorrelation function of the process 
form a Fourier transform pair. This statement is often referred to as Wiener-Khintchine theorem \cite{STOC1,WK1,WK2} and the possibility of defining a spectral amplitude relies on the stationary nature of the underlying random process. However 
the diffuse backgrounds of relic gravitons are typically non-stationary (see Ref. \cite{AS9} and discussions therein). 

With this caveat, if $S_{h}(\nu,\tau_{0})$ is employed 
(by implicitly enforcing the stationarity of the process) we can write a relation similar to the one of Eq. (\ref{SENS2}):
\begin{equation} 
S_{h}(\nu, \tau_{0}) = 128 \pi \biggl(\frac{\nu}{M_{P}^2}\biggr) \, \overline{n}_{\nu} \Rightarrow \sqrt{S_{h}(\nu,\tau_{0})} = 4.65 \times 10^{-21} \biggl(\frac{\nu}{M_{P}}\biggr)^{1/2} \, \frac{\sqrt{ \overline{n}_{\nu}}}{\sqrt{\mathrm{Hz}}}.
\label{SENS5}
\end{equation}
While $h_{c}(\nu, \tau_{0})$ is dimensionless, $\sqrt{S_{h}(\nu,\tau_{0})}$ is measured 
in units of $1/\sqrt{\mathrm{Hz}}$ so that, eventually, the sensitivities can be expressed in terms
of the dimensionless combination  $\sqrt{S_{h}(\nu,\tau_{0})\, \mathrm{Hz}}$. Once more the results 
of Eq. (\ref{SENS5}) should be contrasted with the sensitivities naively suggested by Eq. (\ref{SENS1}).
We can again recall Eq. (\ref{GGG6}) and rewrite Eq. (\ref{SENS5}) for $\nu < \nu_{\mathrm{max}}$ 
in full analogy with what we already did in the case of the chirp amplitude:
\begin{equation}
\sqrt{S_{h}(\nu,\tau_{0}) \, \mathrm{Hz}} = 4.65 \times 10^{-21} \, \sqrt{\kappa_{\nu}} \, \biggl(\frac{\nu_{\mathrm{max}}}{M_{P}}\biggr)^{1/2}  \, \biggl(\frac{\nu}{\nu_{\mathrm{max}}}\biggr)^{(m_{T} - 3)/2}. 
\label{SENS6}
\end{equation}
From the results of Fig. \ref{FIGURE1} and from the bound deduced 
in Eq. (\ref{GGG13}), Eq. (\ref{SENS6}) suggests that $\sqrt{S_{h}^{(\mathrm{min})}(\nu_{\mathrm{max}},\tau_{0}) \, \mathrm{Hz}} = {\mathcal O}(10^{-37}) \, \sqrt{\kappa_{\nu}}$ 
for $\nu_{\mathrm{max}} = {\mathcal O}(\mathrm{THz})$. For smaller frequencies $\sqrt{S_{h}^{(\mathrm{min})}(\nu,\tau_{0}) \, \mathrm{Hz} }> {\mathcal O}(10^{-37}) \, \sqrt{\kappa_{\nu}}$ when $m_{T}< 3$  while it becomes even smaller when $m_{T} > 3$. Finally for $m_{T} \to 3$ we have that $ \sqrt{S_{h}^{(\mathrm{min})}(\nu,\tau_{0}) }$ 
approximately frequency-independent and always ${\mathcal O}(10^{-37}) \, \sqrt{\kappa_{\nu}}$.  
In the most favourable case (i.e. $m_{T} < 3$) the minimal spectral amplitude can increase in the GHz and MHz regions. For instance, when $m_{T} \to 1$ 
 $\sqrt{S_{h}^{(\mathrm{min})}(\nu,\tau_{0}) \, \mathrm{Hz} } = {\mathcal O}(10^{-31}) \, \sqrt{\kappa_{\nu}}$ when $\nu = {\mathcal O}(\mathrm{MHz})$.
The qualitative arguments based on Eq. (\ref{SENS6}) can be corroborated by a more detailed discussion that is reported in Fig. \ref{FIGUREY}.
In all the four plots the common logarithms of $\sqrt{S_{h}\, \mathrm{Hz}}$ remain constant all along the various curves. From the different slices of the parameter space it follows that $\sqrt{S_{h}^{(\mathrm{min})}(\nu,\tau_{0}) \, \mathrm{Hz} } < {\mathcal O}(10^{-33})$ when the frequencies of the gravitons are larger than the GHz (but always smaller than the THz). Finally, in Fig. \ref{FIGUREZ} we fix $\nu_{\mathrm{max}}$ in the GHz range and compare the values of $h_{c}$ and $\sqrt{S_{h} \, \mathrm{Hz}}$. For different parameters the conclusions remain roughly the same and the minimal detectable square root of the spectral amplitude must always be $2$ or $3$ orders of magnitude smaller than the chirp amplitude computed for the same range of parameters.

\renewcommand{\theequation}{6.\arabic{equation}}
\setcounter{equation}{0}
\section{Concluding remarks}
\label{sec6}
Every variation of the space-time curvature generates bunches of gravitons 
whose statistical properties ultimately reflect the successive stages of the evolution 
of the background geometry. If the multiplicity distributions will be observationally assessed  
in the future, the timeline of the expansion rate will also be at hand in all its relevant details. Since the physical properties of the production process are encoded in the Hamiltonian of the tensor modes of the geometry, the goal has been to derive systematically the multiplicity distributions of the relic gravitons. Although the production probabilities are obtained in full generality by squaring the quantum amplitudes, their quantitive features are affected by the ranges of the comoving frequencies and by the underlying cosmological parameters through the averaged multiplicities, the dispersions and eventually the higher-order factorial moments. This is why two seemingly complementary perspectives have been concurrently explored in this analysis: an indirect approach (where the averaged multiplicities and the dispersions appear as parameters of the distributions) and a more direct viewpoint (where all the statistical quantities depend on the frequency and, ultimately, upon the relevant timelines of the expansion rates). Between these two strategies the former is sometimes more transparent than the latter but they are
equally essential, as stressed throughout this paper.

Thanks to the two interrelated approaches pursued here it is possible to argue, from 
complementary perspectives, that the multiplicity distributions of the relic gravitons are all infinitely divisible. This also means that they can be all obtained as a compound Poisson processes. According to the theory of discrete stochastic processes the total distribution of a random sum of identically distributed random variables follows by compounding the probability generating functions of the random variables and of the sum itself. In the different ranges of the physical parameters the analytical expressions of the multiplicity distributions are controlled by the range of the comoving frequency and encompass the Poisson, Bose-Einstein, Gamma and Pascal probability distributions. For the purposes of a concluding summation the multiplicity distributions can be parametrized by their averaged multiplicity $\overline{n}_{\nu}$ and by their dispersion $D_{\nu}^2$; both quantities actually depend on the comoving frequency and on the post-inflationary timeline. When the gravitons are produced from the vacuum we have that $ D_{\nu}^2 \to (\overline{n}_{\nu}+ \overline{n}_{\nu}^2)$ but in the 
regime $\overline{n}_{\nu} \gg 1$ (which is realized below the maximal frequency) the quadratic term is effectively dominant and $D_{\nu}^2 \to \overline{n}_{\nu}^2$. This is the Bose-Einstein limit where the produced gravitons exhibit a strong degree of super-Poissonian correlation. In the opposite physical regime the gravitons are produced independently and $D_{\nu}^2 \to  \overline{n}_{\nu}$: this is the case where the dispersion is of the order of the averaged multiplicity and the gravitons are produced independently as it happens for standard Poisson process. The intermediate 
physical situations ranging between the Bose-Einstein and the Poisson limits are all described 
by the Pascal distribution. The Gamma distribution is finally recovered in the limit of large averaged multiplicities. 

The shapes and the correlation properties of the multiplicity distributions are determined 
by the frequency range and by the timeline of the post-inflationary expansion rate.
Although these two aspects are closely interlocked, the existence of an absolute bound on the maximal frequency 
of the spectrum simplifies the discussion. Since the largest frequency of the relic gravitons should comply with the absolute bound $ \nu_{\mathrm{max}} \leq 
{\mathcal O}(\mathrm{THz})$, for typical frequencies $\nu = {\mathcal O}(\nu_{\mathrm{max}})$ 
the averaged multiplicity is ${\mathcal O}(1)$ whereas $\overline{n}_{\nu}$ gets exponentially 
suppressed for $\nu \gg \nu_{\mathrm{max}}$. If a Poisson shape would be eventually 
observed close to the THz domain, then we could conclude that the gravitons have been produced from a multiparticle initial state. In the same frequency range we might eventually measure a Bose-Einstein distribution: this would indicate that the gravitons 
have been produced from the vacuum. From the phenomenological viewpoint the most relevant ranges involve however frequencies 
that are smaller than $\nu_{\mathrm{max}}$. The spectrum of relic gravitons notoriously 
extends from the  aHz domain up to the THz: in the intermediate region we expect large averaged multiplicities of the relic gravitons. In the concordance scenario the post-inflationary timeline is dominated 
by radiation and in the audio band the average multiplicity of the relic gravitons is ${\mathcal O}(10^{20})$ for $\nu = {\mathcal O}(\mathrm{kHz})$. If the post-inflationary expansion rate is faster than radiation the spectral energy density is comparatively smaller at higher frequencies while it gets larger if the post-inflationary expansion rate is slower than radiation. In spite of the post-inflationary timeline, for $\nu < \nu_{\mathrm{max}}$ the average multiplicity exceeds the particle content of the initial state so that the Gamma multiplicity distribution is quickly recovered.  A number of illustrative examples of these conclusions have been provided with specific attention to the limiting shapes of the distributions.

For putative instruments reaching spectral amplitudes ${\mathcal O}(10^{-35})/\sqrt{\mathrm{Hz}}$ in the MHz or GHz bands, the Bose-Einstein correlations could be employed as direct probe of the properties of cosmic gravitons and of their statistics. It would be however incorrect to postulate in the GHz or THz the same sensitivities achievable in the kHz (i.e. ${\mathcal O}(10^{-21})/\sqrt{\mathrm{Hz}}$) just to ease the potential detection of a purported high frequency signal of fundamental nature. The minimal sensitivities have in fact a physical rationale that comes from the properties of the potential sources.
It is equally unjustified to use, by fiat,  the diffuse backgrounds as a sound scientific case for 
ultra-high frequency instruments without bothering about the specific physical properties of the relic gravitons. All in all, the features of the multiplicity distributions of the relic gravitons have been deduced on a general ground and corroborated by various explicit examples coming from known physical situations. If ever assessed  they will  bear the mark of the multiparticle final state of graviton production. In this sense their r\^ole could be essential if, in the future, the diffuse backgrounds of relic gravitational radiation will be  detected either at high or at intermediate frequencies. They could be eventually used to disentangle the signals of the relic gravitons from other diffuse foregrounds of astrophysical origin.

\section*{Acknowledgements}
I wish to acknowledge relevant discussions with the late Ph. Bernard, G. Cocconi and E. Picasso on the relevance of microwave cavities for high frequency gravitons. It is also a pleasure to thank A. Gentil-Beccot,  A. Kohls,  L. Pieper, S. Rohr and J. Vigen of the CERN Scientific Information Service for their kind assistance. 
\newpage 

\begin{appendix}
\renewcommand{\theequation}{A.\arabic{equation}}
\setcounter{equation}{0}
\section{Irreducible representations and multiplicity distributions}
\label{APPA}
The Hamiltonian of the relic gravitons derived in Eqs. (\ref{QA1})--(\ref{QA2}) contains implicitly 
the generators of the $SU(1,1)$ Lie algebra whose explicit form is given by:
\begin{equation}
[\widehat{K}_{-} , \widehat{K}_{+}] = 2 \widehat{K}_{0},\qquad [\widehat{K}_{0}, \widehat{K}_{\pm}] = \pm \widehat{K}_{\pm}.
\label{TSQ3}
\end{equation}
In full analogy with the Schwinger observation (customarily employed 
for a direct evaluation of the Wigner matrix elements of the $SU(2)$ group \cite{BIE})
the operators $\widehat{K}_{\pm}$ and $\widehat{K}_{0}$ appearing in Eq. (\ref{TSQ3}) can be represented as quadratic combinations of two  
harmonic oscillators $[\widehat{b}_{+} , \widehat{b}_{-}] = 0$ (with $[\widehat{b}_{+},\widehat{b}_{+}^{\dagger}] =1$ and 
$[\widehat{b}_{-}, \widehat{b}_{-}^{\dagger}] =1$):
\begin{equation}
\widehat{K}_{+} = \widehat{b}_{+}^{\dagger} \,\widehat{b}_{-}^{\dagger},\qquad \widehat{K}_{-} = \widehat{b}_{-} \,\widehat{b}_{+}, \qquad \widehat{K}_{0} = [ \widehat{b}_{+}^{\dagger} \, \widehat{b}_{+} + \widehat{b}_{-} \widehat{b}_{-}^{\dagger}]/2.
\label{TSQ4}
\end{equation}
By comparing Eqs. (\ref{QA2}) and (\ref{TSQ4}), the connection between the $SU(1,1)$ generators 
and the Hamiltonian dynamics is immediate. This why we could also expect 
a direct relation between the Wigner matrix elements of the $SU(1,1)$ group and the class 
of infinitely divisible distributions. To clarify this statement we start by noting that a standard basis  
for the irreducible representations of $SU(1,1)$ is given by the states that diagonalize simultaneously  
of $\widehat{K}_{0}$ and the Casimir operator\footnote{In Eq. (\ref{TSQ5}),
the two forms of the Casimir follow from the definition of the ladder operators $\widehat{K}_{\pm} = \widehat{K}_{1}\pm i \widehat{K}_{2}$ and from the first of the two commutation relations of Eq. (\ref{TSQ3}).}
\begin{equation}
\widehat{K}^2 = \widehat{K}_{0}^2 - \widehat{K}_{1}^2 - \widehat{K}_{2}^2 =  \widehat{K}_{0}( \widehat{K}_{0} -1) - \widehat{K}_{+} \, \widehat{K}_{-}, 
\label{TSQ5}
\end{equation}
that commutes, by definition, with all the group generators (and therefore, in particular, with $\widehat{K}_{0}$). In the conventional Bargmann notation \cite{BARG}, the states  $|k;\,m\rangle$  are the standard basis of the irreducible representations $T^{+k}$ of $SU(1,1)$ where $k$ is the principal quantum number (i.e. the eigenvalue of the Casimir operator, not to be confused here with the modulus of the three-momentum) and $m$ is the magnetic quantum number (i.e. the eigenvalue of $\widehat{K}_{0}$). To relate the Wigner matrix elements and the multiplicity distributions investigated in this paper the most convenient 
basis for the irreducible representations is however provided by
\begin{equation}
| n_{+};\, n_{-} \rangle = \frac{(\widehat{b}^{\,\dagger}_{+})^{n_{+}}}{\sqrt{n_{+}!}} \, \frac{(\widehat{b}^{\dagger}_{-})^{\,n_{-}}}{\sqrt{n_{-}!}} | 0_{+};\,0_{-}\rangle.
\label{REP1}
\end{equation}
Since, by definition, $[\widehat{n}_{+}, \widehat{n}_{-}] =0$, the states (\ref{REP1}) diagonalize simultaneously $\widehat{n}_{+} = \widehat{b}_{+}^{\dagger} \, \widehat{b}_{+}$ and $\widehat{n}_{-} = \widehat{b}_{-}^{\dagger} \, \widehat{b}_{-}$. From the Eq. (\ref{REP1}) we can introduce a further basis $|Q,\, n\rangle$ where 
$Q$ and $n$ are, respectively,  the eigenvalues of $\widehat{Q}= \widehat{n}_{+} - \widehat{n}_{-}$ and of $\widehat{n} = \widehat{n}_{+} + \widehat{n}_{-}$. This means that the states $|n_{+},\, n_{-}\rangle$ form a basis for the irreducible representations $T^{k}$ when $k = (1 + Q)/2$ is fixed; the connection between the three different sets of eigenvalues is given by: 
\begin{equation}
n = n_{+} + n_{-} = 2 m -1, \qquad Q = n_{+} - n_{-} = 2 k -1. 
\label{REP2}
\end{equation}
Equation (\ref{REP2}) can also be inverted and this step implies that  $n_{+} = (Q + N)/2 = k + m -1$ and that $n_{-} = (N - Q)/2 = m - k$. The negative series \cite{BARG} $T^{-k}$ is symmetric under the exchange $n_{+} \to n_{-}$ while the principal (continuous) series does not play a specific role in the present considerations. The Wigner matrix 
element of the positive discrete series is therefore defined as
\begin{equation}
T^{+k}_{m\,m^{\prime}}(z) = \langle m^{\prime};\, k| \widehat{\Sigma}(z) | k;\, \, m\rangle,
\label{REP4}
\end{equation}
where $\widehat{\Sigma}(z)$ is the analog of the operators appearing in Eqs. (\ref{SD1})--(\ref{SD3}) and 
it can also be expressed in a factorized form \cite{BKH1,BKH2,BKH3}, modulo some secondary differences involving the phases:
\begin{equation}
\widehat{\Sigma}(z) = e^{ -\frac{z}{|z|} \tanh{|z|} \widehat{K}_{+}} \times e^{ - 2 \ln{( \cosh{|z|})} \, \widehat{K}_{0} } \times e^{ \frac{z^{*}}{|z|} \tanh{|z|} \widehat{K}_{-}},
\label{TSQ6}
\end{equation}
where $z = r\, e^{i \vartheta}$. In the basis of Eq. (\ref{REP1}) we can first evaluate 
\begin{eqnarray}
\langle n_{-}';\, n_{+}' | \widehat{\Sigma}(z) | n_{+};\, n_{-} \rangle &=&  \frac{1}{\cosh{r}^{n_{+}+ n_{-} + 1}}
\sum_{\ell =0}^{\infty} \sum_{j=0}^{\infty} \sqrt{ n_{+}!}\, \sqrt{ n_{-}!} \frac{ \sqrt{(n_{+} - \ell + j)!}  \sqrt{(n_{-} - \ell + j)!}}{\ell!\, j!\, (n_{+} -\ell)!\, (n_{-} - \ell)!}
\nonumber\\
&\times& ( - e^{ i \vartheta} \tanh{r})^{j} \,(  e^{ -i \vartheta} \tanh{r})^{\ell} \,\, \delta_{n_{+}', n_{+}  -\ell + j} 
\,\,  \delta_{n_{-}', n_{-}  -\ell + j}.
\label{REP13}
\end{eqnarray}
To obtain the matrix element of Eq. (\ref{REP4}) the eigenvalues must be redefined as 
$n_{+} = k + m -1$, $n_{-} = m - k$ and as $n_{+}' = k + m' -1$, $n_{-}'  = m' - k$.
From this dictionary between the different irreducible representations, the two delta functions of Eq. (\ref{REP13}) lead actually to the same condition, i.e. $\delta_{n_{+}',\,\, n_{+}  -\ell + j} 
\,\,  \delta_{n_{-}',\,\, n_{-}  -\ell + j} = \delta_{j,\,\,\ell + (m' - m)}$ which means that the only contribution to the sum over $j$ comes from $ j= \ell + (m' -m)$. 
The explicit form of the matrix element then becomes
\begin{eqnarray}
T_{m\,\,m'}^{+k}(z) &=& \frac{1}{\cosh{r}^{2 m}} \sum_{\ell=0}^{\infty} \sum_{j=0}^{\infty} 
 ( - e^{ i \vartheta} \tanh{r})^{j} \,(  e^{ -i \vartheta} \tanh{r})^{\ell}  (\cosh{r})^{2 \ell}\,\, \delta_{j,\,\, \ell + (m' - m)} 
 \nonumber\\
 &\times& \sqrt{ ( k + m -1)!} \, \sqrt{( m - k)!} \frac{\sqrt{ ( k + m -1 - \ell + j)! \,\,  ( m - k  - \ell + j)!}}{\ell !\,\, j!
  ( k + m - 1 -\ell)!\,\, ( m - k - \ell)!}
  \nonumber\\ 
 &=&   \frac{\sqrt{ \Gamma(k + m') !\,\, \Gamma(m' - k+1)}}{\sqrt{ \Gamma(k + m)\,\, \Gamma(m - k+1)}} 
\frac{( - e^{ i \vartheta} \tanh{r})^{m' - m} }{ (m' - m)!\,\, (\cosh{r})^{2 m}}  \times 
\nonumber\\
&\times&  \sum_{\ell =0}^{\infty} 
\,( - \sinh^2{r})^{\ell} \frac{ ( m + k -1)!\, (m - k)!\,\, (m' - m)!}{\ell!\,\, (\ell + m' - m)!\,\, (k + m - 1- \ell)!\,\,( m- k -\ell)!}.
 \label{REP16}
 \end{eqnarray}
 After reshuffling the various Euler Gammas  in the three Prochhammer symbols 
defining the general expression of the hypergeometric function\footnote{In the second expression of Eq. (\ref{REP16}) we introduced the Gamma functions 
 (i.e. $ y! = \Gamma(y +1)$); furthermore the sum appearing in Eq. (\ref{REP16}) is actually 
 given by a suitable hypergeometric function $F(\alpha,\beta ; \gamma; z) = \sum_{\ell=0}^{\infty} (\alpha)_{\ell} (\beta)_{\ell}/[\ell!\,\, (\gamma)_{\ell}] z^{\ell}$, 
where, as usual,  the Prochhammer symbols are $(\alpha)_{\ell} = \Gamma(\alpha + \ell)/\Gamma(\alpha)$ \cite{abr1,abr2} (and 
similarly for $(\beta)_{\ell}$ and  $(\gamma)_{\ell}$).},
the final form of $T_{m\,\,m'}^{+k}(z)$ becomes
\begin{eqnarray}
T_{m\,\,m'}^{+k}(z) &=&  \frac{\sqrt{ \Gamma(k + m') \,\, \Gamma(m' - k+1)}}{\sqrt{ \Gamma(k + m)\,\, \Gamma(m - k+1)}} 
\frac{( - e^{ i \vartheta} \tanh{r})^{m' - m} }{ (m' - m)!\,\, (\cosh{r})^{2 m}} 
\nonumber\\
&\times& F[ 1 - m - k,\, k - m;\, m' - m +1; -\sinh^2{r}],
\label{REP22}
\end{eqnarray}
where $k = 1/2,\, 1,\, 3/2\, .\,.\,.$ is the principal quantum number of the series whereas $m$ and $m^{\prime}$ are the 
magnetic quantum numbers; both $m$ and $m^{\prime}$ take the values $k, \, k+1,\, k+2\,.\,.\,.$.
From Eq. (\ref{REP22}) we therefore have that
\begin{eqnarray}
&& \biggl| T^{1/2}_{1/2\,\, n+ 1/2}\biggr|^2 = \frac{1}{\overline{n} + 1} \biggl(\frac{\overline{n}}{\overline{n} +1}\biggr)^{n} \equiv P_{n}^{(\mathrm{BE})}(\overline{n}) , \qquad 
\sinh^2{r}= \overline{n},
\label{REP23}\\
&&  \biggl| T^{\ell/2}_{\ell/2\,\, n +\ell/2}\biggr|^2 = \frac{\Gamma(\ell + n)}{\Gamma(n+1) \Gamma(\ell)} \biggl(\frac{\overline{n}}{\overline{n} +\ell}\biggr)^{n} \biggl(\frac{\ell}{\ell + \overline{n}}\biggr)^{\ell}\equiv P_{n}^{(\mathrm{PA})}(\overline{n},\ell), \qquad \sinh^2{r} = \frac{\overline{n}}{\ell}, 
\label{REP24}
\end{eqnarray}
where $P_{n}^{(\mathrm{BE})}(\overline{n})$ and $P_{n}^{(\mathrm{PA})}(\overline{n},\ell)$ denote, respectively, the 
Bose-Einstein and the Pascal distributions. Equations (\ref{REP22}) and (\ref{REP23})--(\ref{REP24}) demonstrate the explicit connection of the (infinitely divisible) multiplicity distributions with the discrete representations of $SU(1,1)$. Further considerations 
along this line are per se interesting but will not be pursued here.

\renewcommand{\theequation}{B.\arabic{equation}}
\setcounter{equation}{0}
\section{Estimates of the averaged multiplicity for $\nu \ll\nu_{\mathrm{max}}$}
\label{APPB}
For an estimate of the averaged multiplicity in the regime  $\nu \ll\nu_{\mathrm{max}}$ we first need to solve 
Eqs. (\ref{FFF1})--(\ref{FFF1a}) and this can be done either exactly or approximately.
The different approaches eventually agree and for the present ends the Wentzel–Kramers–Brillouin (WKB) approximation is the most convenient. Since the $\oplus$ and $\otimes$ polarizations evolve in the same manner (see however \cite{MGPOL} for different situations), the corresponding index can be suppressed (i.e. $u_{k,\,\alpha}(\tau) =u_{k}(\tau)$ and  $v_{k,\,\alpha}(\tau) =v_{k}(\tau)$). The WKB solutions depend, in this case, on the time scales $\tau_{ex}(k)$ and $\tau_{re}(k)$ 
corresponding, respectively, to the moments where a given scale either exits and to the reenters the Hubble radius.  After normalizing the solution to a plane wave in the limit $\tau \to -\infty$ (corresponding to the inflationary limit) Eq. (\ref{FFF1}) 
can be solved as:
\begin{equation}
u_{k}(\tau) - v_{k}^{\ast}(\tau) = e^{- i k \tau_{ex}} \biggl[ A_{k}(\tau_{ex},\tau_{re}) \, \cos{(k \,\Delta\tau)} + B_{k}(\tau_{ex},\tau_{re}) \, \sin{(k \,\Delta\tau)} \biggr],
\label{APPB1}
\end{equation}
where $\Delta \tau = (\tau- \tau_{re})$ while $A_{k}(\tau_{ex},\tau_{re})$ and $B_{k}(\tau_{ex},\tau_{re})$ are two complex 
functions defined as  
\begin{eqnarray}
A_{k}(\tau_{ex},\tau_{re}) &=& \biggl(\frac{a_{re}}{a_{ex}}\biggr) \, Q_{k}(\tau_{ex}, \tau_{re}),
\label{APPB2}\\
B_{k}(\tau_{ex}, \tau_{re}) &=& \biggl(\frac{{\mathcal H}_{re}}{k} \biggr) \,  \biggl(\frac{a_{re}}{a_{ex}}\biggr) \, Q_{k}(\tau_{ex}, \tau_{re})
- \biggl(\frac{a_{ex}}{a_{re}}\biggr) \biggl(\frac{{\mathcal H}_{ex} + i k}{k}\biggr).
\label{APPB3}
\end{eqnarray}
where ${\mathcal H}_{ex} = {\mathcal H}(\tau_{ex}) = a_{ex} \, H_{ex}$, 
and ${\mathcal H}_{re} = {\mathcal H}(\tau_{re}) = a_{re} \, H_{re}$. In Eqs. (\ref{APPB2})--(\ref{APPB3}) there also appear the complex function $Q_{k}(\tau_{ex}, \tau_{re})$ which is defined in terms of an integral over the conformal time coordinate 
\begin{equation}
Q_{k}(\tau_{ex}, \tau_{re}) = 1 - ({\mathcal H}_{ex} + i k)\, {\mathcal W}(\tau_{ex}, \tau_{re}), \qquad {\mathcal W}(\tau_{ex}, \tau_{re})= \int_{\tau_{ex}(k)}^{\tau_{re}(k)} \frac{a_{ex}^2}{a^2(\tau^{\prime})} \,\, d \tau^{\prime},
\label{APPB4}
\end{equation}
and that solves Eq. (\ref{FFF1}) in the limit $k^2 \ll ({\mathcal H}^2 + {\mathcal H}^{\prime})$. If Eq. (\ref{APPB1}) is now inserted into Eq. (\ref{FFF1a}) we can obtain the explicit form of the combination $u_{k}(\tau) + v_{k}^{\ast}(\tau)$
\begin{equation}
u_{k}(\tau) + v_{k}^{\ast}(\tau) = i\,e^{- i k \tau_{ex}}\, \sqrt{\frac{k}{2}}\, \biggl[ - A_{k}(\tau_{ex},\tau_{re}) \, \sin{(k \,\Delta\tau)} + B_{k}(\tau_{ex},\tau_{re}) \, \cos{(k \,\Delta\tau)} \biggr].
\label{APPB5}
\end{equation}
Putting together Eqs. (\ref{APPB1}) and (\ref{APPB5}) the explicit expression of $u_{k}(\tau) $ and $v_{k}^{\ast}(\tau)$ becomes:
\begin{eqnarray}
u_{k}(\tau) =  \frac{1}{2} \bigl[A_{k}(\tau_{ex},\tau_{re}) + i\, B_{k}(\tau_{ex},\tau_{re})\bigr] e^{- i \, k (\Delta \tau + \tau_{ex})},
\label{APPB6}\\
v_{k}^{\ast}(\tau) = - \frac{1}{2}  \bigl[A_{k}(\tau_{ex},\tau_{re}) - i\, B_{k}(\tau_{ex},\tau_{re})\bigr] e^{i \, k (\Delta \tau + \tau_{ex})}.
\label{APPB7}
\end{eqnarray}
From Eqs. (\ref{APPB6})--(\ref{APPB7}) and  Eqs. (\ref{APPB2})--(\ref{APPB3}) it follows that $|u_{k} (\tau) |^2 - |v_{k}(\tau)|^2 =1$.
The values of $\tau_{ex}(k)$ and $\tau_{re}(k)$ coincide with the turning points of the WKB solution; both turning points 
must approximately correspond to $ k^2 \simeq a^2\, H^2 [ 2 - \epsilon(\tau)]$ where $\epsilon = - \dot{H}/H^2$ is the slow-roll parameter. When $\epsilon \neq 2$ the previous condition implies  $k \tau_{re} ={\mathcal O}(1)$ and $k \tau_{ex} = {\mathcal O}(1)$. However, 
if the reentry takes place during radiation, we would have instead that $\epsilon_{re}\to 2$ in the vicinity of the turning point: this would mean that $k \tau_{re}  \simeq \sqrt{| 2 - \epsilon_{re}|} \ll 1$. Besides the properties of the turning points, for a direct evaluation of Eqs. (\ref{APPB6})--(\ref{APPB7}) the second observation is that $(a_{re}/a_{ex}) \gg 1$ provided the background expands between $a_{ex}$ and $a_{re}$. If we would simply keep all the terms $(a_{re}/a_{ex}) \gg 1$ and neglect the ones containing powers of $(a_{ex}/a_{re})$ the resulting expressions of $u_{k}(\tau)$ and $v_{k}(\tau)$ would violate unitarity. So the best strategy is to first expand $u_{k}(\tau)$ and $v_{k}(\tau)$ in powers of $(a_{ex}/a_{re}) < 1$ by keeping both the leading and the subleading terms in such a way that the unitarity condition $|u_{k} (\tau) |^2 - |v_{k}(\tau)|^2 =1$ is always enforced. To account for the potentially different nature of the turning points we introduce the dimesnionless ratios $q_{ex} = a_{ex} \, H_{ex}/k$ and $q_{re} = a_{re} \, H_{re}/k$: if the turning point is regular both at the exit and at the reentry we have $q_{ex} \simeq q_{re} = {\mathcal O}(1)$. Conversely, in case the reentry occurs during radiation we would have $q_{re} \gg 1$. With these notations Eqs. (\ref{APPB6})--(\ref{APPB7})
\begin{eqnarray}
u_{k}(\tau_{ex},\tau_{re}) = \frac{e^{- i k\tau}}{2} \biggl[ i +  q_{ex}(1 - i) \,\,  I(\tau_{ex}, \tau_{re}) \, \biggr] (q_{re} - \, i) \biggl(\frac{a_{re}}{a_{ex}}\biggr) + ( 1 - i \, q_{ex}) \biggl(\frac{a_{ex}}{a_{re}}\biggr)+ {\mathcal O}\biggl[ \biggl(\frac{a_{ex}}{a_{re}}\biggr)^{5}\biggr],
\nonumber\\
v_{k}(\tau_{ex},\tau_{re}) = \frac{e^{- i k \tau}}{2} \biggl[ - 1 +  (q_{ex} - i) \,\,I(\tau_{ex}, \tau_{re}) \biggr] (q_{re} - \, i) \biggl(\frac{a_{re}}{a_{ex}}\biggr) + ( 1 + i \, q_{ex}) \biggl(\frac{a_{ex}}{a_{re}}\biggr)+ {\mathcal O}\biggl[ \biggl(\frac{a_{ex}}{a_{re}}\biggr)^{5}\biggr].
\label{APPB8}
\end{eqnarray}
The approximate expressions of Eq. (\ref{APPB8}) imply $|u_{k}(\tau_{ex},\tau_{re})|^2 - |v_{k}(\tau_{ex},\tau_{re})|^2=1$ within the accuracy of the expansion. Unitarity is not lost while keeping the leading terms of the expansion and by taking approximately into account the structure of the turning points we may simplify the expressions even further and write
\begin{equation}
u_{k}(\tau_{ex},\tau_{re}) \,\simeq\,  \frac{1}{2}\biggl[ \biggl(\frac{a_{re}}{a_{ex}}\biggr) + \biggl(\frac{a_{ex}}{a_{re}}\biggr)\biggl], \qquad 
v_{k}(\tau_{ex},\tau_{re}) \,\simeq\, \frac{1}{2}\biggl[ \biggl(\frac{a_{ex}}{a_{re}}\biggr) - \biggl(\frac{a_{re}}{a_{ex}}\biggr)\biggl],
\label{APPB9}
\end{equation}
where, again, we can verify that $|u_{k}(\tau_{ex},\tau_{re})|^2 - |v_{k}(\tau_{ex},\tau_{re})|^2 =1$. Depending on the specific estimate 
either Eq. (\ref{APPB8}) or Eq. (\ref{APPB9}) can be employed but what matters for the present discussion 
are the scaling properties of the averaged multiplicity as a function of the comoving frequency. This quantity is  
controlled by $(a_{re}/a_{ex})^2$ and if the given wavelength 
reenters during a decelerated expanding stage with $a(\tau) \propto (\tau/\tau_{1})^{\delta}$ 
\begin{equation}
|v_{k}(\tau_{ex}, \tau_{re})|^2 \simeq \biggl(\frac{a_{re}}{a_{ex}}\biggr)^2 = \frac{[(\tau_{re}/\tau_{1} +1)/r +1]^{2 \delta}}{(\tau_{ex}/\tau_{1})^{- 2/(1 - \epsilon)}} = r^{-2 \delta} \, \biggl(\frac{k}{a_{1} \, H_{1}}\biggr)^{-4 + m_{T}},
\label{APPB10}
\end{equation}
where $r = (1 - \epsilon) \delta$ is a numerical factor required by the continuity of the scale factor across $\tau = - \tau_{1}$ marking the final stage of the inflationary epoch. The 
exponent $m_{T}$ appearing in Eq. (\ref{APPB10}) is given by:
\begin{equation}
m_{T} = \frac{2 - 4 \epsilon}{1 - \epsilon} - 2 \delta \simeq \frac{32 - 4\, r_{T}}{16 - r_{T}} - 2 \delta \simeq (2 - 2 \delta) + {\mathcal O}(r_{T}).
\label{APPB11}
\end{equation}
The second approximate equality in Eq. (\ref{APPB11}) follows from the consistency condition while the third approximate 
equality is the result of an expansions in powers of $r_{T} \ll 1$. Strictly speaking the results of Eqs. (\ref{APPB10})--(\ref{APPB11}) 
are derived in the limit $\delta \neq 1$. However they also hold, with some modifications, in the case $\delta \to 1$, corresponding to a post-inflationary radiation stage where $\epsilon_{re} \to 2$. To express $k/(a_{1}\, H_{1})$ in terms of the comoving frequency we must recall that
\begin{equation}
\frac{k}{a_{1} H_{1}} = \frac{k}{a_{0} \, H_{0}} \biggl(\frac{a_{0} \, H_{0}}{a_{eq} \, H_{eq}}\biggr) \biggl(\frac{a_{eq} \, H_{eq}}{a_{r} \, H_{r}}\biggr) \biggl(\frac{a_{r} \, H_{r}}{a_{1} \, H_{1}}\biggr).
\label{APPB12}
\end{equation}
Since $k = 2 \pi \nu$ we also have that $k/(a_{1}\, H_{1}) = (\nu/\nu_{\mathrm{max}})$ where 
\begin{equation}
\nu_{\mathrm{max}} = \frac{\sqrt{H_{0} \, M_{P}}}{2 \pi} \, (2 \, \Omega_{R\,0})^{1/4} \sqrt{\frac{H_{1}}{M_{P}}}
\biggl(\frac{g_{\rho,\, r}}{g_{\rho,\,eq}}\biggr)^{1/4} \biggl(\frac{g_{s,\, eq}}{g_{s,r}}\biggr)^{1/3} 
\biggl(\frac{H_{r}}{H_{1}}\biggr)^{\frac{\delta -1}{2(\delta +1)}}.
\label{APPB13}
\end{equation}
As already mentioned in Eqs. (\ref{FFF1c})--(\ref{FFF1h}) $g_{\rho}$ and $g_{s}$ indicate, respectively, the number of relativistic degrees of freedom in the 
energy and in the entropy densities. It can be checked from Eq. (\ref{APPB13}) that $\nu_{\mathrm{max}}= (H_{r}/H_{1})^{(\delta -1)/[2(\delta +1)]} \, \overline{\nu}_{\mathrm{max}}$ where $\overline{\nu}_{\mathrm{max}}$ has been already introduced in Eq. (\ref{FFF1c}). We than have from Eq. (\ref{APPB10}) that $|v_{\nu}(\tau_{ex}, \tau_{re})|^2 = (\nu/\nu_{\mathrm{max}})^{-4 + m_{T}}$; 
it also follows from Eqs. (\ref{APPB12})--(\ref{APPB13}) that the averaged multiplicity for $\nu< \nu_{\mathrm{max}}$ is correctly estimated from Eqs. (\ref{FFF1d})--(\ref{FFF1g}) and (\ref{GGG6}).
\end{appendix}

\end{document}